\documentclass[useAMS,usenatbib]{mn2e}
\newcommand{\beq}{\begin{equation}}
\newcommand{\eeq}{\end{equation}}
\newcommand{\bea}{\begin{eqnarray}}
\newcommand{\eea}{\end{eqnarray}}

\newcommand{\bfA}{\mbox{\boldmath $A$}}
\newcommand{\bfssC}{\mbox{\boldmath $\mathsf{C}$}}
\newcommand{\bfssD}{\mbox{\boldmath $\mathsf{D}$}}
\newcommand{\bfF}{\mbox{\boldmath $F$}}
\newcommand{\bfG}{\mbox{\boldmath $G$}}
\newcommand{\bfssM}{\mbox{\boldmath $\mathsf{M}$}}
\newcommand{\bfssI}{\mbox{\boldmath $\mathsf{I}$}}
\newcommand{\bfssO}{\mbox{\boldmath $\mathsf{O}$}}
\newcommand{\bfssP}{\mbox{\boldmath $\mathsf{P}$}}
\newcommand{\bfssQ}{\mbox{\boldmath $\mathsf{Q}$}}
\newcommand{\bfssR}{\mbox{\boldmath $\mathsf{R}$}}
\newcommand{\bfS}{\mbox{\boldmath $S$}}
\newcommand{\bfssW}{\mbox{\boldmath $\mathsf{W}$}}
\newcommand{\bfssL}{\mbox{\boldmath $\mathsf{L}$}}
\newcommand{\bfL}{\mbox{\boldmath $L$}}

\newcommand{\bfe}{\mbox{\boldmath $e$}}
\newcommand{\bff}{\mbox{\boldmath $f$}}
\newcommand{\bfh}{\mbox{\boldmath $h$}}
\newcommand{\bfr}{\mbox{\boldmath $r$}}
\newcommand{\bfs}{\mbox{\boldmath $s$}}
\newcommand{\bft}{\mbox{\boldmath $t$}}
\newcommand{\bfu}{\mbox{\boldmath $u$}}
\newcommand{\bfv}{\mbox{\boldmath $v$}}
\newcommand{\bfw}{\mbox{\boldmath $w$}}
\newcommand{\bfx}{\mbox{\boldmath $x$}}
\newcommand{\bfy}{\mbox{\boldmath $y$}}
\newcommand{\bfz}{\mbox{\boldmath $z$}}
\newcommand{\del}{\mbox{\boldmath $\nabla$}}
\newcommand{\cross}{\mbox{\boldmath $\times$}}
\newcommand{\bcdot}{\mbox{\boldmath $\cdot$}}

\newcommand{\pc}{\mbox{\,pc}} 
\newcommand{\half}{\textstyle{1\over 2}}
\newcommand{\third}{\textstyle{1\over 3}}
\newcommand{\Order}{{O}}
\def\ffrac#1#2{{\textstyle\frac{#1}{#2}}}

\newenvironment{narrow}[2]{%
\begin{list}{}{%
    \setlength{\topsep}{0pt}%
    \setlength{\leftmargin}{#1}%
    \setlength{\rightmargin}{#2}%
    \setlength{\listparindent}{\parindent}%
    \setlength{\itemindent}{\parindent}%
    \setlength{\parsep}{\parskip}}%
  \item[]}{\end{list}}
\usepackage{epsfig}
\usepackage{psfig}
\title[Gauss's Method, Softened]{Gauss's Method for Secular Dynamics, Softened}
\author[J.\ Touma et al.]{J.\ R.\ Touma$^{1}$\thanks{E-mail:
jihad.touma@gmail.com}, S.\ Tremaine$^{2}$\thanks{E-mail:tremaine@ias.edu}, and
M.\ V.\ Kazandjian$^{1}$\thanks{E-mail:mvk00@aub.edu.lb}\\
$^{1}$Department of Physics, American University of Beirut, PO Box 11-0236,
Riad El-Solh, Beirut 11097 2020, Lebanon\\
$^{2}$School of Natural Sciences, Institute for Advanced Study, Einstein
Drive, Princeton, NJ 08540, USA}
\onecolumn
\begin{document}

\pubyear{2008}

\maketitle

\label{firstpage}

\begin{abstract}
\noindent
We show that the algorithm proposed by Gauss to compute the secular evolution
of gravitationally interacting Keplerian rings extends naturally to softened
gravitational interactions. The resulting tool is ideal for the study of the
secular dynamical evolution of nearly Keplerian systems such as stellar
clusters surrounding black holes in galactic nuclei, cometary clouds, or
planetesimal discs. We illustrate its accuracy, efficiency and versatility on
a variety of configurations. In particular, we examine a secularly unstable
unstable system of counter-rotating disks, and follow the unfolding and
saturation of the instability into a global, uniformly precessing, lopsided
($m=1$) mode.
\end{abstract}

\begin{keywords}
galaxies: kinematics and dynamics -- galaxies: nuclei -- stellar dynamics --
celestial mechanics 
\end{keywords}

\section{Introduction}

\noindent
The theory of secular variations of the planetary orbits was
introduced by Lagrange over two centuries ago. This approach to
celestial mechanics is based on the observation that the orientation
of eccentric orbits in a Keplerian potential is fixed: in a coordinate
representation, the argument of periapsis $\omega$ and the longitude
of the node $\Omega$ are stationary. Because of these conserved
quantities, the time-averaged torque between two planets on fixed
orbits is generally non-zero if the orbits are both eccentric and/or
mutually inclined. Thus, if the typical planetary mass is $m$ and the
mass of the host star is $M_\star$, the inter-planetary gravitational
forces lead to both short-term oscillations, having periods of order
the orbital period $P$ and fractional amplitudes of order $m/M_\star$,
and long-term or secular oscillations, having periods of order
$(M_\star/m)P$, fractional amplitudes in eccentricity and inclinations
of order unity, and constant semi-major axis.  The large amplitude of
the secular oscillations means that in most cases they dominate the
evolution of the planetary system.

In the solar system, secular variations are usually computed as power series
in the eccentricities $e$, inclinations $i$, and mass ratios $m/M_\star$ of
the planets, which are all small. Current secular theories employ differential
equations of motion containing $\sim 10^5$ terms up to order
$(m/M_\star)^2$, $e^5$, and $i^5$ \citep{La1986}.

Expansions of the equations of motion in powers of eccentricity and
inclination work well for the orbits of planets in the solar system
(median eccentricity and inclination 0.05 and $1.8^{\circ}$), but not so well
for orbits with larger eccentricities and inclinations, such as those
of asteroids (median eccentricity and inclination $0.14$ and
$7^{\circ}$), extrasolar planets (median eccentricity 0.2), the stars
that surround the central black holes found in many galaxies, and
planets or binary stars executing Kozai oscillations excited by a
companion star \citep{Inn1997,Ma1997,T1997,WM2003,FT2007}.

The primary goal of this paper is to develop numerical methods for secular
dynamics that are valid for large eccentricities and inclinations. In so doing
we shall restrict ourselves to equations of motion that are first order in the
ratio $m/M_\star$. This approximation does not work well in the solar system,
mostly because of near-resonances between the outer planets (e.g., the `great
inequality' between Jupiter and Saturn; see also \citealt{MH1999}), but should
be more useful in systems where the mass ratio $m/M_\star$ is smaller
and/or we are seeking a qualitative understanding of the secular evolution
rather than an accurate ephemeris.

Although the results we describe in this paper are applicable to any
near-Keplerian system, for conciseness we shall refer to the central body as a
`star' and the satellites in orbit around it as `planets', except in
\S\ref{sec:test} where we examine the performance of the algorithm with
parameters appropriate for stars orbiting a black hole.  To first order in the
planetary masses, the equations of motion for secular dynamics are derived
from a governing Hamiltonian that represents the sum of the time-averaged
interaction potential energies between all pairs of planets in the system.
Therefore the central ingredient in any numerical method for
secular dynamics is the efficient evaluation of the average interaction energy
between two fixed Kepler orbits, each specified by its semi-major axis $a$ and
shape parameters $e$ (eccentricity), $\omega$ (argument of periapsis), $I$
(inclination), and $\Omega$ (longitude of node). The averaging can be thought
of as smearing each planet into a `ring' in which the mass per unit length
is inversely proportional to the speed of the planet in its orbit. We call a
code that follows the secular gravitational dynamics of $N$ planets an
`N-ring code' by analogy with N-body codes that follow the full
gravitational dynamics.

The simplest way to calculate the interaction energy between rings labelled by
$\alpha$ and $\beta$ is to choose $K$ points equally spaced in mean anomaly on
each ring and then compute $-\sum_{i,j=1}^K Gm_\alpha
m_\beta/(K^2\Delta_{ij})$, where $m_\alpha$ and $m_\beta$ are the masses of
the two bodies and $\Delta_{ij}$ is the distance between point $i$ on ring
$\alpha$ and point $j$ on ring $\beta$ (e.g., 
\citealt{hop2007}). This approach is rather inefficient, both because it
requires $\Order(K^2)$ computations per ring pair and because it converges
slowly, especially if the orbits have a close approach
($\hbox{min\,}\Delta_{ij}\ll\hbox{max\,}\Delta_{ij}$). For example,
\cite{hop2007} typically find that $K\sim 10^{3.5}$ is needed for 1 per cent
relative accuracy in the potential.

A better approach, which we pursue in this paper, is to compute the potential
from ring $\alpha$ at an arbitrary point $\bfr$ by integrating analytically
over the mass elements of ring $\alpha$, and then to numerically integrate
this potential over the locus of points $\bfr$ that lie on ring $\beta$. The
rather involved analysis needed to do this was accomplished by Gauss and Hill
in the nineteenth century\footnote{It would be better, of course, to integrate
  over {\em both} rings analytically but even Gauss was apparently not able to
  do this.}. We shall call this approach `Gauss's method', and the use of
Gauss's method in an N-ring code the `Gaussian ring algorithm'.

When he wrote his classic memoir on the force exerted by an elliptic ring on a
mass point, \cite{G1818} achieved at least three lasting outcomes: he provided
dynamics with a vivid illustration of the averaging principle; he introduced
the arithmetic-geometric mean as an efficient method for computing Legendre's
elliptic functions; and he showed how the evaluation of the first-order
average of the planetary equations can be dramatically accelerated with the
help of such functions.  \cite{H1971} gives a thorough review of Gauss's
method, its algorithmic manifestations and its applications. Its main
attraction of course is the efficient averaging of the equations of motion
over the fast, orbital, time scale. The resulting differential equations, in
the case of the solar system, can be evolved with a timestep of at least 500
yr-over $10^4$ times larger than the timestep needed to follow the
unaveraged equations of motion. The main drawback of Gauss's method, as
already noted, is that it neglects perturbations that are second-order or
higher in the planetary masses, some of which are near-resonant terms that, in
the solar system, can be almost as strong as the secular terms that are first
order in the masses.

An implementation of Gauss's method by one of us (J.T.)  was used to explore
\cite{K1962} oscillations of the then newly discovered planet 16 Cygni Bb
\citep{T1997}. In most applications, Kozai oscillations arise from the
influence of a distant companion body. In the case of 16 Cygni Bb, this is a
stellar companion to the host star with a projected separation of $\sim 850$
AU, 500 times the planetary semi-major axis. At such large separations the
perturbing potential from the stellar companion is dominated by its quadrupole
moment, and in this case the averaged interaction potential between the
companion and the planet can be computed analytically. Nevertheless, Gauss's
method offers important insight into this problem. The reason is that in the
quadrupole approximation the averaged interaction potential is axisymmetric in
the reference frame whose equator is the orbital plane of the companion, even
if the companion orbit is eccentric. Thus both the energy and the component of
angular momentum along the axis of the companion orbit are integrals of
motion, and the secular dynamics is described by a Hamiltonian with only one
degree of freedom. The existence of this second integral is an accidental
consequence of the properties of the quadrupole potential, and without the
quadrupole approximation the integral is not present. Gauss's method allows us
to explore the two degree-of-freedom problem defined by the full secular
dynamics, which for example shows that in many cases initially circular
planetary orbits are chaotic and allows us to estimate the width of the
chaotic zone\footnote{An alternative approach is to expand the potential to
  octupole order, which also destroys the second integral; see
  \cite{For2000}.}.

What drives much of the current interest in Gauss's method is the wish to
study the evolution of a distribution of stars within the sphere of influence
of the central black hole in a galaxy. In this situation the central object
enforces Keplerian dynamics, perturbed by the mutual gravitational attraction
between the stars. Here then is a natural context for Gauss's method.
However, in such problems, one immediately faces close encounters between
particles, which translate into ring crossings. At a crossing, the force
between rings changes discontinuously\footnote{Consider a vertical ring with
  mass per unit length $\sigma$ lying on the $z$-axis, and a similar
  horizontal ring lying parallel to the $y$-axis and passing through the point
  $(x,y,z)=(x_0,0,0)$. The force exerted by the vertical ring on the
  horizontal ring is $-2\pi G\sigma^2\,\hbox{sgn}\,(x)\hat{\bfx}$.}, a jump
that is difficult for numerical integrators to follow. Such was the experience
of \cite{RT1996}, who used an unsoftened version of Gauss's method to simulate
resonant relaxation around black holes and found that the speed-up resulting
from averaging is virtually cancelled by the slow-down caused by ring
crossings.

In N-body simulations, one often resorts to softened gravitational
interactions to alleviate the cost of resolving near-collisions between
particles (and to better approximate collisionless behavior). It is natural to
ask whether Keplerian rings that perturb each other with softened
gravitational forces can be evolved with the help of a softened variant of
Gauss's method. We were pleased to learn that the answer is yes, and we will
proceed to tell you how. We start in \S\ref{sec:gm} by reviewing the
underpinnings of Gauss's algorithm, and how they are modified for the softened
case. We thereby derive a formulation of Gauss's method for softened
gravitational interactions. We discuss how to implement the method numerically
in \S\ref{sec:imp}. In \S\ref{sec:test} we describe numerical examples that
illustrate its accuracy, its strengths, and its limitations (limitations of
secular averaging in general), by direct comparison with simulations of the
analogous unaveraged system. These tests lead us to \S\ref{sec:cr}, where we
present simulations of a (linearly) unstable configuration of two
counter-rotating annuli, which track the instability to its (nonlinear)
saturation as a global, uniformly precessing, lopsided ($m=1$) mode. We
conclude in \S\ref{sec:disc} with a discussion of the main features, potential
applications, and future improvements of our algorithm.

\section{Gauss's method revisited}

\label{sec:gm}

\noindent
We argued in the preceding section that to compute the interaction potential
between two Keplerian rings using $K$ equally spaced points per ring required
$\Order(K^2)$ calculations. \cite{G1818} showed that the work can be reduced to
$\Order(K)$ calculations by integrating analytically over one ring with the help
of Legendre's elliptic functions.  \cite{H1882} elaborated the arguments in
Gauss's paper, and described an algorithm for the exact calculation of
secular evolution which is accurate to first order in the masses and all
orders in eccentricity and inclination. \cite{H1888}, and \cite{G1937} after
him, followed an alternative, geometric, route to elliptic functions, which
was later described in tensorial language by \cite{Mu1963}. While Gauss's
exposition is purely algebraic in flavor, Halphen exploits the geometry of the
cone (with base on the disturbing ring and apex at the point where the force
is evaluated), to motivate a transformation to the cone's principal axes
that makes Gauss's insight transparent. While the algebraic manipulations work
through effortlessly for softened interactions, it is less obvious how to
soften things geometrically, since the softening scale breaks a
crucial symmetry. We choose to cover the algebraic derivation in the body
of the text, relegating the geometrical treatment to Appendix \ref{sec:halphen}.

The algorithm, softened or not, has two independent components: on one
hand, the expression of the averaged equations of motion in a
flexible, singularity-free coordinate system, while presuming a means,
efficient or not, for computing averages of interest (the subject of
this section); on the other, an efficient method for computing these
averages (the subject of \S\ref{sec:imp}). Taken together, these
maneuvers produce a regularized vector field which one can evolve with
giant strides using an appropriate numerical integrator.

The calculations below largely follow \cite{H1882}; they are repeated here
partly to include the generalization to softening, partly to update Hill's
derivation to modern notation and style, and partly to express the
results in a vectorial, coordinate-free notation that is much more efficient for
numerical integration algorithms.

\subsection{Regularized equations}

\label{sec:regular}

\noindent
We consider a particle (called the `planet', though it could be a
star, moon, planetesimal, comet, etc.) bound to a massive central body
(called the `star', though it could be a black hole, planet,
etc.). The planet follows a Keplerian orbit and is subject to
relatively weak external forces, which cause time variations in the
(osculating) Keplerian orbital elements. We express the equations
governing these variations in terms of non-singular vectorial
variables, then recover their secular average for the case of
interest, namely when particles perturb each other's Keplerian orbits
with their mutual gravitational attraction (Newtonian or softened).

We denote by $M_\star$ the mass of the star, and adopt the
following notation for properties pertaining to the planet:
\begin{itemize}
\item $m$ mass of the planet,
\item $\bfr$ radius vector from the central mass, with $r=|\bfr|$,
\item $\bfv\equiv \dot{\bfr}$ velocity vector, 
\item $a$ the semi-major axis,
\item $e$ the eccentricity,
\item $n = \sqrt{G (M_\star+ m)/a^3}$ the mean motion, with
  $P=2\pi/n$ the orbital period,
\item $l$, $E$, and $f$ the mean, eccentric and true anomalies, respectively.
\end{itemize}
When considering perturbations between planets, unprimed variables
refer to the perturbing planet, and the primed counterparts to the
perturbed planet. 

In the planetary context, it is common to describe the perturbations
to a Keplerian orbit using formulae that describe the variation of the
orbital elements eccentricity, inclination, etc. (e.g.,
\citealt{MD1999}).  These equations suffer from singularities at zero
eccentricity and inclination, as well as at unit eccentricity. The
singularities at zero eccentricity and inclination are traditionally
handled by a canonical transformation to appropriate Cartesian
coordinates for the orbital elements (the familiar trick for dealing
with the singularity of cylindrical polar coordinates at the
origin). Such transformations work reasonably well in `cold'
settings where the eccentricity and inclination are small. Near unit
eccentricity (nearly radial orbits) the singularities can also be
removed by an appropriate canonical transformation
\citep{tre01}. However, in hot environments such as the stellar cusp
around a black hole the eccentricities and inclinations take on all
values so one must regularly switch between coordinate charts if one
insists on working with elements such as eccentricity and inclination
or even their Cartesian analogs. We initially implemented such
switches, and found them costly and cumbersome. Instead, we chose to
parametrize an orbit with the coordinate-free (though redundant) angular
momentum and eccentricity (or Laplace, or Runge-Lenz) vectors (see
\citealt{H1971} or \citealt{MD1999}).

We associate with an orbit three orthogonal unit vectors: $\hat{\bfz}$, along
the rescaled angular momentum vector $\bfL= \sqrt{1 - e^2}\, \hat{\bfz}$
(normal to the orbital plane, scaled by $m n a^2$), $\hat{\bfx}$, pointing
toward periapsis, and $\hat{\bfy} = \hat{\bfz}\cross \hat{\bfx}$,
perpendicular to both. The eccentricity vector is 
\beq
\bfA={\bfv\cross\bfL\over na}-\hat{\bfr}=e\hat{\bfx}.  \eeq

The position vector of the planet is $\bfr = x\hat{\bfx}+ y\hat{\bfy}$ where $x =
r\cos f=a(\cos E-e)$, and $y = r\sin f = a\sqrt{1-e^2}\sin E$, with the radius
$r=a(1-e\cos E)$.  We introduce $\hat{\bfr}$, a radial unit vector, and the
tangential vector $\hat{\bft}=\hat{\bfz}\cross\hat{\bfr}$; explicitly
\beq
\begin{array}{rl}
\hat{\bfr} =  \cos f\,\hat{\bfx} + \sin f\, \hat{\bfy} & = 
\frac{\displaystyle\phantom{\Big[}(\cos E - e) \hat{\bfx} + \sqrt{1-e^2}\sin E
  \,\hat{\bfy}}{\displaystyle\phantom{\Big[} 1 - e \cos E} \\
\hat{\bft} =  -\sin f\, \hat{\bfx} + \cos f\, \hat{\bfy} & =
\, \frac{\displaystyle\phantom{\Big[} -\sqrt{1-e^2}\sin E\,\hat{\bfx} + (\cos
  E - e) \,\hat{\bfy}}{\displaystyle\phantom{\Big]} 1 - e \cos E}.
\end{array}
\label{eqs:radtan}
\eeq
We will also need the velocity vector $\bfv = \dot x\hat{\bfx} + \dot
y\hat{\bfy}$, where $\dot x = - n a^2\sin E/r$ and $\dot y = n a^2
\sqrt{1-e^2}\cos E/r$ (using $\dot E = na/r$).

The semi-major axis, angular momentum vector, and
eccentricity vector of the disturbed (primed) planet respond to a
perturbing acceleration $\bff'$ on that planet as 
\bea
\frac{{\rm d} a'}{{\rm d}t} & = & \frac{2}{n'^2 a'} \bfv' \bcdot \bff'
\nonumber \\ 
\frac{{\rm d}(n'a'^2\, {\bfL}')}{{\rm d}t} & = & \bfr'\cross \bff' \nonumber \\
\frac{{\rm d} {\bfA}'}{{\rm d}t} & =&  \frac{1}{G(M_\star+m)}
\left[2\bfr'(\bff'\bcdot\bfv')\bfv' -
  \bfv'(\bfr'\bcdot\bff')-\bff'(\bfr'\bcdot\bfv')\right].
\label{eq:vectorial}
\eea
The first equation expresses the change in the Keplerian energy
resulting from the perturbing force, the second the change in angular
momentum resulting from the associated torque, and with it the change in 
orientation of the orbital plane; the third governs the eccentricity
and periapsis direction of the orbit, hence dictating shape and orientation in
the orbital plane. 

We shall work with the radial ($R$), tangential ($S$) and normal
components ($W$) of the perturbing acceleration:
\beq
\begin{array}{lll}
R  =  \hat{\bfr}' \bcdot \bff', &  S  =  \hat{\bft}' \bcdot \bff' 
           & W  =  \hat{\bfz}' \bcdot \bff'.
\end{array}
\label{eq:rswdef}
\eeq
Substitute into equation (\ref{eq:vectorial}) to get 
\bea
\frac{{\rm d} a'}{{\rm d}t} &=& \frac{2 (R\,e'\sin E'+S\,\sqrt{1-e'^2})}
{n'(1-e' \cos E')}, \nonumber \\
\frac{{\rm d} (n'a'^2\,{\bfL}')}{{\rm d}t} & = &
N'_{x'}\,\hat{\bfx}'+N'_{y'}\,\hat{\bfy}'  
+ N'_{z'}\,\hat{\bfz}', \nonumber \\
\frac{{\rm d} {\bfA}'}{{\rm d}t} & = & \dot A'_{x'}\,\hat{\bfx}'+\dot
A'_{y'}\,\hat{\bfy}' 
+\dot A'_{z'}\, \hat{\bfz}',
\label{eqs:vecteq}
\eea
where
\bea
N'_{x'}  & = & W a' \sqrt{1-e'^2} \sin E', \nonumber \\
N'_{y'}  & = & - Wa' (\cos E' - e'), \nonumber \\
N'_{z'}  & = &  Sa' (1 - e' \cos E'), 
\label{eqs:Leq}
\eea
and 
\bea
\dot A'_{x'} &=& \frac{\sqrt{1-e'^2}[(4\cos E'-e'\cos 2E'-3e')S+2\sqrt{1-e'^2} 
\sin E'\, R]}{2n'a'(1-e'\cos E')},  \nonumber \\
\dot A'_{y'} &=& \frac{[2(2 - e'^2)\sin E' - e'\sin 2 E']\, S - 2
\sqrt{1-e'^2} \,(\cos E'-e')\, R} {2 n' a'\, (1-e' \cos E')},  \nonumber \\
\dot A'_{z'} &=& -{e'\over n'a'}\,W\sin E'.
\label{eqs:Aeq}
\eea These equations govern the secular evolution of an osculating
Keplerian orbit in the vectorial representation. They are general
enough to accommodate a great variety of forces, conservative (external
perturbers, such as satellites, planets, stars, or black holes, on
asteroids, planets, or stars) or not (drag, tides, dynamical
friction), that may affect the motion of a particle whose motion is
otherwise dominated by a massive central body. In this work, we are
concerned with analytic averaging over softened mutual gravitational
interactions in a near-Keplerian cluster of particles, to which
we shall restrict our attention in the following.

\subsection{Averaged equations}

\noindent
The softened gravitational force on the primed planet, $m' \bff' =
-\del'\Phi$, derives from the disturbing function 
\beq 
\Phi = -G m m' \left [ \frac{1}{{\Delta}_b} - \frac{x x' + y y' +
    zz'}{r^3} \right ]. 
\eeq 
where ${\Delta}_b =\sqrt{(x -x')^2 + (y-y')^2 + (z-z')^2 + b^2}$, and
$b$ is the softening length (primed and unprimed refer to perturbed
and perturbing respectively). The first term is Plummer's potential
(which reduces to Newton's for $b=0$) coupling the two particles, the
second (the indirect term) is a non-inertial contribution resulting
from pinning the reference frame to the central mass. Note that we
soften only the interplanetary potential, not the potential between
each planet and the central star. We are interested in the secular
evolution of the osculating Keplerian orbital elements, which we
obtain by averaging the equations of motion (\ref{eqs:Aeq}) over the
mean anomalies $l$ and $l'$ of the disturbing and disturbed
planets. In other words we are after
\bea
\left[\frac{{\rm d} a'}{{\rm d}t}\right]_{ll'} &=&  \frac{1}{4\pi^2} \int_0^{2\pi}
\!\!\int_0^{2\pi} \frac{{\rm d} a'}{{\rm d}t}\, {\rm d}l'\,{\rm d}l, \nonumber \\ 
\left[\frac{{\rm d} {\bfL}'}{{\rm d}t}\right]_{ll'} &=& \frac{1}{4\pi^2} \int_0^{2
  \pi}\!\!\int_0^{2\pi} \frac{{\rm d} {\bfL}'}{{\rm d}t}\,{\rm d}l\, {\rm d}l', \nonumber \\
\left [\frac{{\rm d} {\bfA}'}{{\rm d}t}\right]_{ll'}  &=&  \frac{1}{4\pi^2} \int_0^{2
  \pi}\!\!\int_0^{2\pi} \frac{{\rm d} {\bfA}'}{{\rm d}t} \,{\rm d}l'\,{\rm d}l. 
\label{eq:poiss}
\eea

Before proceeding further, we make a couple of simplifying remarks. (i) The
indirect term in $\Phi$ contributes a force $\bfr/r^3$ that averages to zero
over $l$. Therefore we may drop the indirect term. (ii) Another classical
result of secular dynamics is that ${\rm d}a'/{\rm d}t$ averages to zero over $l'$, since
the average is proportional to the work of a conservative force over a closed
path. Thus we may treat the semi-major axis $a'$ and mean motion $n'$ of the
disturbed planet as constants of motion when examining conservative
perturbations in secular dynamics (in the case of unsoftened gravitational
forces this is Poisson's theorem, but the result requires only that the
perturbing force is conservative and time-independent and hence holds for
softened forces as well).

Below, we shall demonstrate that the average of the direct force over $l$ is
analytic and expressible in terms of elliptic functions. For now, we denote by
$[R]_l$, $[S]_l$, and $[W]_l$ the $l$-averaged radial, tangential and normal
components of the direct force, and note that the $l$-averaged equations can
be directly translated from their unaveraged counterparts in equations
(\ref{eqs:Leq}) and (\ref{eqs:Aeq}), by replacing $R$, $S$ and $W$ by their
respective $l$ averages.

Now consider the average over the orbit of the disturbed planet. For
any function $X(E')$,
\beq
     [X]_{l'}={1\over 2\pi}\int_0^{2\pi}X(E')(1-e'\cos E')\,{\rm d}E'.
\eeq
Inspecting equations (\ref{eqs:Leq}) and (\ref{eqs:Aeq}) we find that
the average over $l'$ will at most involve the second-order term of
the Fourier expansion of $[R]_l$, $[S]_l$ and $[W]_l$ in their
periodic dependence on the eccentric anomaly $E'$, leading to:
\bea 
n'a'^2{\left[\frac{{\rm d} {\bfL}'}{{\rm d}t}\right]}_{ll'}&=&{[N'_{x'}]}_{ll'}\hat{\bfx}'
  +{[N'_{y'}]}_{ll'}\hat{\bfy}' + {[N'_{z'}]}_{ll'}\hat{\bfz}', \nonumber \\
{\left[\frac{{\rm d} {\bfA}'}{{\rm d}t}\right]}_{ll'} & = &  {[{\dot A'}_{x'}]}_{ll'}
\hat{\bfx}' + {[{\dot A'}_{y'}]}_{ll'}\hat{\bfy}' + {[{\dot A'}_{z'}]}_{ll'}
\hat{\bfz}' 
\label{eqs:aveq}
\eea
where
\bea
{[N'_{x'}]}_{ll'} & = & a'\sqrt{1-e'^2} (W^1_s-\half e'W_s^2), \nonumber \\
{[N'_{y'}]}_{ll'} & = & -a'[(1+e'^2)W_c^1 -\textstyle{3\over 2} e'
  W_c^0-\half e'W_c^2],  \nonumber \\
{[N'_{z'}]}_{ll'} & = & a'[(1+\half e'^2)S_c^0-2e'
S_c^1+\half e'^2 S_c^2],
\label{eqs:avl}
\eea
and
\bea
{[{\dot A'}_{x'}]}_{ll'}  &=& \frac{\sqrt{1-e'^2}}{2n'a'}
[(4S_c^1-e'S_c^2-3e'S_c^0)+2\sqrt{1-e'^2}R_s^1],  \nonumber \\
{[{\dot A'}_{y'}]}_{ll'}  &=& \frac{1}{2n'a'}[2(2-e'^2)S_s^1-e'
S_s^2-2\sqrt{1-e'^2}(R_c^1-e'R_c^0)], \nonumber \\
{[{\dot A'}_{z'}]}_{ll'}  &=& -\frac{e'}{n'a'}(W^1_s-\half W_s^2). 
\label{eqs:ava}
\eea
Here
\beq
R_c^n={1\over 2\pi}\int_0^{2\pi}[R]_l\cos nE'\,{\rm d}E',\qquad
R_s^n={1\over 2\pi}\int_0^{2\pi}[R]_l\sin nE'\,{\rm d}E',
\label{eq:avtwo}
\eeq
with similar definitions for $S_c^n$, $W_c^n$, etc.

For conservative forces, $[{\rm d}a'/{\rm d}t]_{ll'}=0$
and this gives us the identity
\beq
e'R_s^1+\sqrt{1-{e'}^2}S_c^0=0\quad\hbox{if $\bff'$ is conservative.}
\label{eq:adot}
\eeq

The cost of non-singular coverage of the full range in eccentricity
and inclination using the vectors $\bfL'$ and $\bfA'$ has been to
expand the phase space of an N-ring system from $4N$ (reduced from
$6N$ as a result of averaging over the mean anomaly $l'$ and
conservation of the semi-major axis $a'$), to the redundant
$6N$-dimensional phase space spanned by the angular momentum and
eccentricity vector of each ring; $N$ of the redundant components are
removed by noting that ${\bfL}'$ and ${\bfA}'$ are orthogonal, while
the other $N$ reflect the relation between the magnitudes of the two
vectors, $L'^2+A'^2=1$. It is straightforward to show that
$[{\rm d}\bfL'\bcdot\bfA'/{\rm d}t]_{ll'}=0$ and $[{\rm d}({L'}^2+{A'}^2)/{\rm d}t]_{ll'}=0$
are satisfied by equations (\ref{eqs:aveq})-(\ref{eqs:ava}) if the
perturbing forces are conservative.

\subsection{Analytic averaging}

\noindent
We shall go over the crux of our exercise, namely the demonstration that
Gauss's averaging works out perfectly for softened potentials. 

The average of the gravitational potential $\Phi$ over the orbit of the
perturbing (unprimed) planet can be written
\beq
[\Phi]_l(\bfr')=-{Gm\over 2\pi}\int_0^{2\pi}{\rm d}E\,{1-e\cos E\over\Delta_b},
\eeq
where
\beq
\Delta_b^2=A_b-2B\cos(E-\epsilon)+C\,\cos^2E,
\label{eq:wwwqqq}
\eeq
with
\bea
A_b            & = & r'^2 + a^2 + b^2 + 2ae\bfr'\bcdot\hat{\bfx}, \nonumber \\
B \cos\epsilon & = & a\bfr'\bcdot\hat{\bfx} + a^2 e, \nonumber \\
B \sin\epsilon & = & a\sqrt{1-e^2}\, \bfr'\bcdot\hat{\bfy}, \nonumber \\
C              & = & a^2e^2.
\label{eq:abcdef}
\eea
Note the inequalities 
\beq
A_b > a^2(1-e^2)+b^2 > 0.
\label{eq:cccddd}
\eeq

We pause briefly to emphasize the seemingly innocuous absorption of the
softening $b^2$ in $A_b$. As far as Gauss's method is concerned, this is
the only direct change that is brought about by softening.

The average perturbing acceleration is
\bea
[\bff']_{l} &=& -\del'[\Phi]_{l}=-{Gm\over 2\pi}\int_0^{2\pi}{\rm d}E\,
      {1-e\cos E\over2\Delta_b^3}\del'\Delta_b^2 \nonumber \\
               &=& -{Gm\over 2\pi}\int_0^{2\pi}{\rm d}E\,
      {1-e\cos E\over\Delta_b^3}(\bfr'+ae\hat{\bfx}-a\sqrt{1-{e}^2}\,
      \hat{\bfy}\sin E-a\hat{\bfx}\cos E) \nonumber \\         
               &=& {Gm\over 2\pi}\int_0^{2\pi}{\rm d}E\,
      {1-e\cos E\over\Delta_b^3}({\bfF}_0+{\bfF}_1\sin E+{\bfF}_2\cos E),
\label{eq:ffdeff}
\eea
where 
\beq
{\bfF}_0=-\bfr'-ae\,\hat{\bfx}, \quad {\bfF}_1=a\sqrt{1-e^2}\,\hat{\bfy}, 
\quad {\bfF}_2=a\hat{\bfx}.
\label{eq:fdef}
\eeq

At this point, we write
\beq
\sin E\equiv {x_1\over x_0}, \quad \cos E={x_2\over x_0};
\label{eq:xdef}
\eeq
introduce a new angular variable $T$ (the perspective anomaly) defined by 
\beq
\sin T\equiv {y_1\over y_0}, \quad \cos T={y_2\over y_0};
\label{eq:ydef}
\eeq
and demand that
\beq
\bfx=\bfssQ\bfy
\label{eq:qdef}
\eeq
where $\bfssQ$ is a $3\times3$ matrix with elements $Q_{ij}$, $i,j=0,1,2$. 

The geometric motivation for this change of variable is described more fully
in Appendix \ref{sec:halphen}. Briefly, the orbit average $[\cdot]_{l}$ can be
thought of as a line integral around the orbit, with $r$ and $E$ the polar
variables in the orbit plane for this line. We show in the Appendix that the
orbit average is unchanged if we shift the line integral to a new contour,
defined by the intersection of {\it any\/} plane with the cone that contains
the orbit and has its apex at the disturbed planet. The perspective anomaly is
a polar coordinate in this new plane. The plane is later chosen to make the
line integral as simple as possible.

Since $\sin^2E+\cos^2E=\sin^2T+\cos^2T=1$ we must have
\beq
x_0^2-x_1^2-x_2^2=0, \qquad y_0^2-y_1^2-y_2^2=0.
\eeq
The first of these simply states that $\bfx$ is confined to a cone; the
second restricts $\bfssQ$ to be a matrix that conserves the quadratic form
$\bfx^{\rm T}\bfssM\bfx$, where $\bfx^{\rm T}$ is the transpose of
the column vector $\bfx$ and $\bfssM$ is the diagonal matrix
$\mbox{\textsf{\textbf{diag}\,}}(1,-1,-1)$. This requirement can be written
\beq
\bfssQ^{\rm T}\bfssM\bfssQ=\bfssM.
\label{eq:symp}
\eeq
Thus $\bfssQ$ is a pseudo-orthogonal matrix; in the language of special
relativity, $\bfssQ$ is a Lorentz transformation in two spatial dimensions,
$\bfssM$ is the Minkowski tensor, and the vectors $\bfx$ and $\bfy$
lie on the light cone passing through the origin. 

We define a diagonal matrix with complex coefficients,
$\bfssC\equiv\mbox{\textsf{\textbf{diag}\,}}(1,{\rm i},{\rm i})$. Then   
\beq
{\bfssC}^2=({\bfssC}^\ast)^2=\bfssM, \quad {\bfssC}^\ast\bfssC=\bfssI, \quad 
\bfssC\bfssM\bfssC={\bfssC}^\ast\bfssM{\bfssC}^\ast=\bfssI,
\label{eq:sympa}
\eeq
where $\ast$ denotes complex conjugation and $\bfssI$ is the unit matrix. Let
$\bfssL\equiv {\bfssC}^\ast\bfssQ{\bfssC}^\ast$. Then using equations
(\ref{eq:symp}) and (\ref{eq:sympa}) we have 
\beq
{\bfssL}^{\rm T}\bfssL\equiv {\bfssC}^\ast\bfssQ^{\rm
  T}\bfssM\bfssQ{\bfssC}^\ast={\bfssC}^\ast\bfssM{\bfssC}^\ast=\bfssI.
\eeq
Thus $\bfssL$ is orthogonal if $\bfssQ$ is pseudo-orthogonal, and any
pseudo-orthogonal $\bfssQ$ can be written in the form
\beq
\bfssQ=\bfssC\bfssL\bfssC
\label{eq:sympc}
\eeq
where $\bfssL$ is orthogonal. Note that even though $\bfssL$ may have complex
coefficients, it is orthogonal ($L_{ij}^{-1}=L_{ji}$) rather than Hermitian
($L_{ij}^{-1}=L_{ji}^\ast$). Because we allow complex coefficients
the inner product of a vector with itself,  ${\bfx}^{\rm T}\bfx=
\bfx\bcdot\bfx=\sum_{i=0}^2x_i^2$ is not necessarily positive or even real. 

We may now re-write equation (\ref{eq:wwwqqq}) as
\beq
x_0^2\Delta_b^2=A_bx_0^2 -2Bx_0x_1\sin\epsilon -2Bx_0x_2\cos\epsilon  
+Cx_2^2=\bfx^{\rm T}\bfssP\bfx,
\label{eq:quaddef}
\eeq
where
\beq
\bfssP\equiv \left(\begin{array}{ccc} A_b & -B\sin\epsilon & -B\cos\epsilon
  \\                                   -B\sin\epsilon & 0 & 0 \\
                                       -B\cos\epsilon & 0 & C\end{array}\right).
\eeq
Gauss now asks for the pseudo-orthogonal matrix $\bfssQ$ that diagonalizes 
the quadratic form for $x_0^2\Delta_b^2$, that is, 
\beq
\bfx^{\rm T}\bfssP\bfx=\bfy^{\rm T}\bfssQ^{\rm T}\bfssP\bfssQ
\bfy\equiv\bfy^{\rm T}\bfssD\bfy,
\label{eq:xxx}
\eeq
where $\bfssD$ is some diagonal matrix. Using equation (\ref{eq:sympc}) we 
re-write the quadratic form as 
\beq
\bfy^{\rm T}\bfssC{\bfssL}^{\rm T}\bfssR\bfssL\bfssC\bfy,
\label{eq:diagone}
\eeq
where
\beq
\bfssR\equiv \bfssC\bfssP\bfssC=\left(\begin{array}{ccc} A_b & -{\rm
      i}B\sin\epsilon & -{\rm i}B\cos\epsilon
  \\                             -{\rm i}B\sin\epsilon & 0 & 0 \\
                                 -{\rm i}B\cos\epsilon & 0 & -C\end{array}\right).
\eeq 
Since $\bfssR$ is symmetric, (i) its eigenvectors ${\bfe}^j$ can be chosen
to be orthonormal, ${\bfe}^j\bcdot{\bfe}^k=\delta_{jk}$; (ii) the matrix ${\bf
L}$ defined by $L_{ij}=e_i^j$ is orthogonal; (iii) ${\bfssL}^{\rm
T}\bfssR\bfssL=\mbox{\textsf{\textbf{diag}\,}}(\lambda_i)$, where
$\{\lambda_i\}$ are the eigenvalues of 
$\bfssR$. Thus equation (\ref{eq:diagone}) can be written as
\beq
\bfy^{\rm T}{\bfssC}{\bfssL}^{\rm T}\bfssR\bfssL\bfssC\bfy=
\bfy^{\rm T}\bfssC\,\mbox{\textsf{\textbf{diag}\,}}(\lambda_i)\,\bfssC\bfy=
\bfy^{\rm T}\mbox{\textsf{\textbf{diag}\,}}(\lambda_0,-\lambda_1,-\lambda_2)\bfy,
\label{eq:diagl}
\eeq
which is the desired diagonalization (\ref{eq:xxx}), with $\bfssQ$
related to $\bfssL$ by (\ref{eq:sympc}) and
$\bfssD=\mbox{\textsf{\textbf{diag}\,}}(\lambda_0,-\lambda_1,-\lambda_2)$.  

The eigenvalues of $\bfssR$ are the solutions of the cubic equation 
\beq
y(\lambda)=\lambda^3+(C-A_b)\lambda^2+(B^2-A_bC)\lambda+B^2C\sin^2\epsilon=0.
\eeq
Using the relations (\ref{eq:abcdef}) it is straightforward to show that
\bea
y(-C) &=& -B^2C\cos^2\epsilon < 0, \nonumber \\
y(0)  &=&  B^2C\sin^2\epsilon > 0, \nonumber \\
y[a^2(1-e^2)] &=& -a^4(1-e^2)[(\bfr'\bcdot\hat{\bfz})^2+b^2]<0,
\nonumber \\
y(A_b) &=& B^2(A_b+C\sin^2\epsilon)>0.
\eea
Furthermore from (\ref{eq:cccddd}) ${a}^2(1-{e}^2)<A_b$ so $y(\lambda)$ must have
three real roots, one between $-C$ and 0, one between 0 and
$a^2(1-e^2)$, and one between $a^2(1-e^2)$ and $A_b$.
It proves useful to label these roots so that $\lambda_0>\lambda_1>\lambda_2$;
thus
\beq
-C < \lambda_2 < 0 < \lambda_1 < a^2(1-e^2) < \lambda_0 < A_b.
\label{eq:lamdef}
\eeq
Relations between these roots include
\bea
\lambda_0\lambda_1\lambda_2&=&-B^2C\sin^2\epsilon \nonumber \\
(\lambda_0+C)(\lambda_1+C)(\lambda_2+C)&=&B^2C\cos^2\epsilon \nonumber \\
\lambda_0+\lambda_1+\lambda_2&=& A_b-C \nonumber \\
\lambda_0\lambda_1+\lambda_1\lambda_2+\lambda_2\lambda_0&=&B^2-A_bC.
\label{eq:lamrel}
\eea
Explicit expressions are
\bea
\lambda_0 &=& -2\sqrt{Q}\cos\left(\third\theta+\frac{2}{3}\pi
\right)-\third(C-A_b)\nonumber \\
\lambda_1 &=& -2\sqrt{Q}\cos\left(\third\theta-\frac{2}{3}\pi\right)
-\third(C-A_b) \nonumber \\
\lambda_2 &=& -2\sqrt{Q}\cos\left(\third\theta\right)-\third(C-A_b)
\eea
where
\bea
Q&=&\ffrac{1}{9}(C-A_b)^2-\ffrac{1}{3}(B^2-A_bC), \nonumber \\
R&=&\ffrac{1}{27}(C-A_b)^3-\ffrac{1}{6}(C-A_b)(B^2-A_bC)+\half
B^2C\sin^2\epsilon, \nonumber \\ 
\theta&=&\cos^{-1}\big(R/\sqrt{Q^3}\big). 
\eea

The eigenvectors of $\bfssR$ are
\beq
{\bfe}^k=\alpha_k\left({\rm i},{B\sin\epsilon\over\lambda_k},
{B\cos\epsilon\over\lambda_k+C}\right), 
\eeq
where $\alpha_k$, which may be complex, is chosen so that ${\bfe}^k\bcdot{\bf
 e}^k=1$ or 
\beq
1=\alpha_k^2\left[-1+{B^2\sin^2\epsilon\over\lambda_k^2}+{B^2\cos^2\epsilon
\over(\lambda_k+C)^2}\right]. 
\eeq
Using the relations (\ref{eq:lamrel}) this can be manipulated into a more compact form,
\beq
\alpha_k^2=-{\lambda_k(\lambda_k+C)\over(\lambda_k-\lambda_l)
(\lambda_k-\lambda_m)},
\eeq
where $l$, $m$ are the indices other than $k$ from $\{0,1,2\}$. 
With the ordering (\ref{eq:lamdef}), 
\beq
\alpha_0^2<0, \quad \alpha_1^2 >0, \quad \alpha_2^2 > 0.
\eeq
We choose the phases of the eigenvectors so that 
\beq
\alpha_0=-{\rm i}\sqrt{-\alpha_0^2},\quad \alpha_1=-\sqrt{\alpha_1^2},\quad
\alpha_2=-\sqrt{\alpha_2^2}.
\eeq
Then 
\[
\bfssQ=\bfssC\bfssL\bfssC=\left(\begin{array}{ccc} 
       \sqrt{-\alpha_0^2}  & \sqrt{\alpha_1^2} & \sqrt{\alpha_2^2}\\
       \sqrt{-\alpha_0^2}{B\sin\epsilon/\lambda_0}  &
       \sqrt{\alpha_1^2}{B\sin\epsilon/\lambda_1}  
      &\sqrt{\alpha_2^2}{B\sin\epsilon/\lambda_2} \\                                
\sqrt{-\alpha_0^2}B\cos\epsilon/(\lambda_0+C)  & \sqrt{\alpha_1^2}B\cos\epsilon/(\lambda_1+C) 
      &\sqrt{\alpha_2^2}B\cos\epsilon/(\lambda_2+C) \end{array}\right)
\]
\beq
=\left(\begin{array}{ccc}
  \sqrt{\lambda_0(\lambda_0+C)\over(\lambda_0-\lambda_1)(\lambda_0-\lambda_2)}
 &\sqrt{\lambda_1(\lambda_1+C)\over(\lambda_0-\lambda_1)(\lambda_1-\lambda_2)}
 &\sqrt{|\lambda_2|(\lambda_2+C)\over(\lambda_0-\lambda_2)(\lambda_1-\lambda_2)}\\
  \sqrt{\lambda_0+C\over\lambda_0(\lambda_0-\lambda_1)(\lambda_0-\lambda_2)}
     B\sin\epsilon 
 &\sqrt{\lambda_1+C\over\lambda_1(\lambda_0-\lambda_1)(\lambda_1-\lambda_2)}
  B\sin\epsilon 
 &-\sqrt{\lambda_2+C\over|\lambda_2|(\lambda_0-\lambda_2)(\lambda_1-\lambda_2)}
  B\sin\epsilon \\
  \sqrt{\lambda_0\over(\lambda_0+C)(\lambda_0-\lambda_1)(\lambda_0-\lambda_2)}
  B\cos\epsilon 
 &\sqrt{\lambda_1\over(\lambda_1+C)(\lambda_0-\lambda_1)(\lambda_1-\lambda_2)}
  B\cos\epsilon 
 &\sqrt{|\lambda_2|\over(\lambda_2+C)(\lambda_0-\lambda_2)(\lambda_1-\lambda_2)}
  B\cos\epsilon \\
\end{array}\right). 
\label{eq:qqdeff}
\eeq
The determinant of this matrix is 
\beq
\hbox{det\,}(\bfssQ)=\hbox{det\,}(\bfssL)=\hbox{det\,}(e_i^j)\equiv
s\quad\hbox{where}
\quad s=\hbox{sgn\,}[\sin(2\epsilon)]=\pm1.
\label{eq:sssdef}
\eeq

We now examine the relation between the eccentric anomaly $E$ and the
perspective anomaly $T$ in more detail. From the $(0,0)$ component of the
equations $\bfssQ^{\rm T}\bfssM\bfssQ=\bfssM$ and $\bfssQ\bfssM{\bfssQ}^{\rm
  T}=\bfssM$ we have 
\beq
Q_{10}^2+Q_{20}^2=Q_{01}^2+Q_{02}^2=Q_{00}^2-1.
\eeq
Set
\beq
\Delta=\sqrt{Q_{00}^2-1},\quad \sin\phi={Q_{10}\over\Delta},\quad
\cos\phi={Q_{20}\over\Delta},\quad\sin\psi={Q_{01}\over\Delta},\quad
\cos\psi={Q_{02}\over\Delta};
\eeq
it is straightforward to show that $Q_{00}>1$ so $\Delta$ is real and
positive. Then from equations (\ref{eq:xdef}) and (\ref{eq:ydef})
\beq
\Delta x_0+Q_{10}x_1+Q_{20}x_2 =\Delta x_0[1+\cos(E-\phi)], \quad
\Delta y_0+Q_{01}y_1+Q_{02}y_2 =\Delta y_0[1+\cos(T-\psi)].
\label{eq:xxxccc}
\eeq
From equation (\ref{eq:qdef}) we also have
\beq
\Delta x_0+Q_{10}x_1+Q_{20}x_2 =y_0(\Delta Q_{00}+Q_{10}^2+Q_{20}^2)
+y_1(\Delta Q_{01}+Q_{10}Q_{11}+Q_{20}Q_{21})
+y_2(\Delta Q_{02}+Q_{10}Q_{12}+Q_{20}Q_{22}).
\eeq
Using the $(0,0)$, $(0,1)$ and $(0,2)$ components of the equations
${\bfssQ}^{\rm T}\bfssM\bfssQ=\bfssM$ we can simplify this to  
\beq
\Delta x_0+Q_{10}x_1+Q_{20}x_2 = (Q_{00}+\Delta)(\Delta y_0 +Q_{01}y_1 +
Q_{02}y_2), 
\eeq
so from equations (\ref{eq:xxxccc}) we have
\beq
x_0[1+\cos(E-\phi)]=y_0(Q_{00}+\Delta)[1+\cos(T-\psi)].
\label{eq:tedef}
\eeq

Similarly,
\beq
Q_{20}x_1-Q_{10}x_2 =\Delta x_0\sin(E-\phi), \quad 
Q_{02}y_1-Q_{01}y_2 =\Delta y_0\sin(T-\psi),
\label{eq:xxxddd}
\eeq
and 
\beq
Q_{20}x_1-Q_{10}x_2=y_1(Q_{20}Q_{11}-Q_{10}Q_{21})
                   +y_2(Q_{20}Q_{12}-Q_{10}Q_{22}).
\eeq
Using equations (\ref{eq:lamrel}) and (\ref{eq:qqdeff}) it is straightforward
to show that
\beq
Q_{20}Q_{11}-Q_{10}Q_{21}=sQ_{02}, \quad Q_{20}Q_{12}-Q_{10}Q_{22}=-sQ_{01}
\eeq
where $s=\pm 1$ is defined by equation (\ref{eq:sssdef}). Thus
\beq
Q_{20}x_1-Q_{10}x_2=s(Q_{02}y_1-Q_{01}y_2),
\eeq
so
\beq
x_0\sin(E-\phi)=sy_0\sin(T-\psi).
\label{eq:ssindef}
\eeq
Combining this result with (\ref{eq:tedef}),
\beq
\tan\half(T-\psi)=s(Q_{00}+\Delta)\tan\half(E-\phi)
=s\left(Q_{00}+\sqrt{Q_{00}^2-1}\right)\tan\half(E-\phi).
\eeq
This result relates the perspective and eccentric anomalies and shows
that the perspective anomaly circulates through $2\pi$ (if $s=1$) or $-2\pi$
(if $s=-1$) when the eccentric anomaly
circulates through $2\pi$. Moreover the differentials are related by
\beq
{\rm d}T=s(Q_{00}+\Delta){\cos^2\half(T-\psi)\over\cos^2\half(E-\phi)}{\rm d}E=
    {\sin(T-\psi)\over\sin(E-\phi)}{\rm d}E=s{x_0\over y_0}{\rm d}E,
\eeq
where the last equality employs (\ref{eq:ssindef}). 

We may now combine equations (\ref{eq:ffdeff}), (\ref{eq:xdef}),
(\ref{eq:ydef}), (\ref{eq:quaddef}), and (\ref{eq:diagl}) into an expression
for the orbit-averaged perturbing acceleration at position $\bfr$:
\bea
[\bff']_{l} &=& {Gm\over 2\pi}\int_0^{2\pi}{\rm d}E\,
      {1-e\cos E\over\Delta_b^3}({\bfF}_0+{\bfF}_1\sin E+{\bfF}_2\cos
      E) \nonumber \\
            &=& {Gm\over 2\pi}\int_0^{2\pi}{\rm d}E\,
      {x_0(x_0-ex_2)({\bfF}_0x_0+{\bfF}_1x_1+{\bfF}_2x_2)\over 
      (A_bx_0^2 -2Bx_0x_1\sin\epsilon-2Bx_0x_2\cos\epsilon+Cx_2^2)^{3/2}}
      \nonumber \\
            &=& {Gm\over 2\pi}\int_0^{2\pi}{\rm d}T\,
      {y_0\sum_{j=0}^2(Q_{0j}-eQ_{2j})y_j\sum_{j,k=0,2}{\bfF}_jQ_{jk}y_k\over 
         (\lambda_0y_0^2-\lambda_1y_1^2-\lambda_2y_2^2)^{3/2}} \nonumber \\
               &=& {Gm\over 2\pi}\int_0^{2\pi}{\rm d}T\,
      {\sum_{j=0}^2(Q_{0j}-eQ_{2j})y_j/y_0\sum_{j,k=0,2}{\bfF}_jQ_{jk}y_k/y_0\over 
         [\lambda_0-\lambda_1(y_1/y_0)^2-\lambda_2(y_2/y_0)^2]^{3/2}}.
\eea
Replacing $y_1/y_0$ by $\sin T$ and $y_2/y_0$ by $\cos T$, and dropping terms
that integrate to zero, we have
\beq
[\bff']_{l} = {Gm\over 2\pi}\int_0^{2\pi}{\rm d}T
      {{\bfF}_U+{\bfF}_V\sin^2T\over 
      [\lambda_0-\lambda_2 -(\lambda_1-\lambda_2)\sin^2T]^{3/2}},
\eeq
where
\beq
{\bfF}_U = \sum_{j=0}^2 {\bfF}_jU_j, \quad
{\bfF}_V = \sum_{j=0}^2 {\bfF}_jV_j;
\eeq
the ${\bfF}_j$ are defined by equation (\ref{eq:fdef}), and
\bea
U_0 &\equiv & Q_{00}^2     -eQ_{00}Q_{20}+Q_{02}^2    -eQ_{02}Q_{22} \nonumber \\
U_1 &\equiv & Q_{00}Q_{10} -eQ_{10}Q_{20}+Q_{02}Q_{12}-eQ_{12}Q_{22} \nonumber \\
U_2 &\equiv & Q_{00}Q_{20} -eQ_{20}^2    +Q_{02}Q_{22}-eQ_{22}^2     \nonumber \\
V_0 &\equiv & Q_{01}^2     -eQ_{01}Q_{21}-Q_{02}^2    +eQ_{02}Q_{22} \nonumber \\
V_1 &\equiv & Q_{01}Q_{11} -eQ_{11}Q_{21}-Q_{02}Q_{12}+eQ_{12}Q_{22} \nonumber \\
V_2 &\equiv & Q_{01}Q_{21} -eQ_{21}^2    -Q_{02}Q_{22}+eQ_{22}^2.
\eea
The integrals can be expressed in terms of the complete elliptic integrals
$E(k)=\int_0^{\pi/2}{\rm d}\vartheta(1-k^2\sin^2\vartheta)^{1/2}$, 
$K(k)=\int_0^{\pi/2}{\rm d}\vartheta(1-k^2\sin^2\vartheta)^{-1/2}$:
\beq
\int_0^{\pi/2}\frac{{\rm d}T}{{(1-k^2\sin^2 T)}^{3/2}}={E(k)\over 1-k^2}, \quad
\int_0^{\pi/2} \frac{\sin^2 T\, {\rm d}T}{{(1-k^2\sin^2 T)}^{3/2}} 
   ={E(k)\over k^2(1-k^2)}-{K(k)\over k^2}.
\label{eq:ellint}
\eeq
Thus
\beq
[\bff']_l =
{2Gm\over\pi}{\sqrt{\lambda_0-\lambda_2}\over(\lambda_0-\lambda_1)
(\lambda_1-\lambda_2)}[(k^2{\bfF}_U+{\bfF}_V)E(k)-(1-k^2){\bfF}_VK(k)], \quad
k^2={\lambda_1-\lambda_2\over\lambda_0-\lambda_2}.
\label{eq:fav}
\eeq
The averaged radial, tangential and normal components of the force
($[R]_l$, $[S]_l$, $[W]_l$) are found by replacing $\bff'$ in equations
(\ref{eq:rswdef}) by its averaged value (\ref{eq:fav}).

Thus, one of the averages is done, the force of a softened ring at a point is
calculated. To get the fully averaged secular equations (\ref{eqs:avl}), a
second average over the perturbed ring itself is required, as given 
by equation (\ref{eq:avtwo}). The second average is accomplished
by numerical quadrature as described in the following section.

\section{Remarks concerning implementation}

\label{sec:imp}

\noindent
To be sure, we are not yet at the point where we have an optimal N-ring code,
partly because the improvements afforded by averaging lessen the urgency to
optimize, partly because we prefer to explore novel dynamical phenomena rather
than novel algorithms. Obviously, a careful optimization will eventually be
necessary, but for now we simply outline the steps involved in the efficient
implementation of the Gaussian ring algorithm, emphasizing
difficulties encountered, and possible ways around them. We are engaged in
optimizing and parallelizing the code, with a view to simulations of stellar
clusters surrounding massive black holes. 

\subsection{Numerical averaging}
\label{sec:numav}

\noindent
In order to perform the numerical average over the orbit of the perturbed
planet, which gives us equations (\ref{eqs:aveq})-(\ref{eqs:ava}) for the
evolution of its angular momentum and eccentricity vectors, we need to compute
the first three Fourier coefficients in the eccentric anomaly $E'$ of the
force vector ($R_c^0, R_c^1,\ldots,W_s^2$; see eq.\ \ref{eq:avtwo}). Direct
summation at points equally spaced in $E'$ (i.e., a discrete Fourier
transform) will do in most cases, and we have adopted this approach in the
calculations presented below.

The required number of points $K$ for the evaluation of these Fourier
coefficients depends both on the error one is willing to tolerate and on the
eccentricities and separations of the rings. The Euler-Maclaurin summation
formula tells us that the accuracy of the evaluation will improve with
increasing $K$ faster than any power of $1/K$, typically as $\exp(-\alpha K)$
for some constant $\alpha>0$. Of course, during close approaches between rings
this desirable asymptotic behavior may only occur at rather large $K$, and the
constant $\alpha$ may be disappointingly small.

\cite{H1882}, who was mainly concerned with planetary motion (nearly circular,
nearly coplanar, well-separated rings), was happy with $K=12$, even 8. More
generally, he pointed out that `if the number of these values be even, the
order of the error$\ldots$will be the same as that of the power of the
eccentricities or mutual inclination of orbits, whose exponent is one less
than the number of these values'. We can re-state Hill's conclusion--which
we have confirmed using expansions of the disturbing functions in
eccentricity and inclination--more precisely as follows: Consider two rings
with eccentricities and semi-major axes $e_j$, $a_j$, $j=1,2$, and mutual
inclination $I$. Let $\delta=|a_2-a_1|/\overline a$ where $\overline
a=\half(a_1+a_2)$, and let $e=\max(e_1,e_2)$. Then if $e\ll\delta$,
$I\ll\delta$, the fractional error in the time derivatives of the eccentricity
and angular-momentum vectors is $\Order[\max(e/\delta,I/\delta)^{K-1}]$ as $e,
I\to 0$.\footnote{Hill limits this result to the case where $K$ is even. In
  fact, for rings with small but non-zero eccentricity the result also holds
  for odd $K$. Hill's qualification apparently relates to the case of
  inclined, circular rings: in this case odd $K$ yields much faster
  convergence than even $K$, particularly if one point lies on the mutual
  node.} This result also implies that the implementation of Gauss's method
using numerical quadrature with $K$ equally spaced points is formally
equivalent to a secular perturbation theory in which the Hamiltonian is
expanded to first order in the planetary masses and $K^{\rm th}$ order in the
eccentricities and inclinations (e.g., \citealt{lh08}, for $K=12$).

Going by this estimate, Hill's use of $K=12$ would yield a relative
error of $10^{-10}$ in the time derivatives of the eccentricity and
angular-momentum vectors of a two-planet system consisting of Jupiter
and Mercury (eccentricity 0.2); numerical experiments on this system,
with varying orientations, yielded relative errors in the range
$10^{-9}$--$10^{-12}$ in reasonable agreement with this estimate. To
achieve a similar accuracy for the system consisting of Jupiter and
Halley's comet our experiments indicate that one needs $K\sim
200$--400, and this again depending on argument of node and
perihelion, while keeping the other elements fixed.  A softening
length $b=0.01a$ or $b=0.1a$ can cut this number by about a factor of
two or four, respectively.  The stellar systems which we will be
dealing with have higher eccentricities and inclinations than anything
Gauss or Hill envisioned, as well as frequent ring crossings. One can
get away with $K=16$--32 for nearly circular rings, while achieving a
relative error of $10^{-10}$ or smaller in the evolution
equations. For nearly radial orbits, we found that we needed up to
$K=512$ for the same level of accuracy.

It is useful to have a criterion for the accuracy of the numerical
average, which can be used to provide an adaptive choice for $K$ for
each ring pair.  Hill's criterion works when $e\ll\delta$,
$I\ll\delta$, but not for near-radial or crossing rings.  We devised
an alternative criterion which is equivalent to Hill's in the limit
where Hill's works. We rely on the observation that, had the averages
been exact, the equation governing the evolution of a ring's
semi-major axis would be $[{\rm d}a'/{\rm d}t]_{ll'}=0$ (see discussion following
eq.\ \ref{eq:poiss}), which in turn implies the identity
(\ref{eq:adot}). Therefore we may simply choose $K$ adaptively so that
the numerical evaluation of the absolute value of the left side of
(\ref{eq:adot}), divided by ${n'}^2a'$ to make a dimensionless
quantity, is less than some specified tolerance $\epsilon_{\rm quad}$
(`quad' for quadrature) for each ring pair. We start with $K=16$
and proceed through doublings. Note that Richardsonian extrapolation
is not useful here, because the asymptotic convergence is faster than
any power of $1/K$. What is an appropriate tolerance?  A tolerance
$\epsilon_{\rm quad}$ translates into a tolerance on the relative
error in semi-major axes per secular or precession period that is no
worse than $(N - 1)\epsilon_{\rm quad}$. This in turn translates into
into a relative error over $M$ secular periods no worse than
$NM\epsilon_{\rm quad}$.  Presuming one is interested in simulating a
system of $10^5$ particles over 100 precession periods, and one can
live with cumulative relative error in the dominant Keplerian energy
of $10^{-4}$, then ${\epsilon}_{\rm quad} \sim 10^{-11}$ should be
amply sufficient for such a bound. For most ring-ring interactions,
such a tolerance is achieved with 64 to 128 sectors per ring at a
softening of 0.01 times the typical size of a ring. There are of
course likely to be some ring pairs for which sampling of 16 to 32
times these values may be necessary, and in these cases adaptive
sampling proves essential.

So far we have described how to carry out numerical averaging over the
perturbed ring using equally spaced points in eccentric anomaly. However,
we are aware that more sophisticated quadrature routines may be better,
particularly at ring crossings when the softening length $b$ is small.  We
have conducted limited experiments with the QAGS adaptive integrator from GSL
(intended to handle integrable singularities by adaptive application of the
Gauss-Kronrod [Gauss again] rule). In the rather extreme case of two coplanar,
intersecting rings, with a softening of only $0.001$ times the orbital radius,
QAGS required 1500 judiciously placed integration points to achieve a relative
error of $10^{-11}$; ten times more points were required by uniform sampling
to achieve the same tolerance. The factor of 10 drops to 2 for a softening
length which is ten times larger.

\subsection{Numerical integration}
\label{sec:numint}

\noindent
The rings are evolved in time using a Bulirsch-Stoer (BS)
integrator. The BS integrator is not tailored to this problem, but for
a reliable, brute force exploration, it is efficient and simple to
implement. The ideal of course would be to exploit the Lie-Poisson
structure of the equations to construct a geometric integrator with an
adaptive timestep that would handle the broad range of dynamical
time-scales in typical stellar and planetary systems. This is certainly
a noble undertaking, which we put off to a later stage, and we will be
more than happy if others beat us to it. In any case, in our BS
integrations, we set a tolerance $\epsilon_{\rm int}$ for the error in
the integrated vector which is similar to the tolerance $\epsilon_{\rm
quad}$ that we demand in the numerical quadratures. In most cases, we
are quite happy with a tolerance of $\epsilon_{\rm int}=10^{-8}$ (at
which we live with step sizes on the order of 0.1 secular periods);
we do push things further in the simulations below, and evolve
some configurations with a tolerance of $\epsilon_{\rm int}=10^{-12}$,
and consequently smaller timesteps on the order of 0.01 secular periods.

Highly eccentric orbits are likely to be common in many nearly
Keplerian systems (e.g., stellar clusters around black holes, cometary
clouds).  In the plane one has the lens orbits of \cite{st99}; related
families exist in three dimensions and were studied in oblate and
prolate potentials by \cite{ss00}.  The averaging process we describe
here is well-behaved for nearly radial orbits ($e\to1$), as are the
equations of motion for the eccentricity and angular-momentum
vectors--which is one reason why we chose to work with vectors rather
than orbital elements, which usually give singular equations of motion
as $e\to1$ \citep{tre01}. Nevertheless we have found that our BS
integrator struggles with nearly radial configurations, slowing the
integration by a factor of 10 to 1000 in some extreme cases.

Part of the slow down is dynamical, and will be faced by any adaptive
integration scheme. The other, and more critical, part is a
consequence of the violation of the constraint $L^2+A^2=1$ in the
course of the numerical evolution; it manifests as a square-root
singularity in the equations of motion
(\ref{eqs:aveq})--(\ref{eqs:ava}), and forces a slowing down in the
integration, as steps are reduced to avoid that singularity. One can
imagine a number of solutions to this problem.  As suggested above,
the best would be to design a geometric integrator which is adapted to
the Lie-Poisson structure of the equations at hand, and hence
preserves the integral in question automatically. Less elegant, but
still effective, would be to enforce the constraint $L^2+A^2=1$ by
hand, renormalizing one or the other vector at regular intervals to
eliminate the small errors introduced by the integration
scheme. Without modifying the BS integrator, one could impose a
reflecting boundary condition at high eccentricity. There are also
physical processes that mitigate the problem in many systems: these
include general-relativistic precession (Appendix \ref{app:gr}) and
tidal disruption, collisions, or swallowing of stars by the central
object. A careful examination of the proper implementation of these
and related alternatives would take us too far afield.  In the ring
simulations described below, which include general-relativistic
precession, we encountered an occasional near-radial ring that slowed
down the integration of the whole cluster. We found that this problem
could be removed either by brute-force removal of highly eccentric
orbits ($e >0.999$) or by exploiting the redundancy in the present
vectorial formulation. In the second solution, we simply used the
eccentricity vector $\bfA$ to solve for the eccentricity at high
eccentricities and the angular-momentum vector $\bfL$ to solve for the
eccentricity otherwise. The resulting algorithm still slowed down at
high eccentricity passages, but only by that dynamical factor which
reflects the higher precession rates at these eccentricities, boiling
down to a factor of 10 decrease in performance during these phases. We
found this intermittent slowing down--about a factor of ten when a ring
(or a population of rings) developed mean eccentricity greater than
$0.8$--tolerable in the cases we studied, and believe it manageable
with larger systems, when the code is eventually parallelized.

\subsection{Evaluation of elliptic integrals} 

\noindent
To compute the elliptic integrals one could either iterate using the
arithmetic-geometric mean (Gauss once again), interpolate in a table, or rely
on Chebyshev polynomial expansions.  We tried all three methods and
found that smart searches through tables were twice as costly as Chebyshev
expansions, which were marginally better than the arithmetic-geometric mean
(for an absolute error goal of $10^{-14}$). In our current implementation,
elliptic integrals are calculated with the arithmetic-geometric mean.

\subsection{Additional physics} 

\noindent
Nearly Keplerian stellar systems are often subject to additional effects with
substantial secular contributions. We can easily accommodate the average
contributions of tides (Galactic for cometary clouds, bulge/halo for stellar
clusters around a central black hole). We can also account for dynamical
friction contributions, in a manner analogous to related effects which one of
us has handled in studies of the history of the lunar orbit \citep{tw98}. Finally,
we can also account for the secular effects of general-relativistic
corrections (Appendix \ref{app:gr}).

\section{The algorithm tested}
\label{sec:test}

\noindent
In this section, we test the Gaussian ring algorithm by comparing its
results to the behavior of the equivalent unaveraged N-body system,
integrated with the same Bulirsch-Stoer (BS) algorithm at equivalent
tolerance. Alternatively, one may match numerically averaged solutions
with results of analytical averaging, an exercise which we leave to a
forthcoming discussion of analytical theories of softened systems of
coupled rings (Mrou\'e \& Touma, in preparation).

The desired accuracy of the calculation is specified by two parameters: the
first is $\epsilon_{\rm quad}$, the fractional accuracy of the numerical
averaging, which was discussed in \S\ref{sec:imp}; and the second is
$\epsilon_{\rm int}$, the tolerance specified in the numerical integrator for
the ordinary differential equations. 

A technical point is that when comparing secular computations to
unaveraged calculations, one should choose the initial conditions of
the unaveraged system to ensure that the comparison is not affected by
high-frequency oscillations \citep{ls88}. This can be done by
experimenting with various initial conditions for the unaveraged
calculations to find one at which the high-frequency oscillation is
near a node, or by techniques such as `warming up' the unaveraged
integration \citep{st92}. In the present paper, however, we simply use
the same initial conditions for both N-body and averaged calculations.

The simulations below involve stars orbiting a $10^7\,{\rm M}_\odot$
black hole.  Distances are measured in parsecs, time in years, and
mass in units of the black-hole mass. In these units, Kepler's law
relating the period $P$ and semi-major axis $a$ is $P=2.96\times
10^4\hbox{\,yr}\,(a/\pc)^{3/2}$. Orbital inclinations are measured
with respect to the $x$-$y$ plane of a right-handed reference frame,
with origin at the central black hole; longitudes of node and
periapsis are referred to the $x$-axis of this frame. The dominant
relativistic correction (apsidal precession due to the central mass)
is sometimes included, as described in Appendix \ref{app:gr}.

\subsection{Softened Kozai oscillations}

\noindent
To start out, consider a system that represents a star orbiting one of
the components of a binary black hole at the centre of a galaxy. The
black-hole companion is represented by a fixed ring having mass
$M_{o}= 1$, semi-major axis $a_{o}=10$, and eccentricity $e_{o} = 0.5$
(recall that semi-major axis is in units of parsecs and mass in units
of the mass of the primary black hole, $10^7\,{\rm M}_\odot$). The fixed
orbital plane of the black hole binary serves as the $x$-$y$ plane of the
reference frame, with the $x$-axis taken perpendicular to the line of
apsides, the $y$-axis pointing to the periapsis, and
the $z$-axis along the orbit normal. The star is represented by a ring
with $M_{i}= 10^{-7}$, $a_{i}=0.1$, and initial eccentricity
$e_{i}=0.01$. Its orbital plane is initially inclined by
$I_{i}=60^{\circ}$, cuts the plane of the binary along the $x$-axis, and
has a periapsis which is $90^{\circ}$ from the node. The softening is
$0.1 a_{i}$; relativistic precession is switched off. We first
carried out a direct (restricted) three-body integration (BS tolerance
${\epsilon}_{\rm int} = 10^{-12}$) with these initial conditions,
supplemented by randomly chosen mean anomalies. Then we performed a
number of ring simulations for comparison. At a fixed sampling of 16
sectors per ring, a maximum error of $10^{-12}$ in the Keplerian
energy was comfortably achieved. The Gaussian ring algorithm follows
the secular evolution with giant strides ranging from $1.4 \times
10^{7}$ ($\epsilon_{\rm int}=10^{-14}$) to $4.5\times10^{7}$ yr
($\epsilon_{\rm int}=10^{-6}$). 

Below, we compare results of the secular ring computation at
$\epsilon_{\rm int} = 10^{-12}$ with the benchmark particle simulation. In
this rather conservative run, the timestep in the averaged simulation
($\sim 1.8 \times10^7$ yr, $1/40$ of the Kozai period) was $2 \times
10^5$ times larger than in the direct three-body integration ($\sim
90$ yr, $1/10$ of the orbital period).

Neither energy nor angular momentum is preserved in the restricted,
eccentric, three-body calculations.  In the averaged dynamics, the
star's energy is conserved but its angular momentum is not. The energy
conservation provides a check on the accuracy of the Gaussian ring
algorithm: the ring's secular energy was conserved to a maximum
fractional error of $4\times10^{-10}$, over a hundred Kozai
cycles\footnote{By {\em secular} energy, we mean the value 
of the averaged Hamiltonian that generates the
secular dynamics.  Relative errors in the {\em total} energy come
out better (by several orders of magnitude) than one might expect for
the given value of $\epsilon_{\rm int}$ and $\epsilon_{\rm quad}$; the
reason is that the denominator is then dominated by the {\em
Keplerian} energy $-\half GM_\star/a$, which is irrelevant to the actual
integration.}. 

\begin{figure}
\begin{center}
\epsfxsize= 8 in 
\epsfysize= 8 in
\begin{tabular}{cc}
\epsfig{file=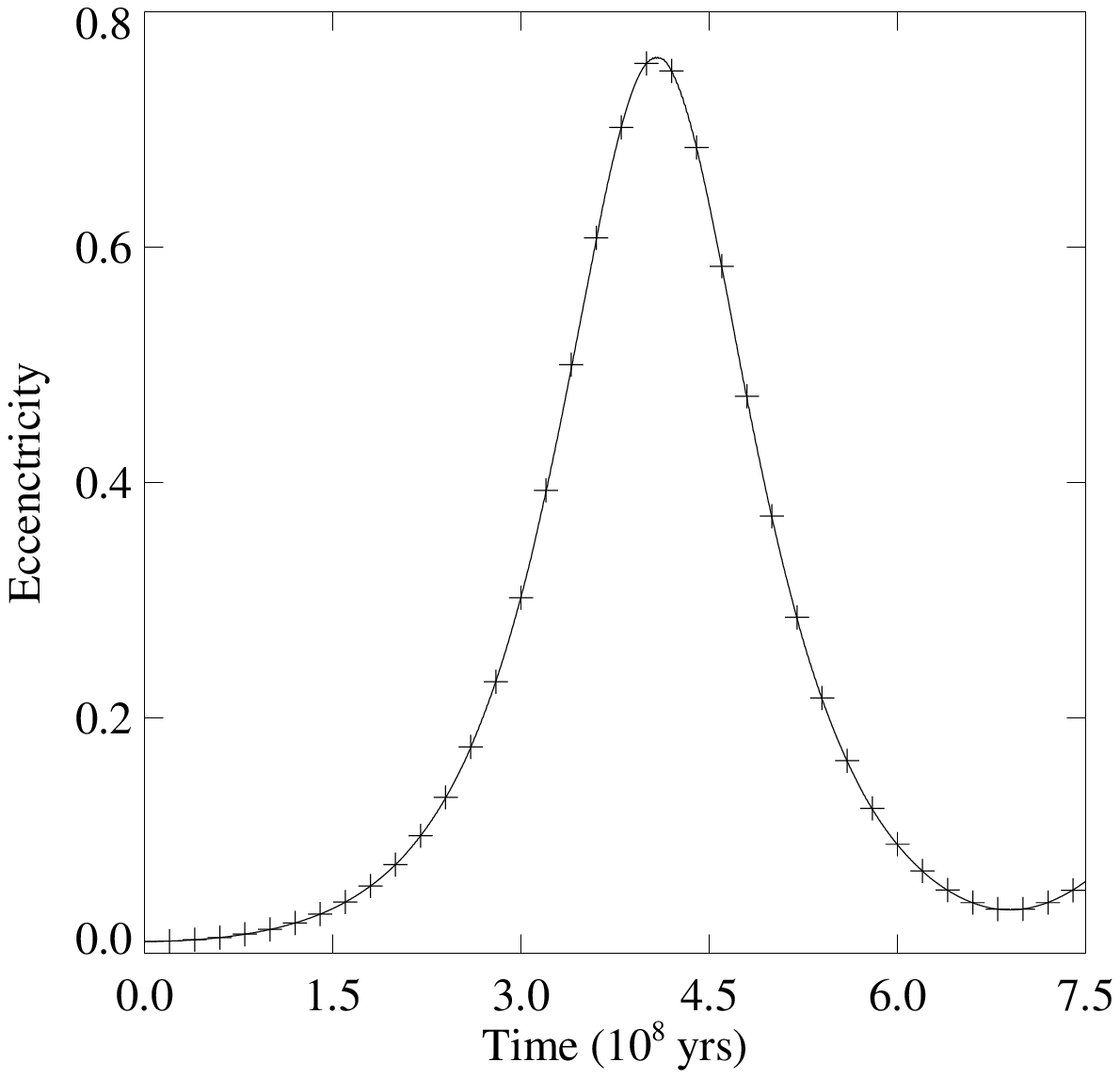,width=0.45\linewidth,clip=} &
\epsfig{file=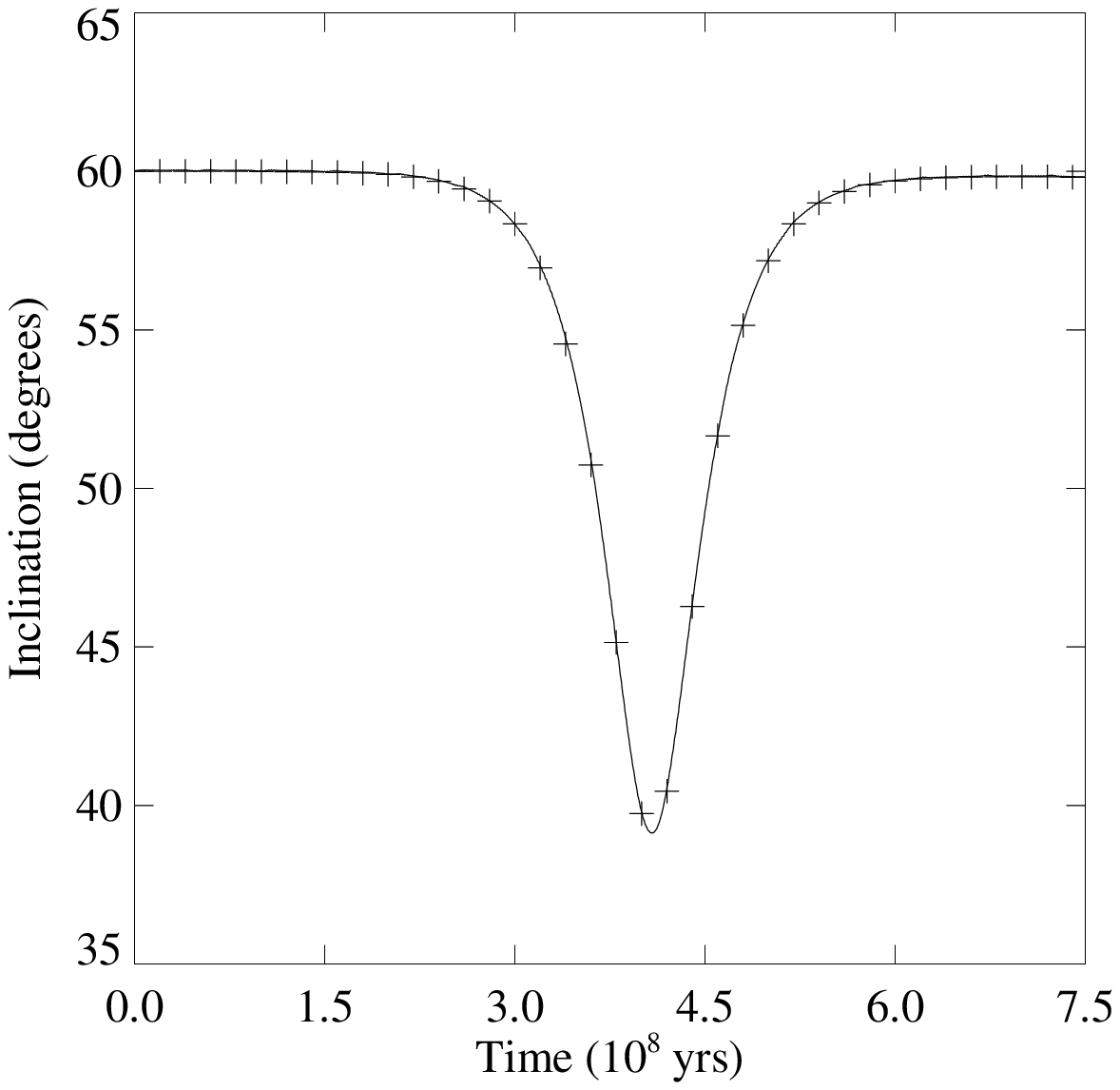,width=0.45\linewidth,clip=} \\ 
\epsfig{file=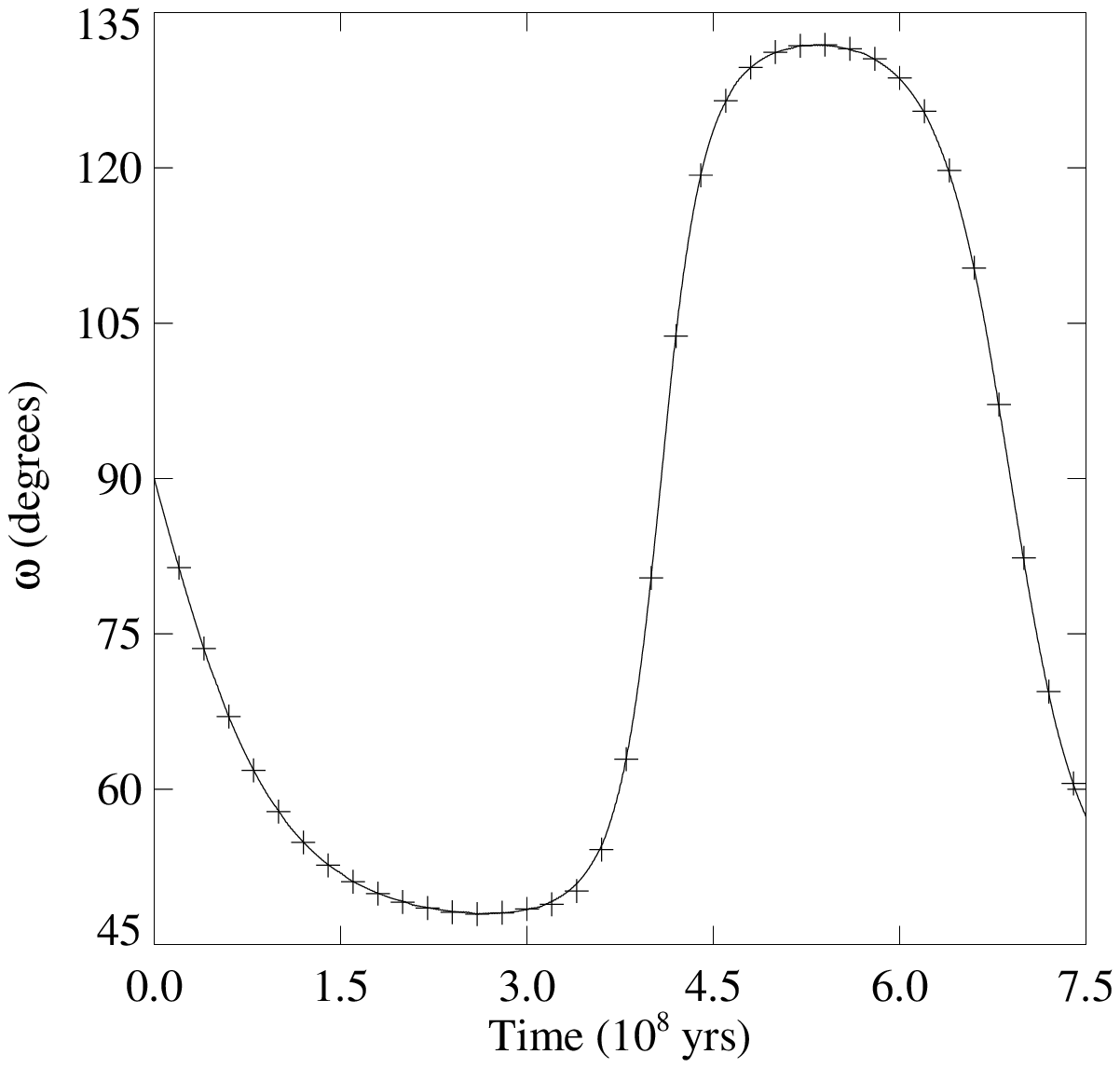,width=0.45\linewidth,clip=} &
\epsfig{file=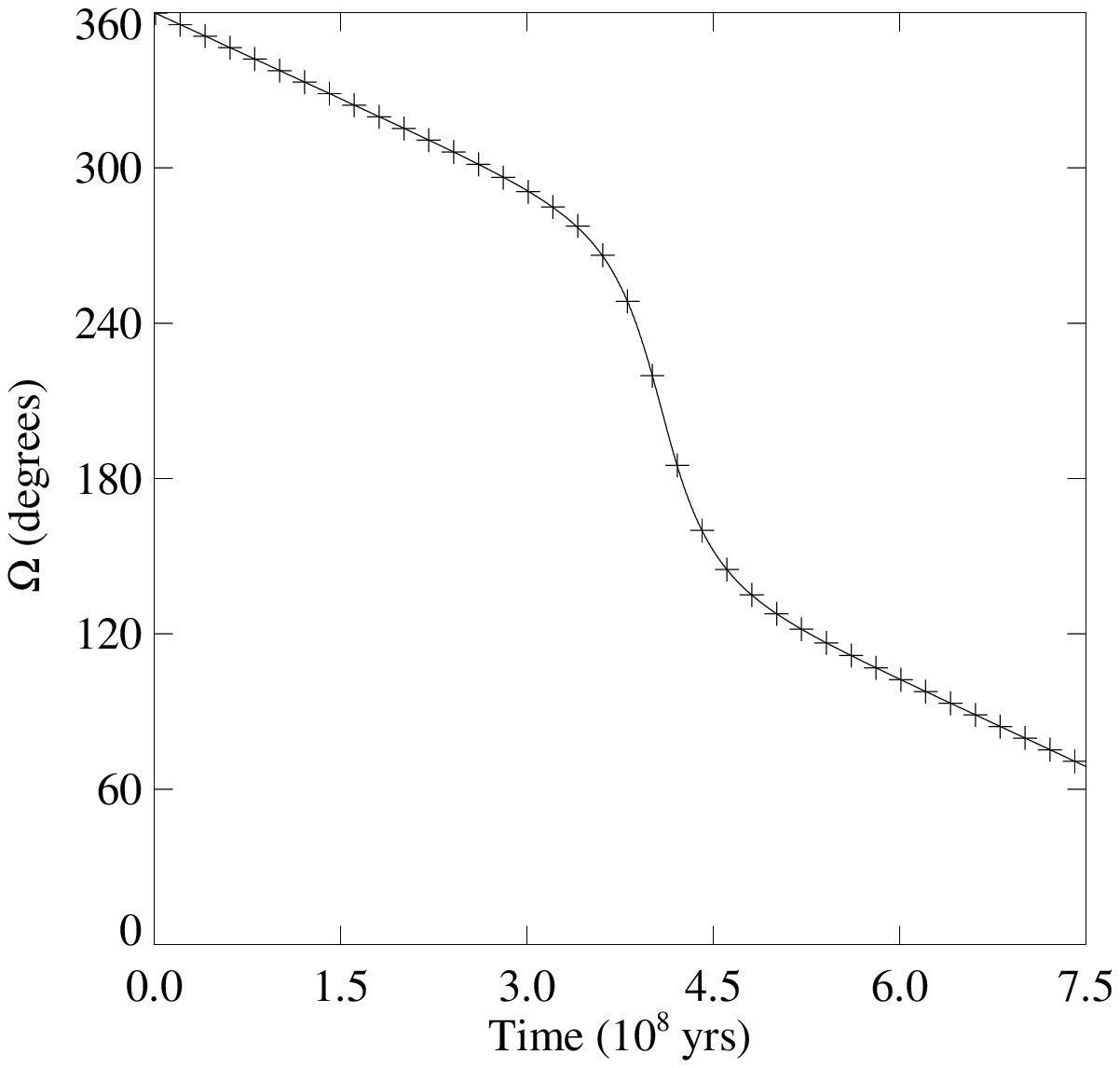,width=0.45\linewidth,clip=}
\end{tabular}
\end{center}
\caption[fig1]{Osculating elements of a star in a black-hole binary, as
  determined from a direct three-body integration (solid curve) and from
  the Gaussian ring algorithm (crosses) over a single
  Kozai cycle.}
\label{fig:t1_elts}
\end{figure}

\begin{figure}
\begin{center}
\epsfxsize= 8 in
\epsfysize= 8 in
\begin{tabular}{cc}
\epsfig{file=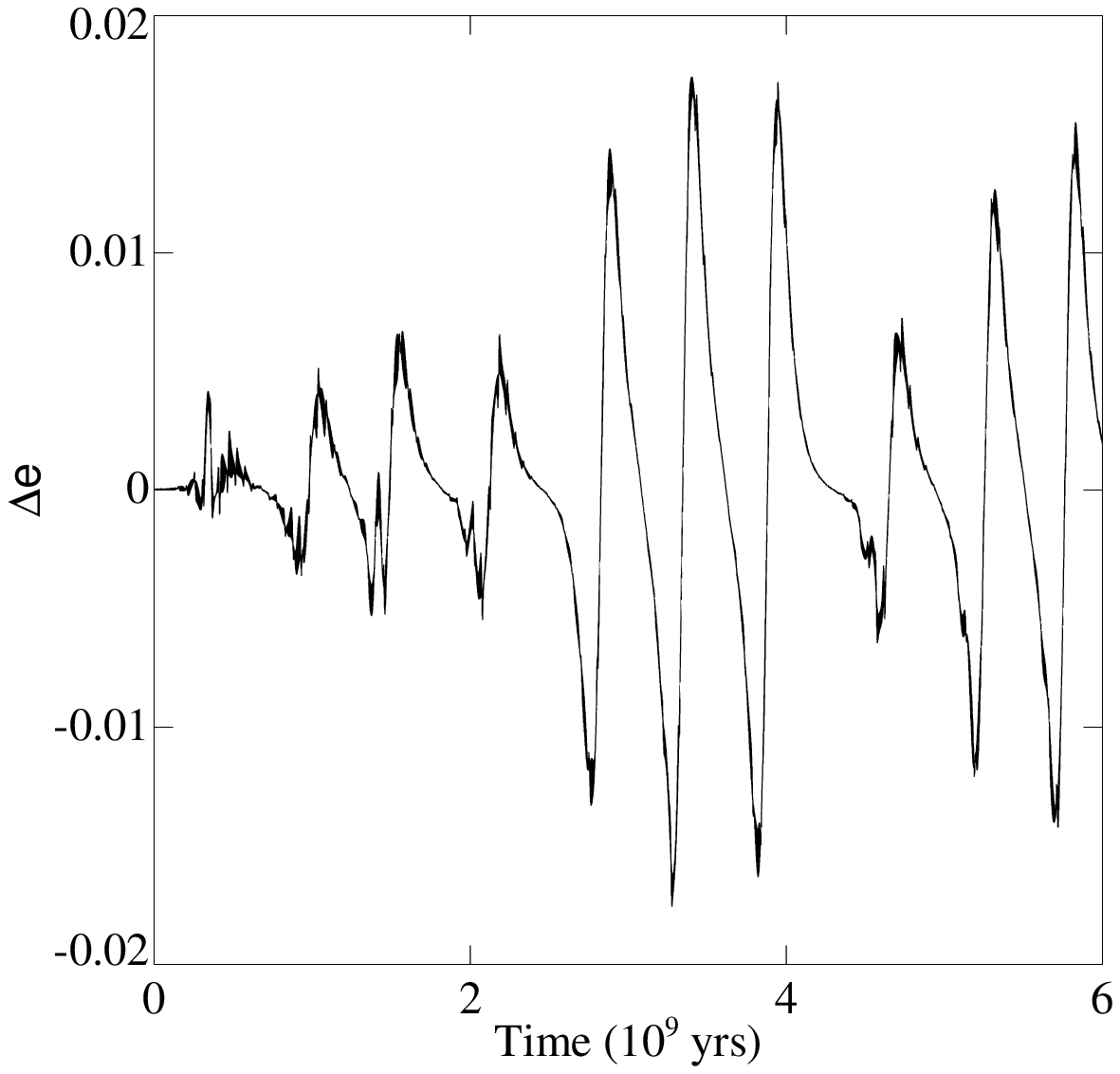,width=0.45\linewidth,clip=} &
\epsfig{file=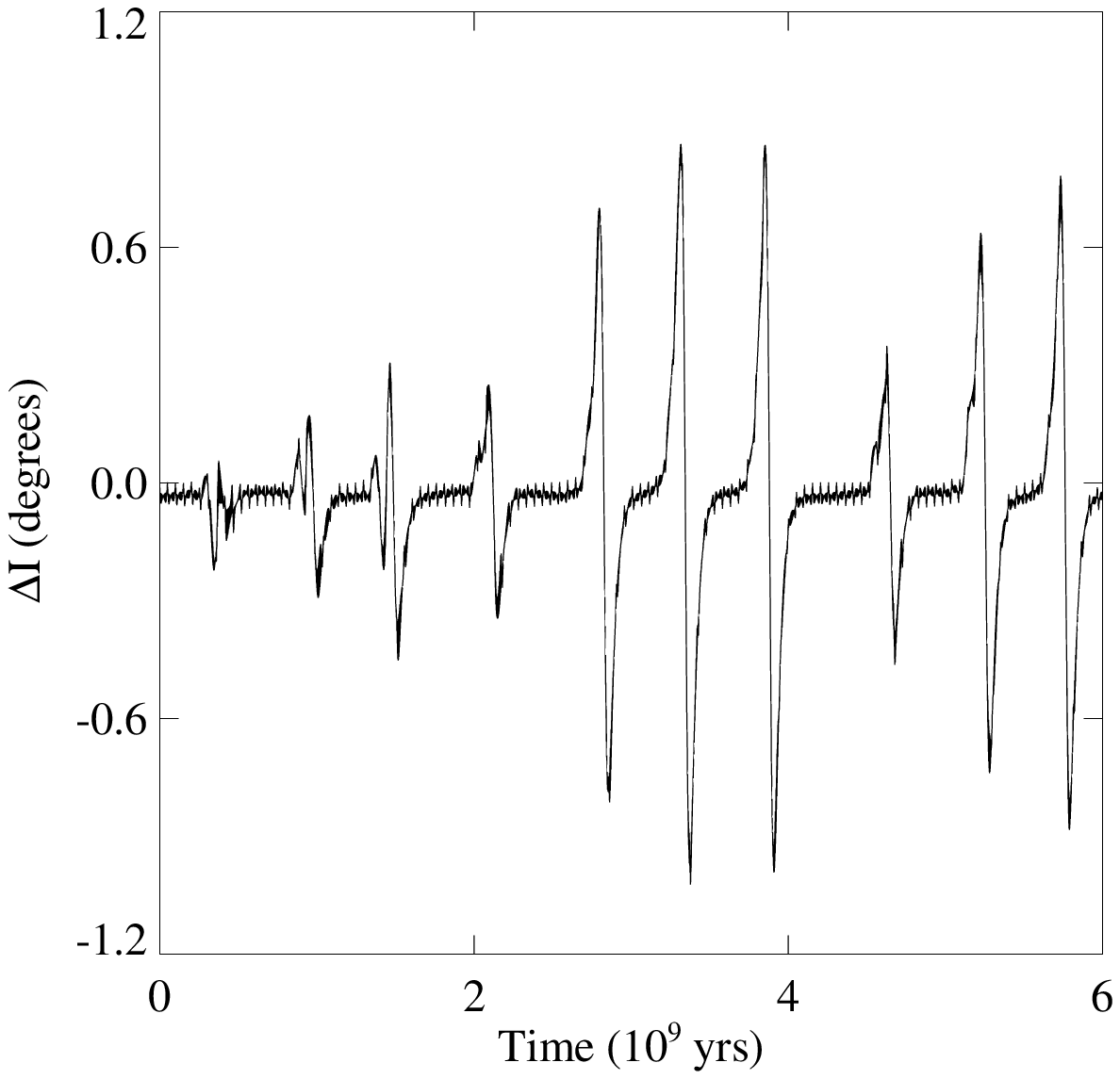,width=0.45\linewidth,clip=} \\ 
\epsfig{file=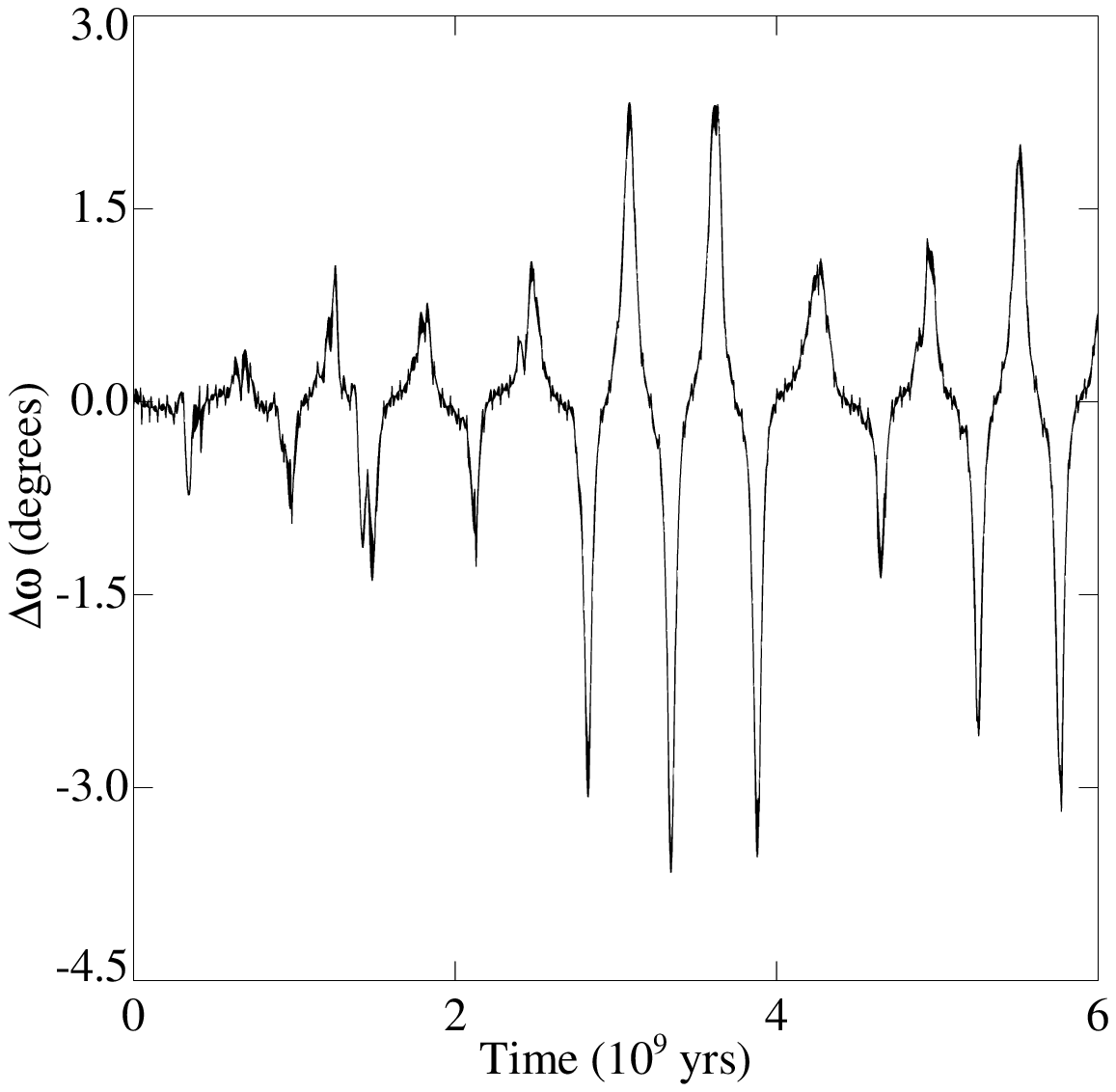,width=0.45\linewidth,clip=} &
\epsfig{file=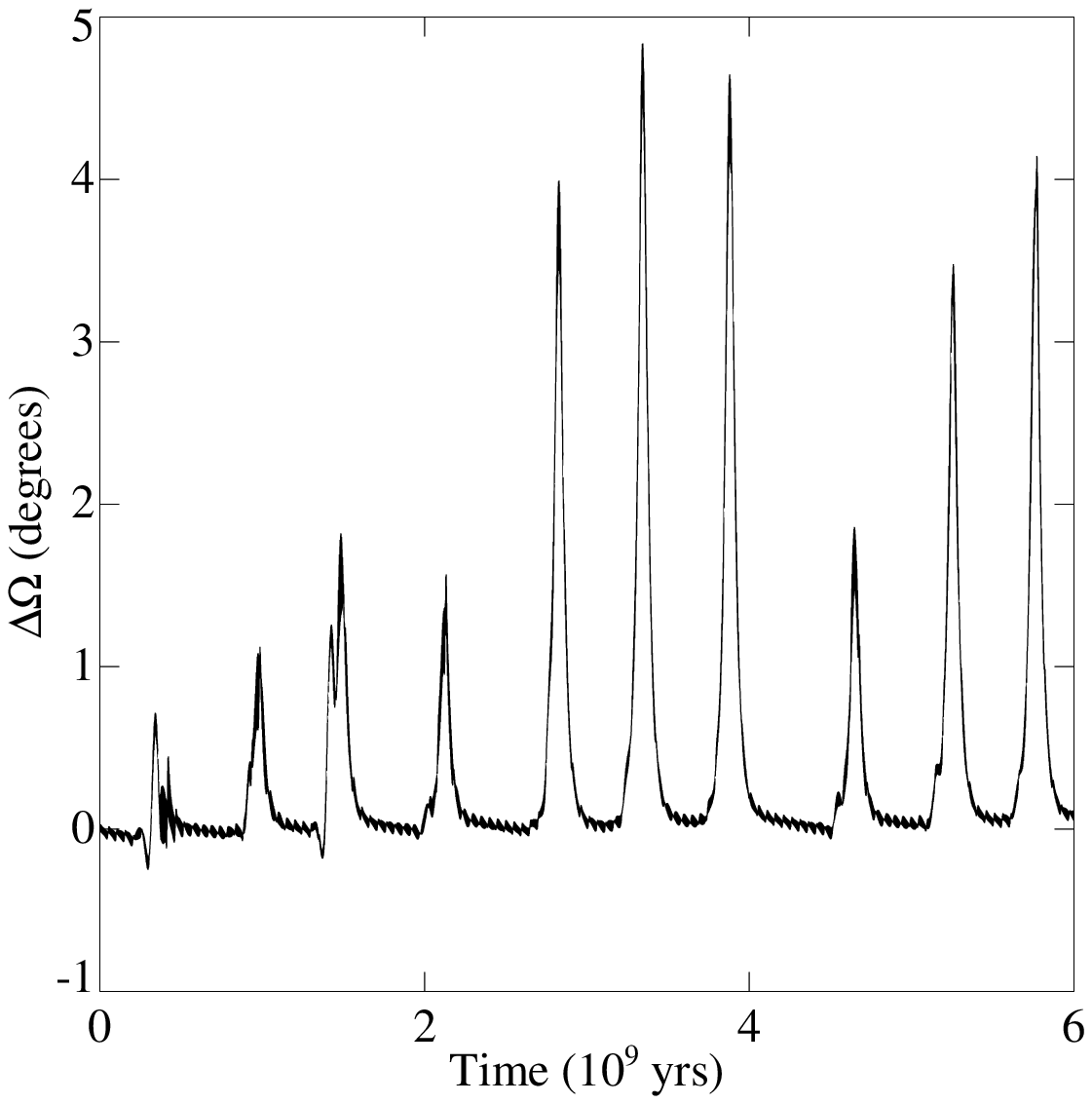,width=0.45\linewidth,clip=}
\end{tabular}
\end{center}
\caption[fig2]{Difference between the osculating elements of a star in a
  black-hole binary, as determined from a direct three-body integration and from
  the Gaussian ring algorithm, over ten Kozai cycles.}
\label{fig:t1_errors}
\end{figure}

The orbital elements of the star from the three-body and secular codes
are shown superimposed in Fig.\ \ref{fig:t1_elts}. The eccentricity
and inclination variations are due to Kozai oscillations. To the naked
eye, the runs are practically indistinguishable in this figure. A
closer look reveals two types of (minor) differences between the two
codes, as shown in Fig.\ \ref{fig:t1_errors}: (i) small-amplitude
variations over the binary's orbital period, which average out in the
secular dynamics; (ii) relatively larger variations with the same
period as the Kozai cycle. The latter are mainly due to differences in
the initial conditions between the three-body and secular codes; we
were able to reduce them significantly by using averaged initial
conditions for the secular dynamics as described above. While the
differences in initial conditions are caused by second-order terms in
the masses, the time evolution of differences is not; in fact, ring
simulations with slight variations in initial conditions revealed
similar (qualitative and quantitative) time variations in the
differences of their orbital elements.

\subsection{Counter-rotating rings}

\noindent
The linearized dynamics of softened, planar, counter-rotating, nearly
Keplerian systems of stars and gas was considered by \cite{tou02}. There it was
shown that (i) such configurations are prone to violent $m=1$ (lopsided)
instabilities; (ii) the bifurcation to instability, the growth rate and the
pattern speed are functions of softening; (iii) the instability is associated
with interaction of modes of opposite energy.

\begin{figure}
\begin{center}
\epsfxsize= 8 in 
\epsfysize= 8 in
\begin{tabular}{cc}
\epsfig{file=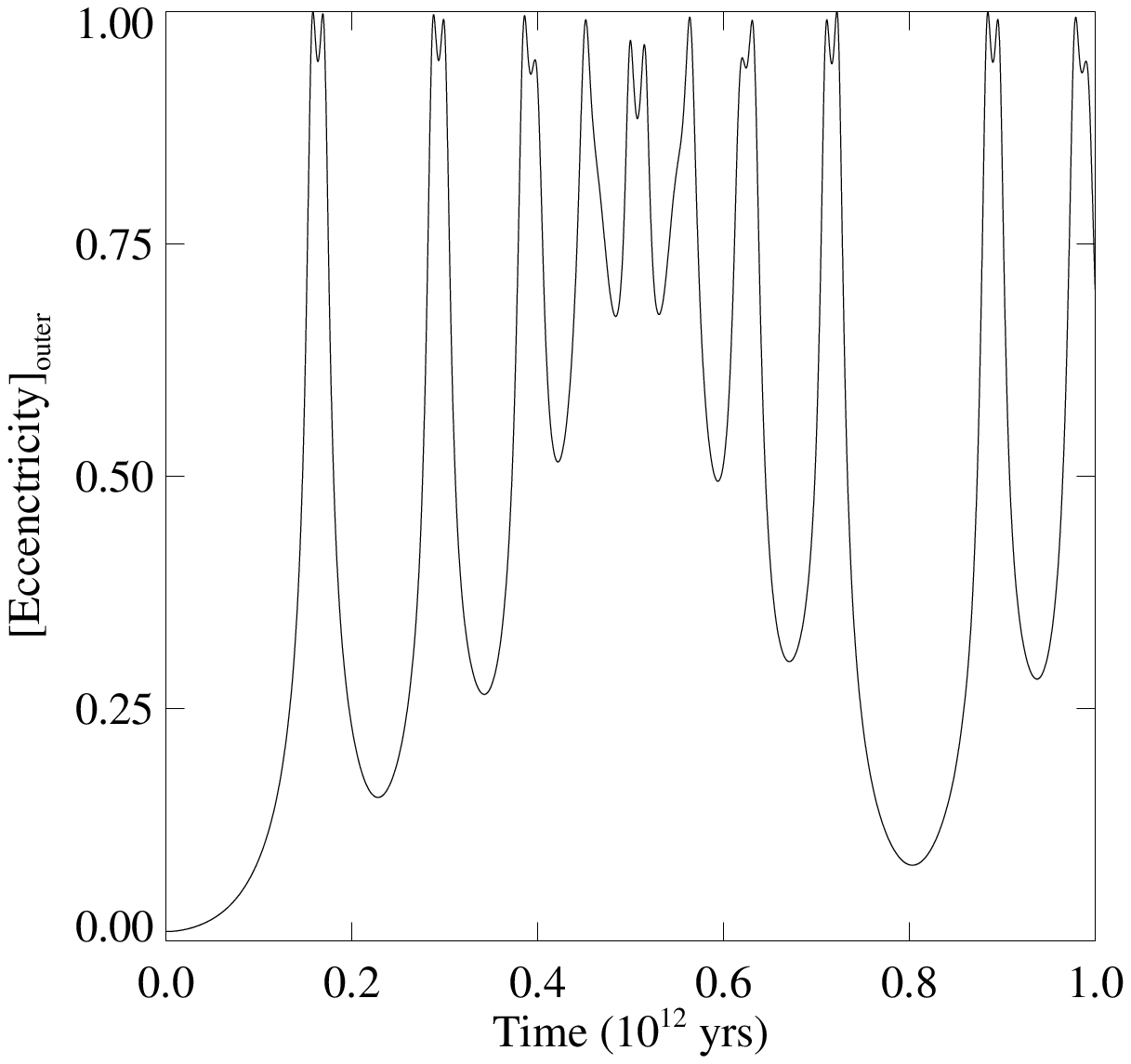,width=0.45\linewidth,clip=} &
\epsfig{file=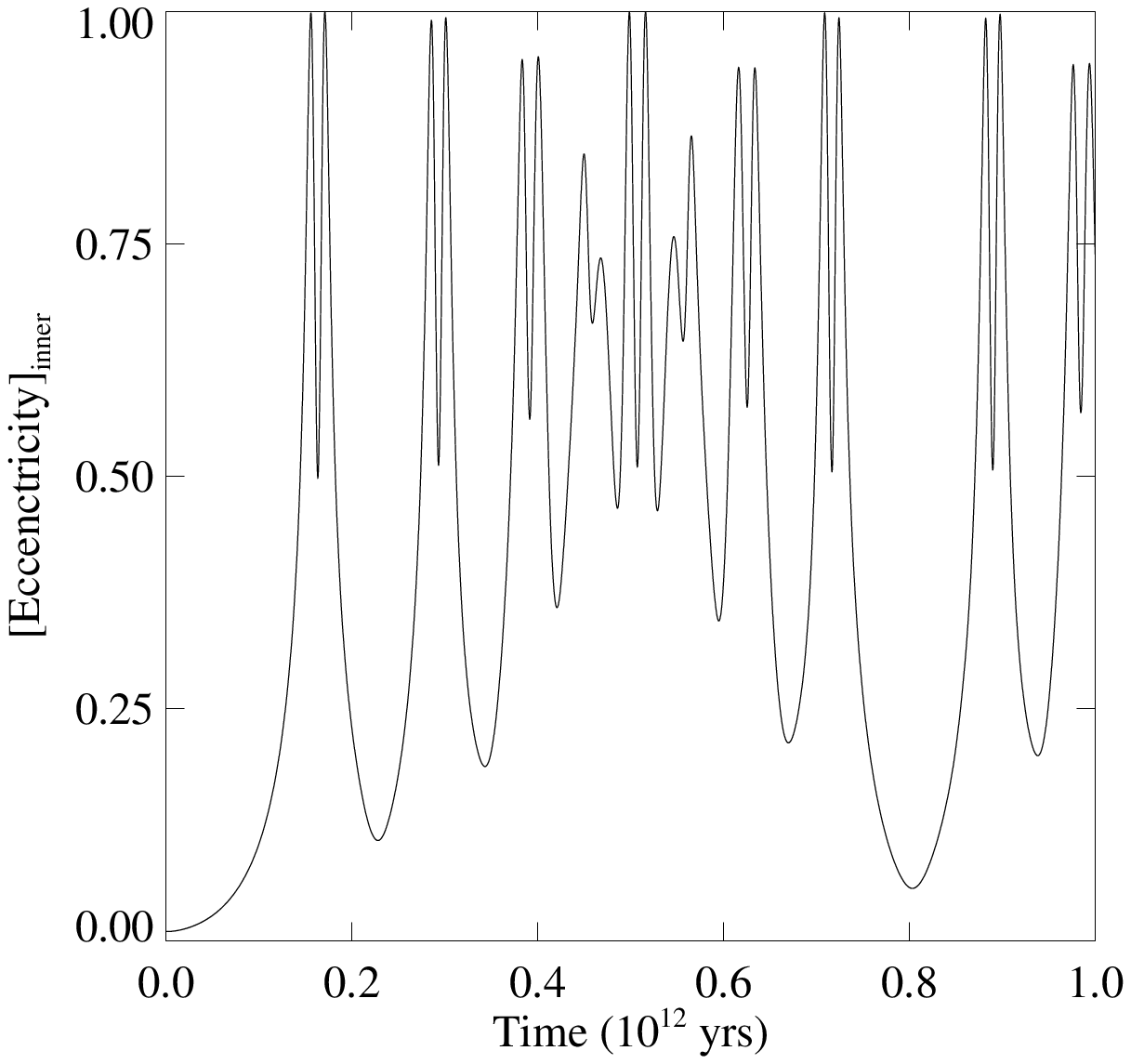,width=0.45\linewidth,clip=} \\ 
\epsfig{file=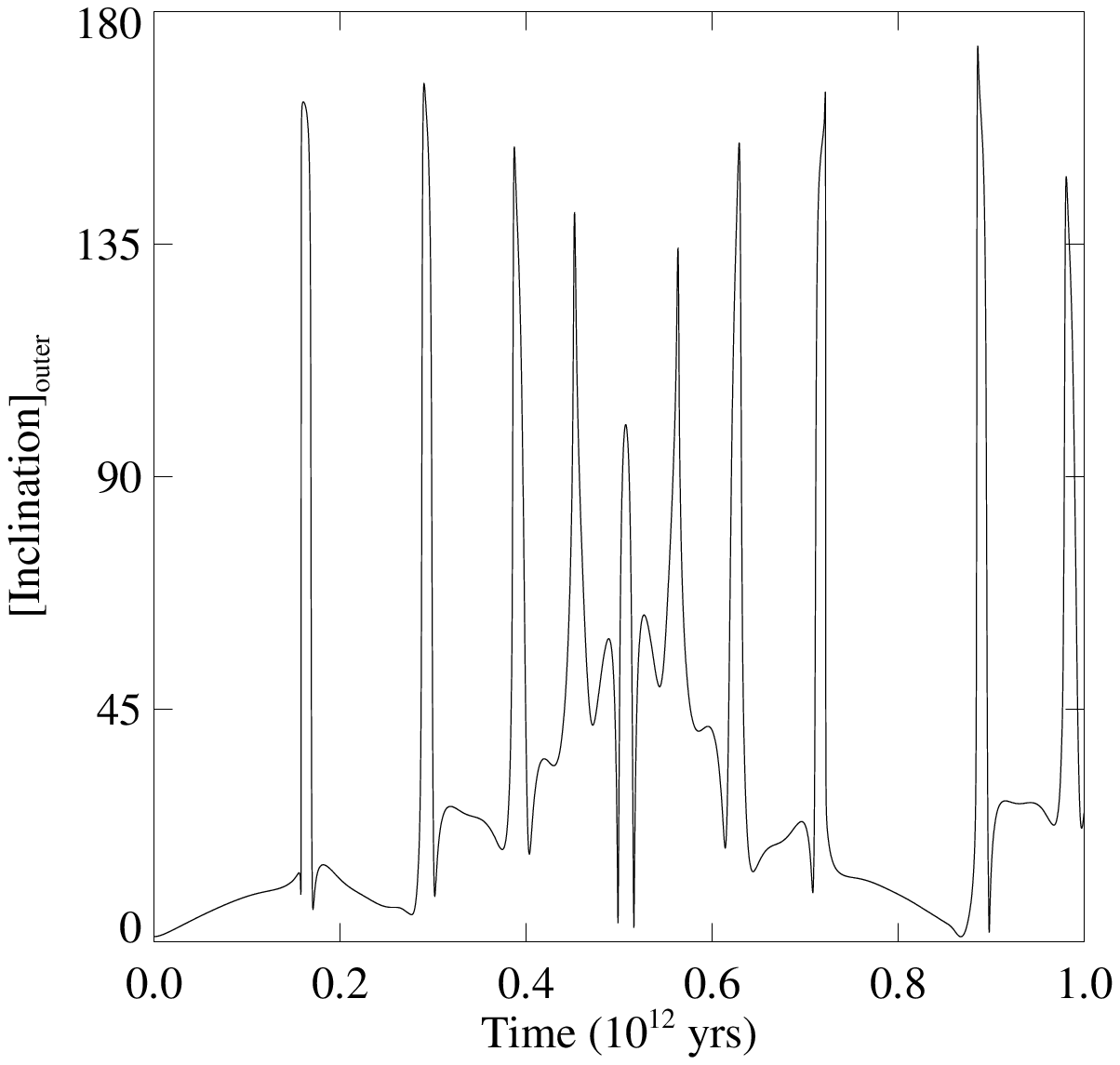,width=0.45\linewidth,clip=} &
\epsfig{file=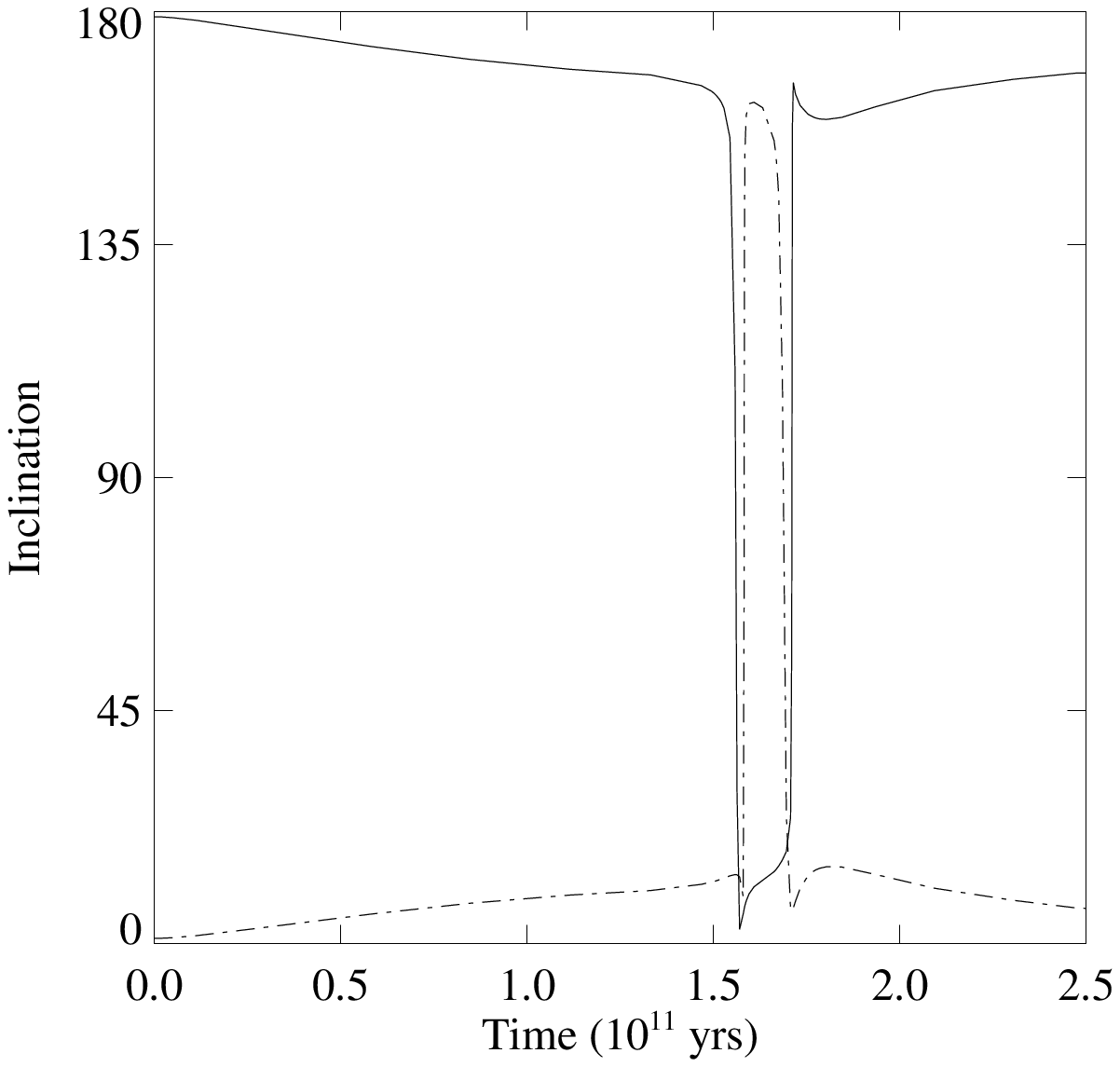,width=0.45\linewidth,clip=}
\end{tabular}
\end{center}
\caption[fig3]{Evolution of a counter-rotating system of two nearly circular,
  nearly coplanar rings. Top: eccentricity oscillations of the outer,
  initially prograde, ring (left), and of the inner, initially retrograde,
  ring (right). Bottom left: variations in the inclination
  of the outer ring. Bottom right: an expanded view of the exchange of angular
  momentum and inclination between the inner (solid) and outer (dashed)
  ring.}
\label{fig:t2_elts}
\end{figure}

The secular theory employed by \cite{tou02} was based on a Hamiltonian that
was truncated at $\Order(e^2)$. Mrou\'e \& Touma (in preparation) extend the
analysis to $\Order(e^4)$, and provide a description of the unfolding of the
bifurcation, the appearance of uniformly precessing ring configurations, and
the secular dynamics around them. The Gaussian ring algorithm offers an ideal
tool for exploring this secular dynamics to all orders in inclination and
eccentricity. Results on the planar two-ring problem will be presented in
Mrou\'e \& Touma. Here, we further test our method by following the
development of a nearly coplanar, counter-rotating system of two rings.

Specifically, we follow two equal-mass rings, in initially nearly circular
($0.01$ eccentricity), nearly coplanar ($1^{\circ}$ inclination),
counter-rotating orbits. We consider a case in which $M_{o} = M_{i} =
10^{-6}$, $a_{o} = 1$ and $a_{i} = 0.5$, with softening $b= 0.3 a_{o}$ and
relativistic precession switched off. This is a linearly unstable 
configuration, in which the eccentricity of both rings is expected to grow
(note that reducing the softening length below $0.26 a_{o}$ switches off the
instability, at this ratio of semi-major axes). The evolution is shown in
Fig.\ \ref{fig:t2_elts}, where one observes oscillations in the eccentricity
from near-circular to near-radial and back. Even more dramatic
is the inclination dynamics: the rings experience repeated flips between
prograde and retrograde near the peaks of the eccentricity oscillations. The
cause of this out-of-plane instability is not fully diagnosed,
though it smells like Kozai cycles. To be sure, this behavior, which may
stimulate bending instabilities in near-Keplerian discs, is markedly different
from the behavior of counter-rotating configurations studied in the context of
the three-body problem. There, it has been understood for decades that 
counter-rotating configurations are rather stable, even more so than their
prograde counterparts. It appears that softening (or heat) modifies the
secular frequency spectrum to a point that angular momentum exchange between
rings leads to an instability, which first promotes eccentricity growth
to a near-radial configuration, then inclination growth to flipped orientations,
and back.

In the course of these acrobatics, the secular algorithm satisfied an
averaging tolerance $\epsilon_{\rm quad}=10^{-14}$, by working its way
up to 128 subdivisions per ring (from a minimum of 16) in the high
eccentricity/inclination phases of the evolution. The dynamics was
followed with a mean timestep of about 1/100 of the oscillation
period, dictated by an integrator tolerance $\epsilon_{\rm
int}=10^{-12}$. At this tolerance, the relative error in the secular
energy was no more than a few times $10^{-9}$ over 100 oscillations;
the fractional error in the total angular momentum was bounded by 
$3\times10^{-11}$ over that same period.

\begin{figure}
\begin{center}
\epsfxsize= 8 in
\epsfysize= 8 in
\begin{tabular}{cc}
\epsfig{file=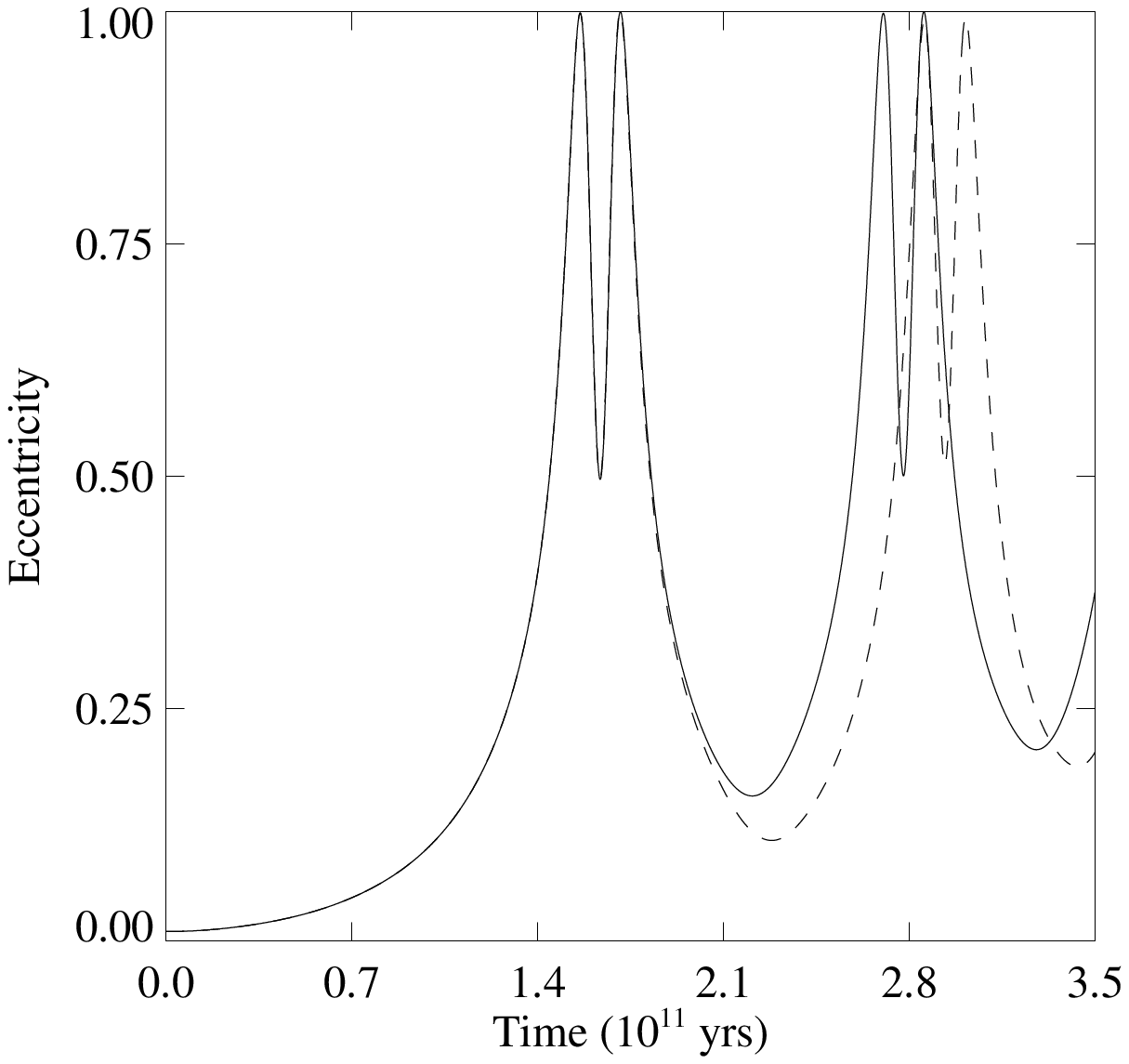,width=0.45\linewidth,clip=} &
\epsfig{file=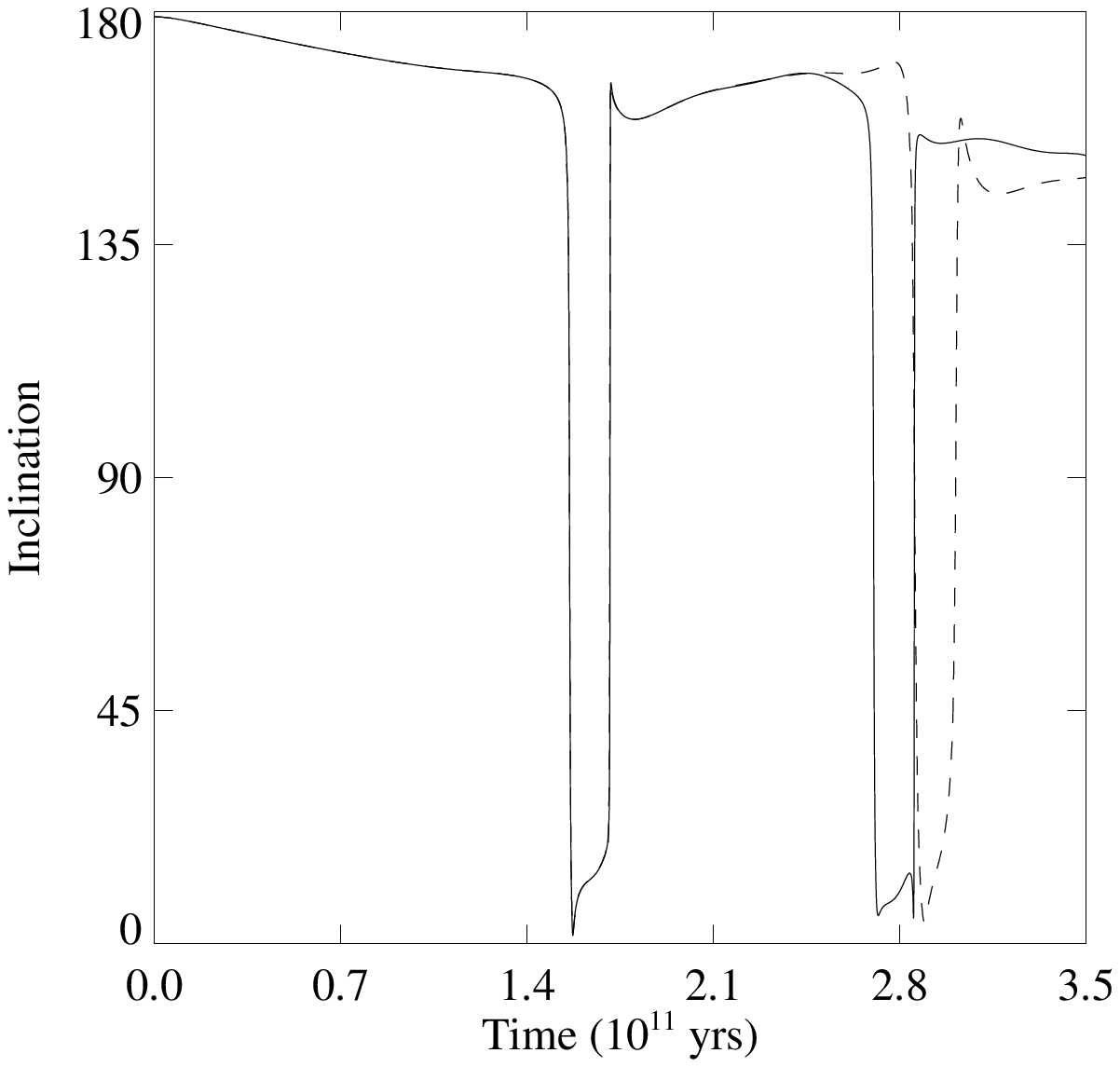,width=0.45\linewidth,clip=}
\end{tabular}
\end{center}
\caption[fig7]{Eccentricity and inclination of inner ring from
three-body integration (dashed) and averaged integration
    (solid). Relativistic precession is not included.}
\label{fig:t2_bs_rg}
\end{figure}

\begin{figure}
\begin{center}
\epsfxsize= 8 in
\epsfysize= 8 in
\begin{tabular}{cc}
\epsfig{file=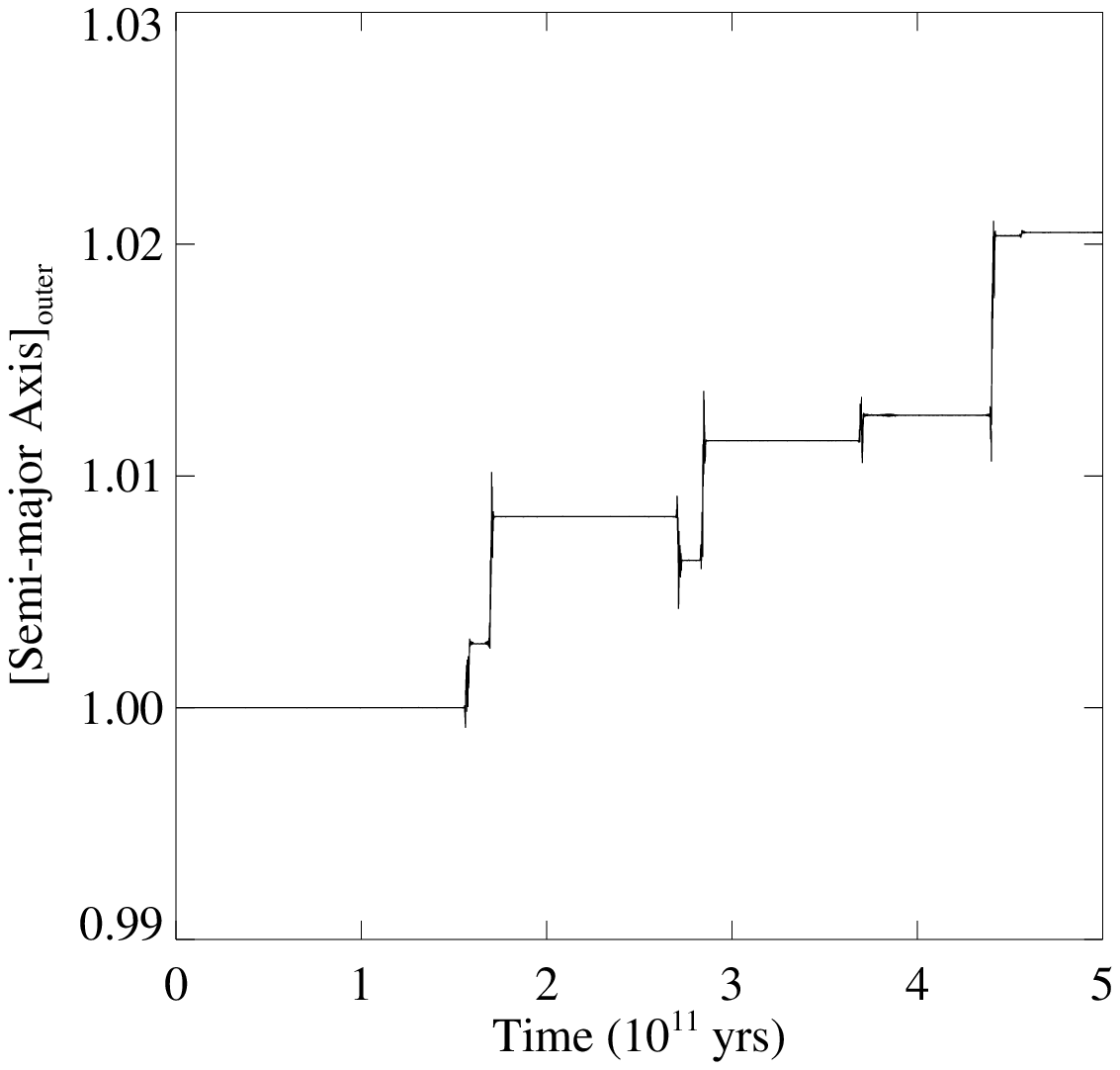,width=0.45\linewidth,clip=} &
\epsfig{file=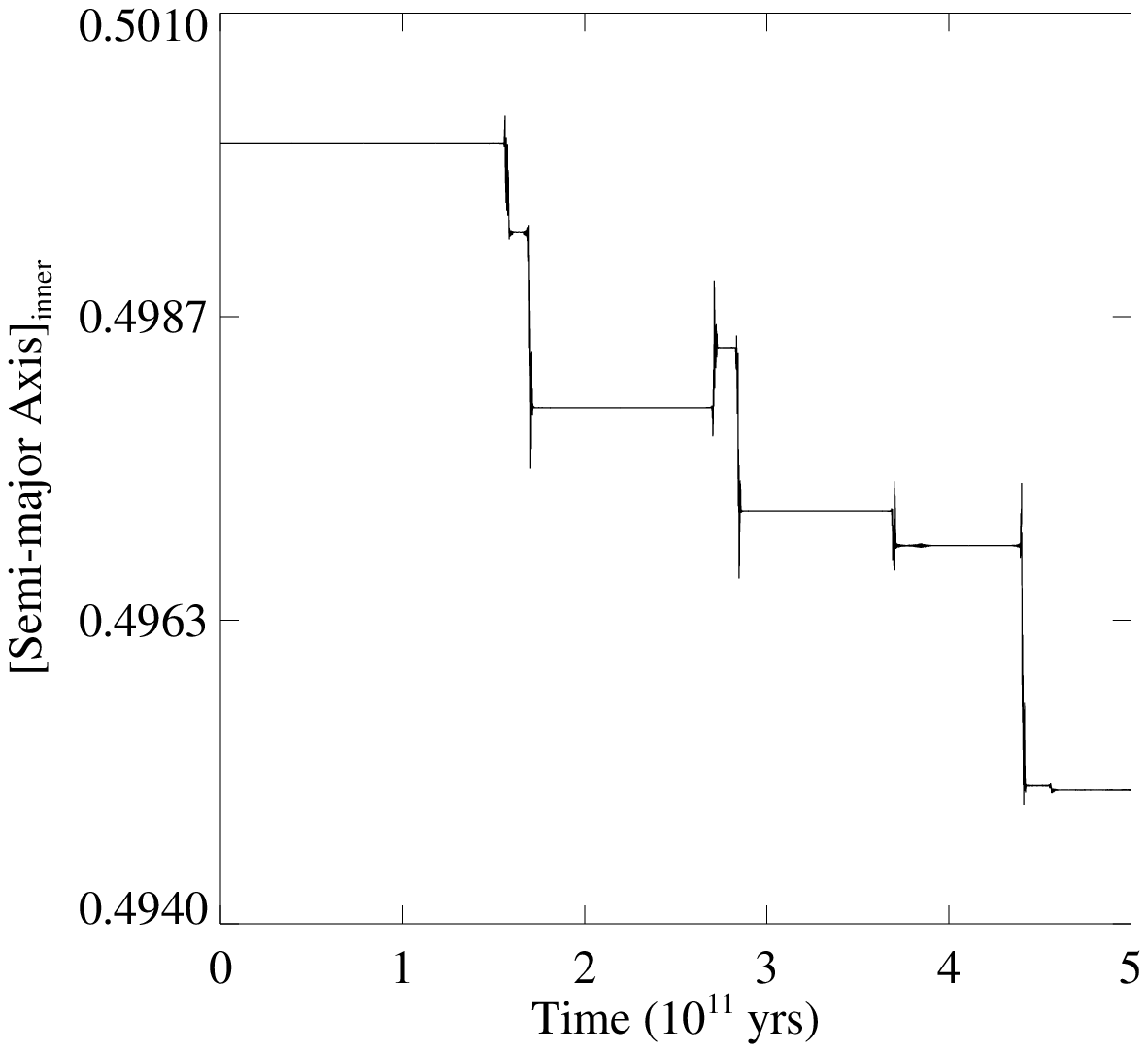,width=0.45\linewidth,clip=} 
\end{tabular}
\end{center}
\caption[fig5]{Jumps in the semi-major axis of particles in the three-body
  integration during close encounters, in the course of oscillations
  caused by the counter-rotating instability.}
\label{fig:t2_bs_a}
\end{figure}

The reader may wonder how well we do against particle integrations in
these vigorously unstable configurations. The situation is illustrated
in Fig.\ \ref{fig:t2_bs_rg}: there is good agreement over a cycle,
giving way to quantitative disagreement but similar qualitative
trends.  This discrepancy has little to do with proper adjustment of
initial conditions. Rather, second-order effects in the masses, which
are neglected in ring simulations, are amplified in close encounters
that result in the large eccentricity and inclination phases of this
system. This is particularly apparent in Fig.\ \ref{fig:t2_bs_a},
where jumps in particle semi-major axis accumulate to a few percent in
less than five cycles. These jumps, which occur at close
encounters/crossings between particles, 
are not accounted for in the secular representation of the dynamics.
They bring about shifts in the secular frequencies which are
responsible for the discrepancies observed in Fig.\ 
\ref{fig:t2_bs_rg}\footnote{In comparisons between ring and particle
  simulations it should be borne in mind that there are several possible
  particle analogs to a ring. Here, we associate a single
  particle with a ring, to highlight the discrepancies that the
  practitioner should keep in mind when studying secular few-body
  dynamics, say in the context of planetary systems. Alternatively,
  one could think of a ring as a guiding-center trajectory that provides
  the mean phase-space position of a distribution of particles, with total mass
  equal to the mass of the ring and spatial extent set by the ring's
  softening; this point of view is more natural to stellar clusters in
  galactic nuclei. Finally, the ring could represent a point in the
  five-dimensional space of actions and resonant angles,
  i.e., a component of a phase-space distribution function that is uniformly
  distributed in mean anomaly. The error
  associated with Gauss's method depends on the analogy being used.}.

\begin{figure}
\begin{center}
\epsfxsize= 8 in
\epsfysize= 8 in
\begin{tabular}{cc}
\epsfig{file=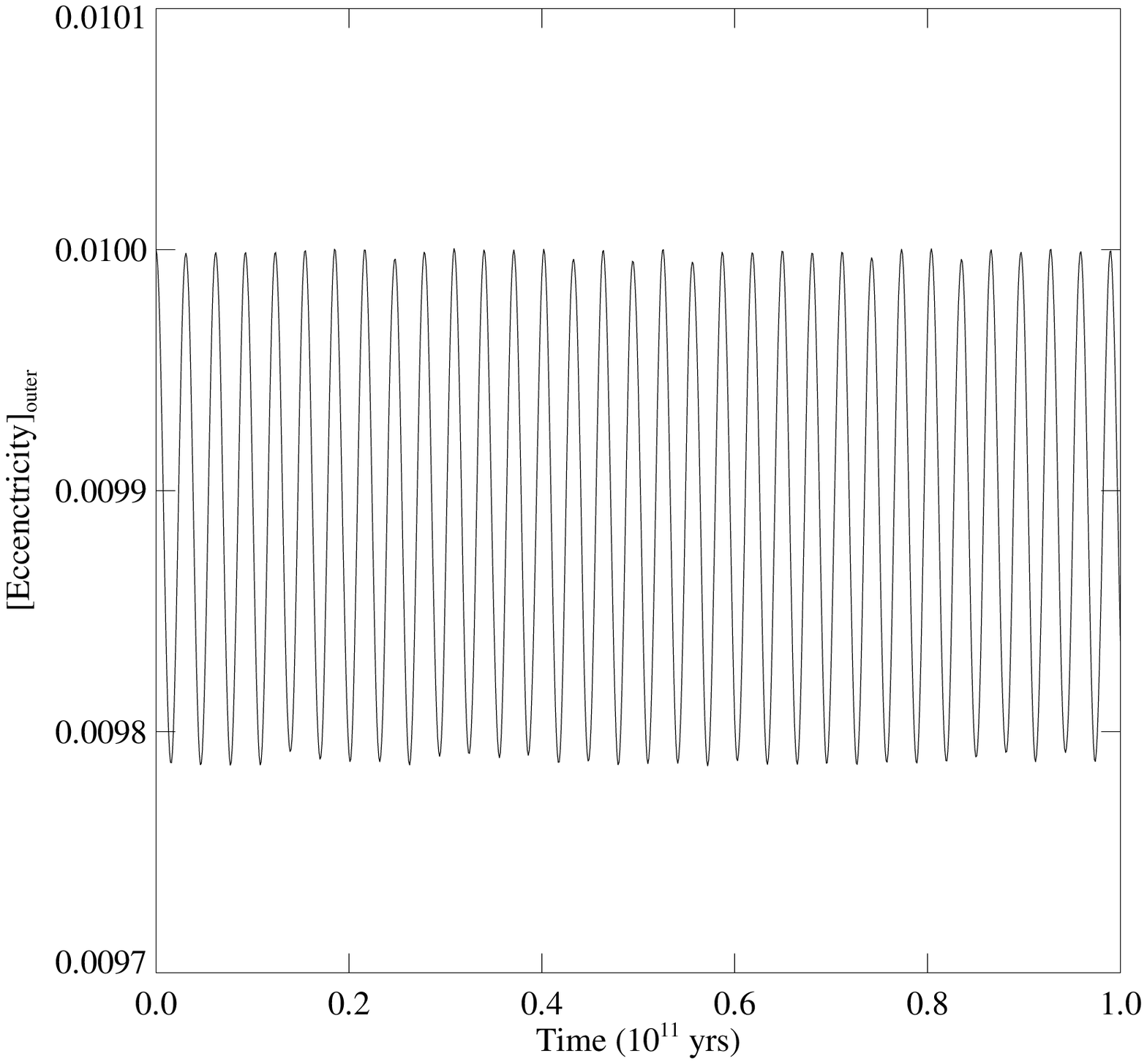,width=0.45\linewidth,clip=} &
\epsfig{file=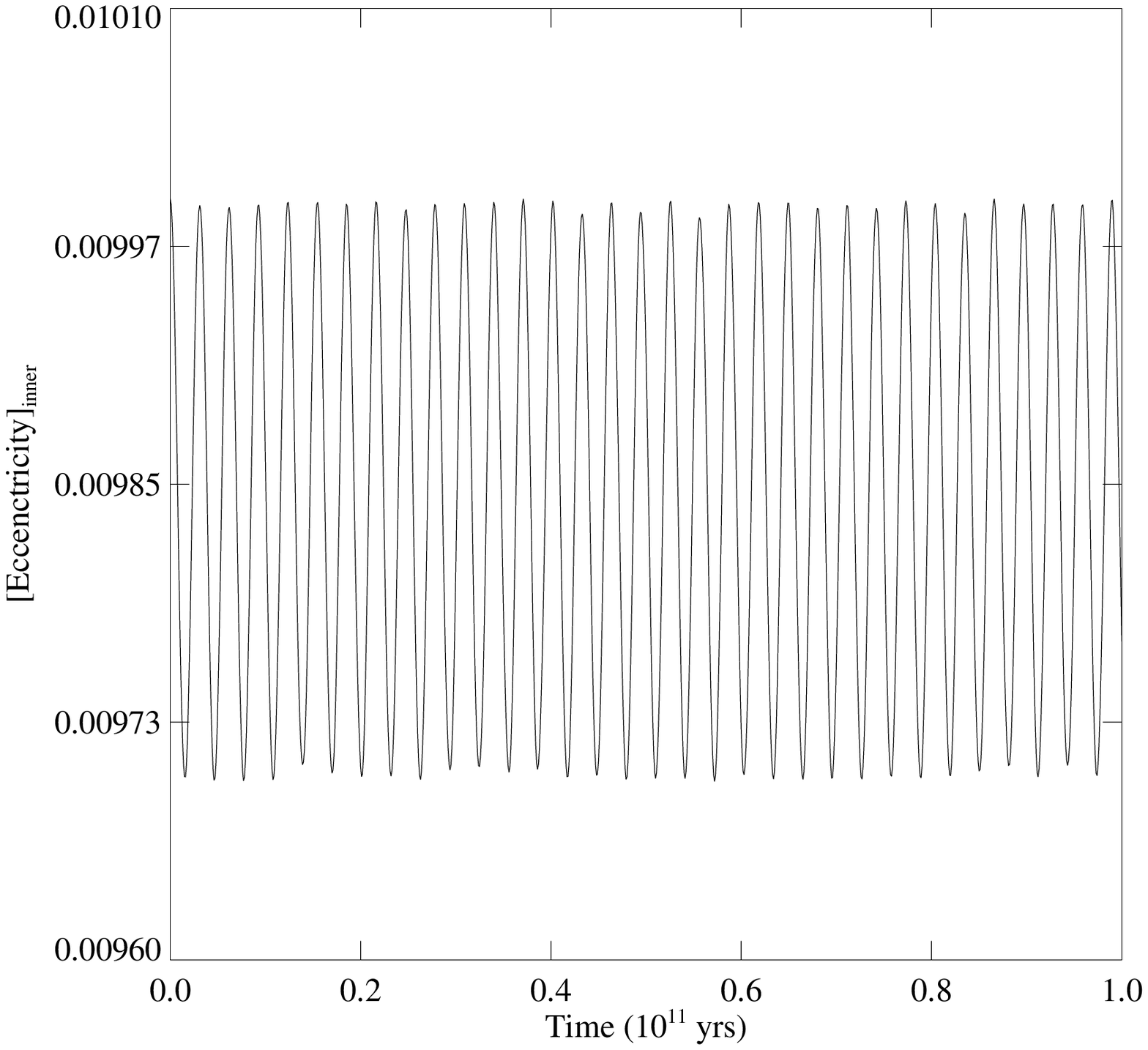,width=0.45\linewidth,clip=} \\ 
\epsfig{file=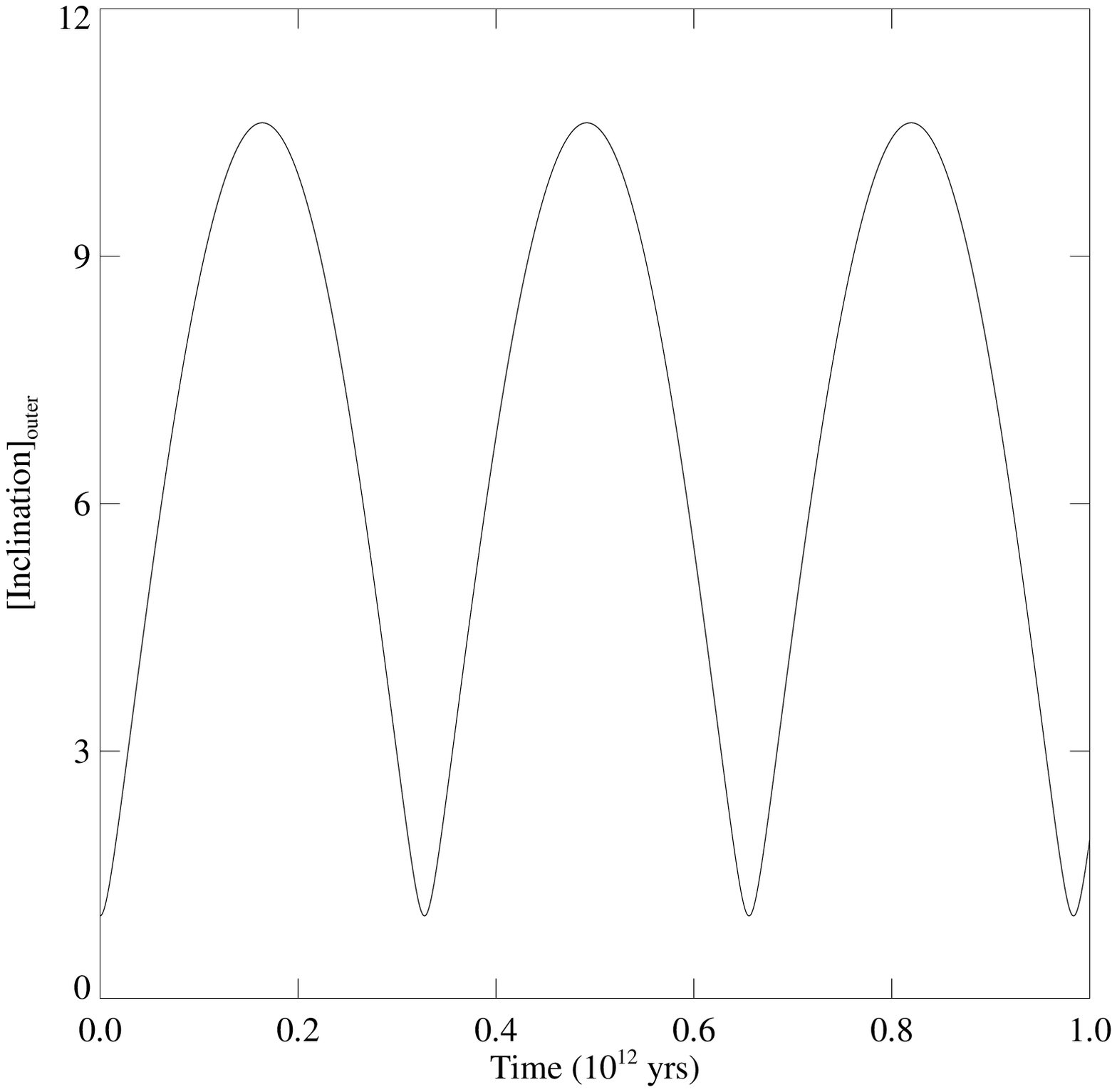,width=0.45\linewidth,clip=} &
\epsfig{file=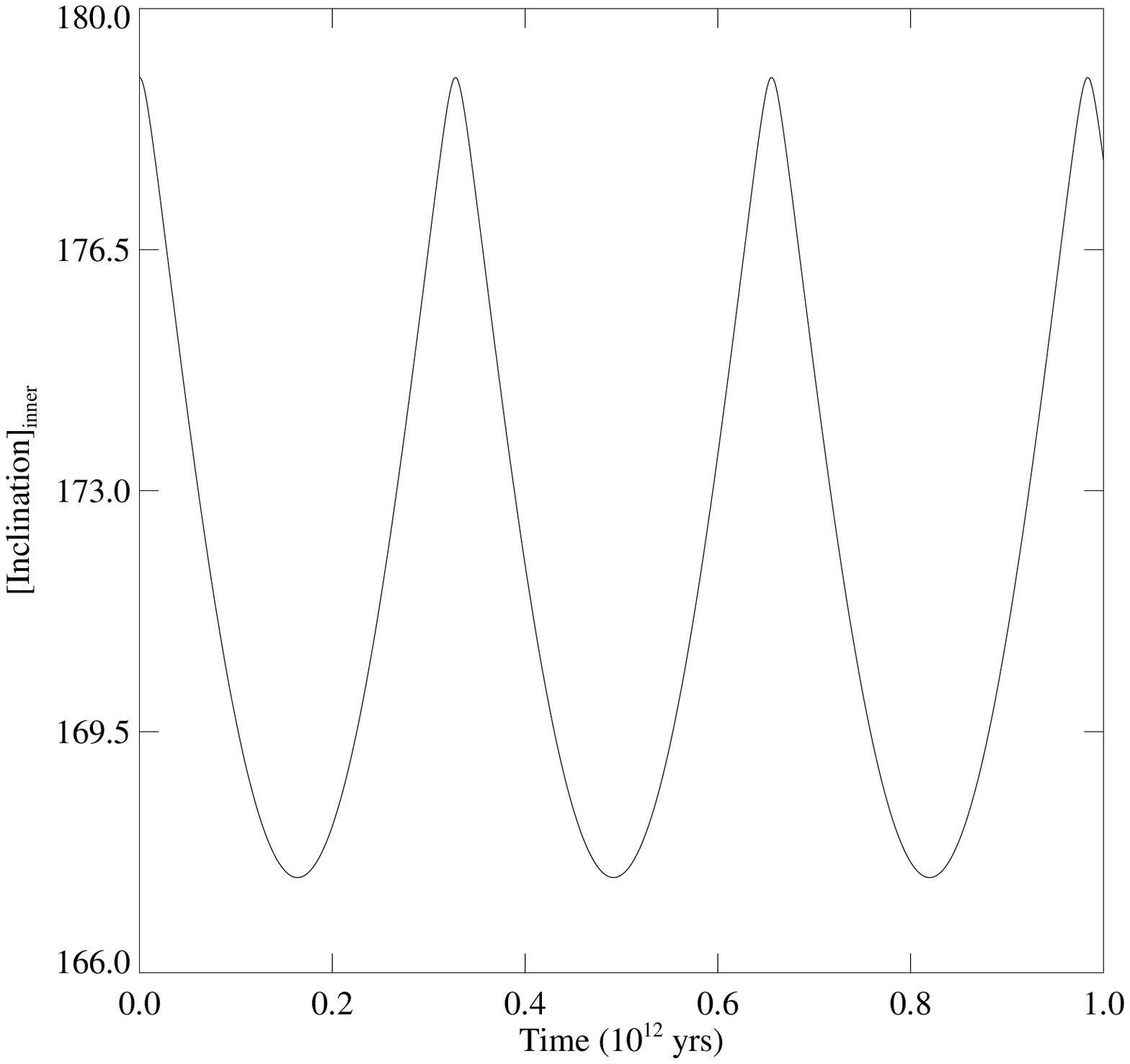,width=0.45\linewidth,clip=}
\end{tabular}
\end{center}
\caption[fig6]{The stabilizing effect of general-relativistic precession as
  evident in the tame behavior of eccentricity and inclination of the same
  ring system.}
\label{fig:t2_gr}
\end{figure}

Finally, while softening/heat can induce instability in a counter-rotating
system of rings, general-relativistic precession can suppress the instability.
The importance of normally negligible relativistic effects is already
appreciated in studies of Kozai oscillations, (e.g., Holman et al.\ 1997), and
we demonstrate that the same applies in this case by switching on relativistic
precession in the same configuration (Fig.\ \ref{fig:t2_gr}). 

\section{Counter-rotating systems of rings}

\label{sec:cr}

\noindent
We conclude our battery of tests with a modest, yet challenging, look at systems of
counter-rotating rings. For some time now, it has been recognized that
counter-rotation is likely to be associated with lopsidedness in
self-gravitating systems of gas and stars. \cite{tou02} pointed out that
counter-rotating instabilities are at work in nearly Keplerian systems and
speculated that non-linear evolution of unstable discs may lead to the
formation of eccentric discs such as the one observed in the `double'
nucleus of M31 \citep{tre95,pei03}. We limit our discussion to a couple of
runs, which are illustrative of both the challenges faced by the algorithm,
and the typical dynamical fate of such systems.

\begin{figure}
\begin{narrow}{-0.6in}{-0.6in}
\begin{tabular}{lll}
\epsfig{file=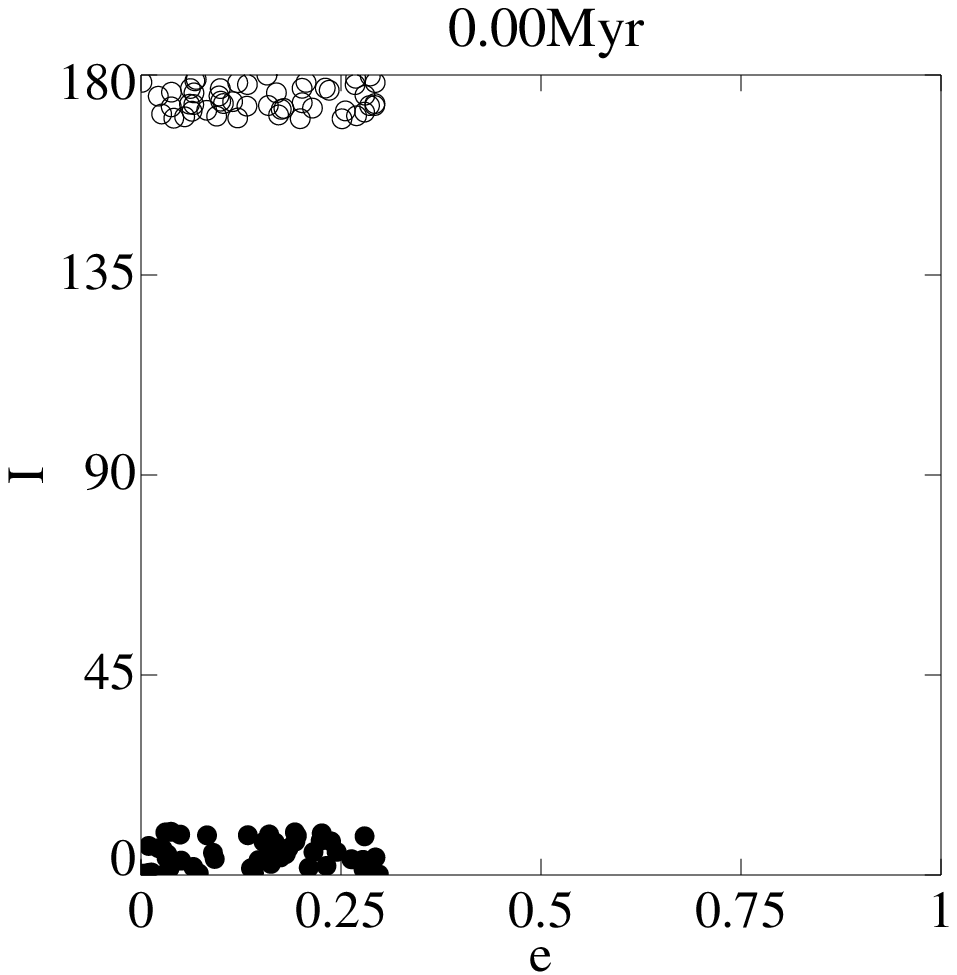,height=0.36\textwidth,   clip=} &
\epsfig{file=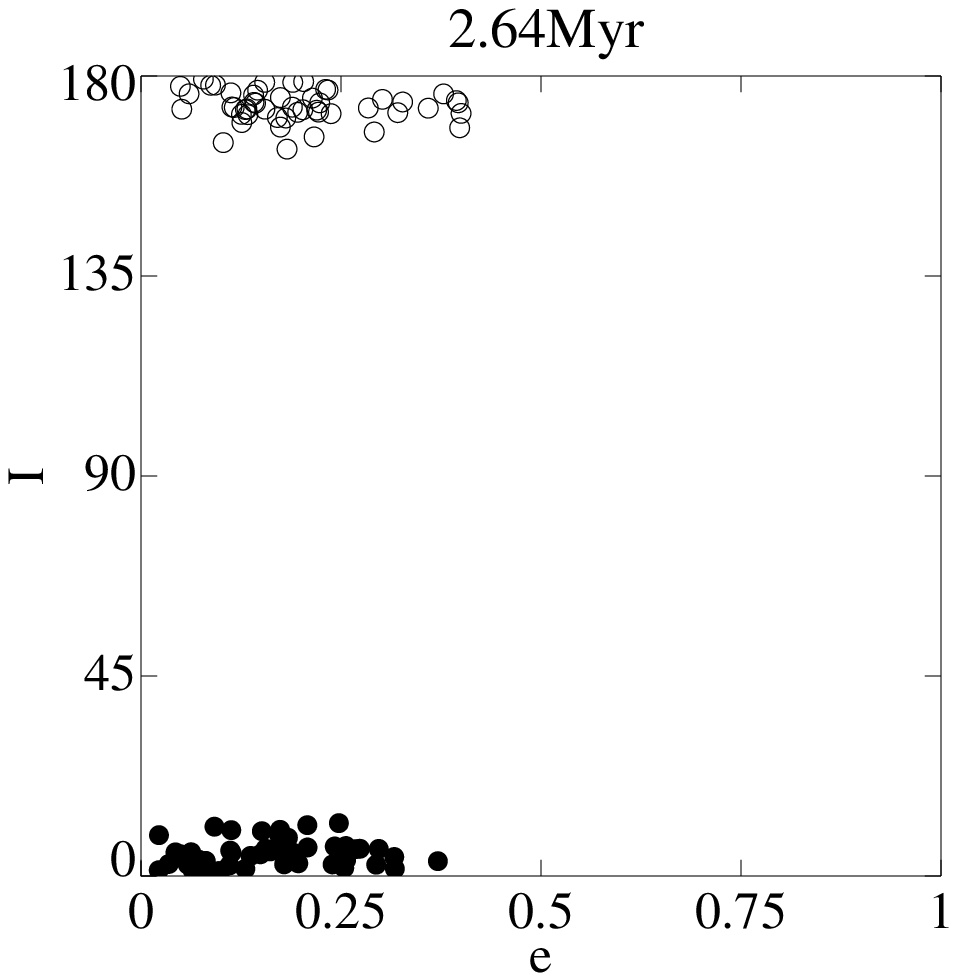,height=0.36\textwidth,   clip=} & 
\epsfig{file=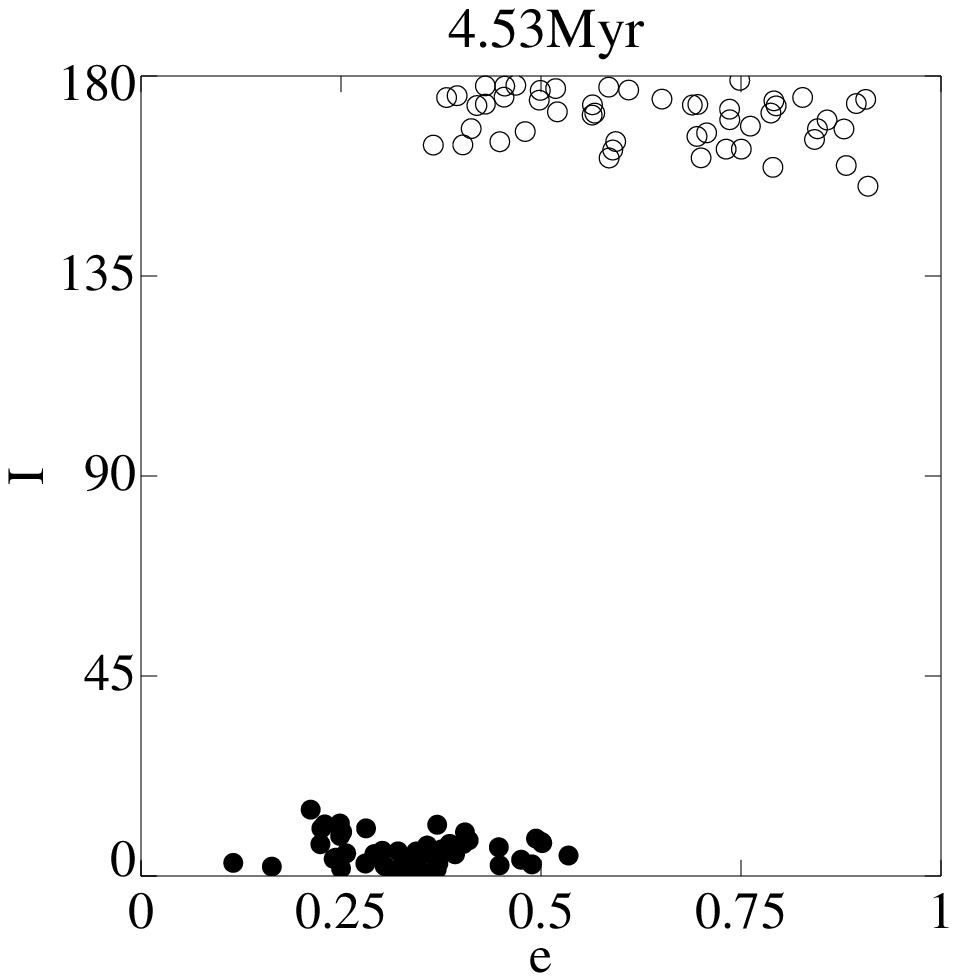,height=0.36\textwidth,
  clip=}
\end{tabular}
\end{narrow}
\begin{narrow}{-0.6in}{-0.6in}
\begin{tabular}{lll}
\epsfig{file=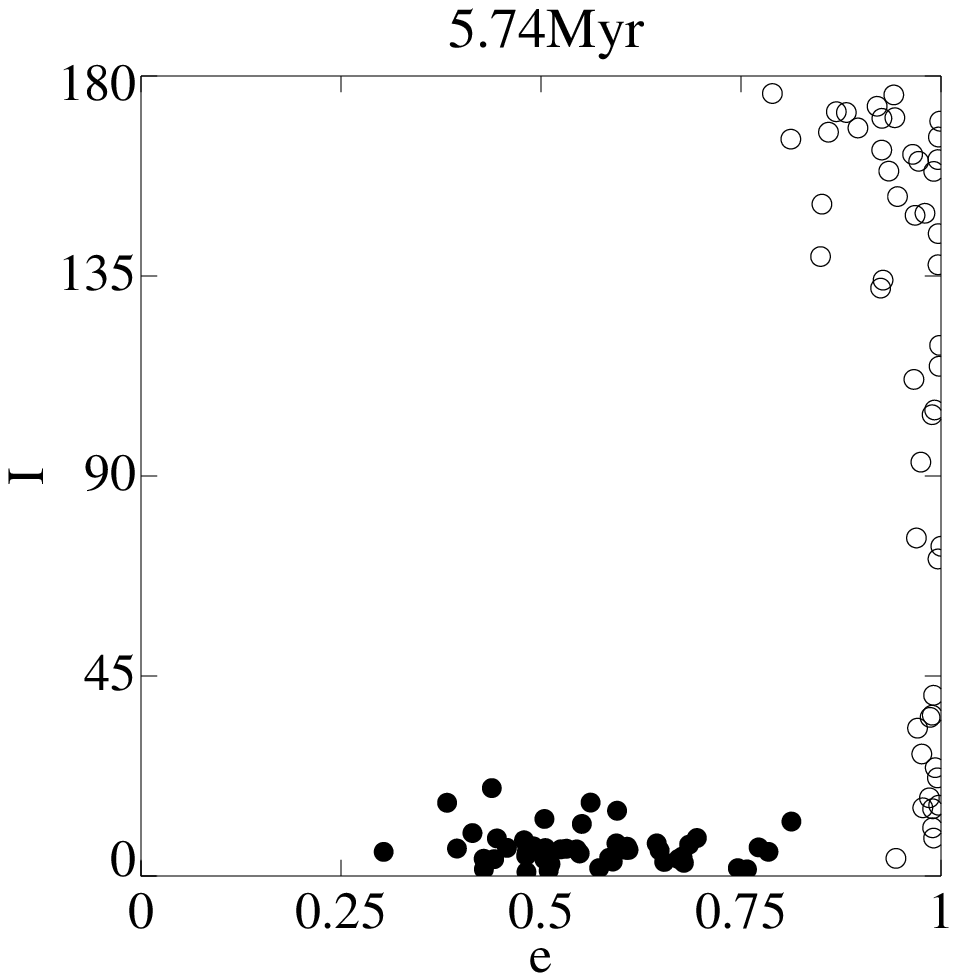,height=0.36\textwidth,   clip=} &
\epsfig{file=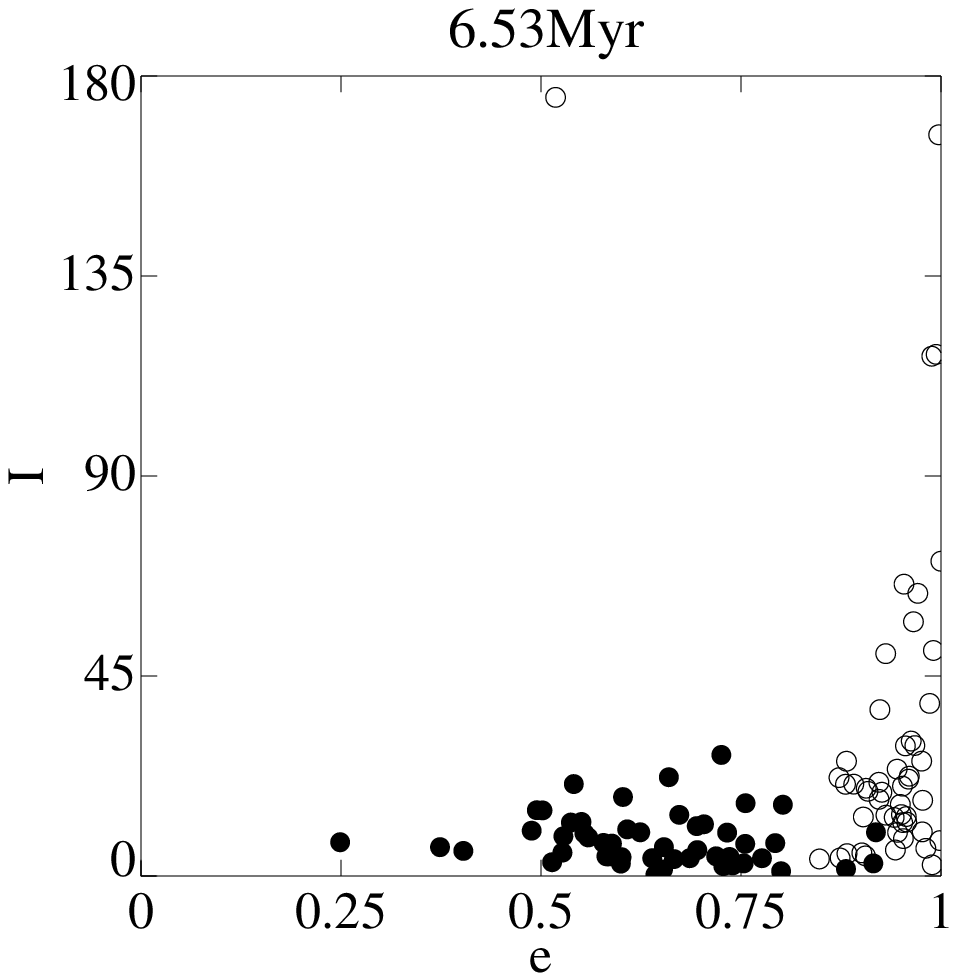,height=0.36\textwidth,   clip=} &
\epsfig{file=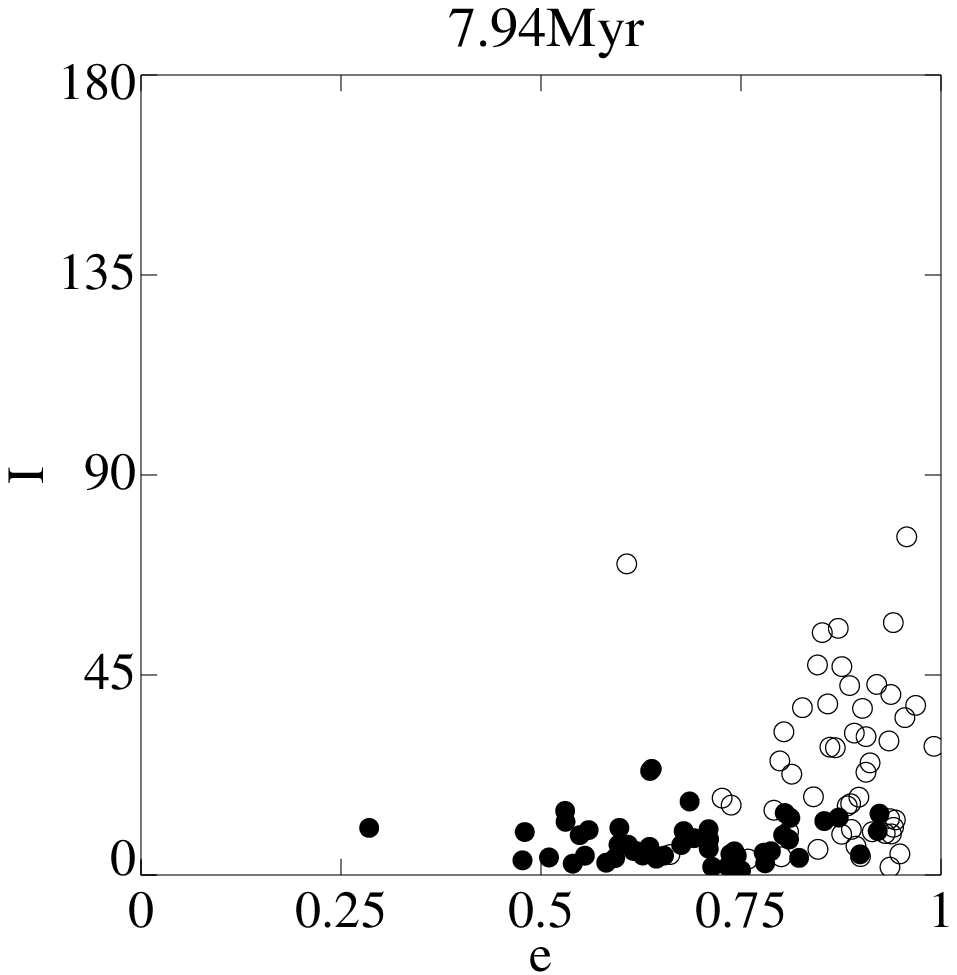,height=0.36\textwidth,
  clip=}
\end{tabular}
\end{narrow}
\begin{narrow}{-0.6in}{-0.6in}
\begin{tabular}{lll}
\epsfig{file=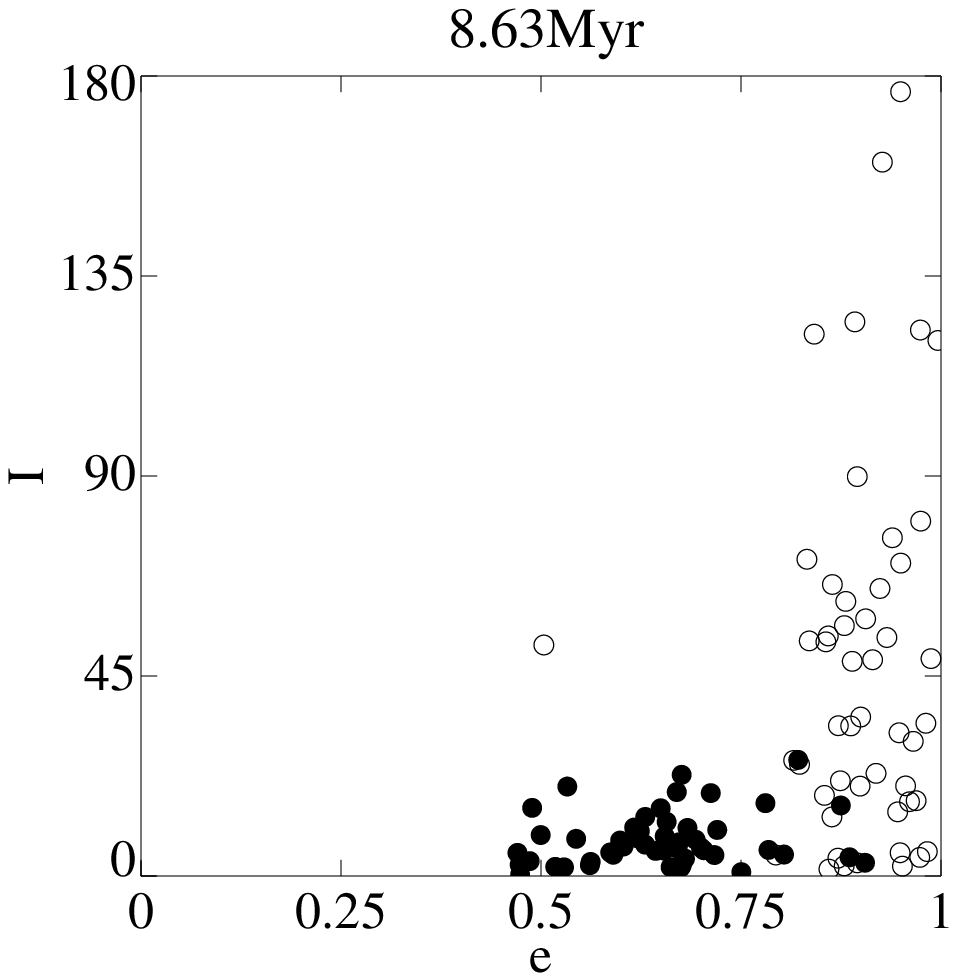,height=0.36\textwidth,   clip=} &
\epsfig{file=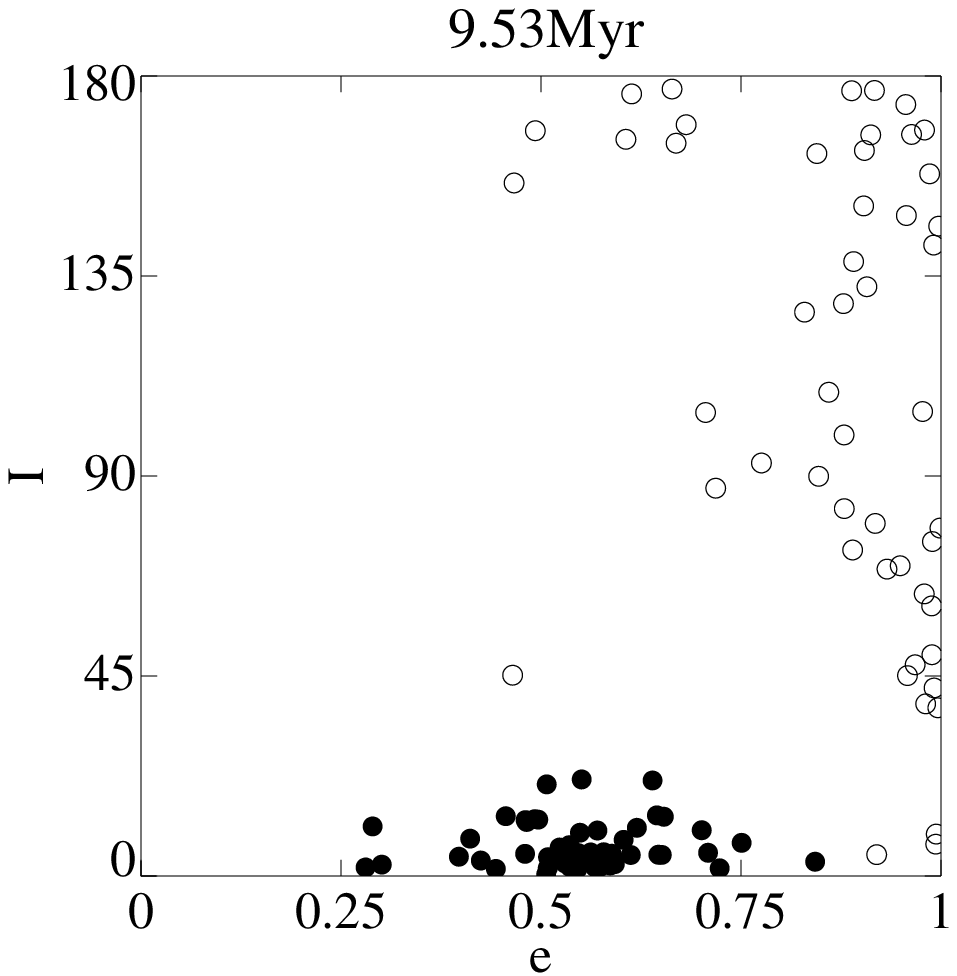,height=0.36\textwidth,   clip=} &
\epsfig{file=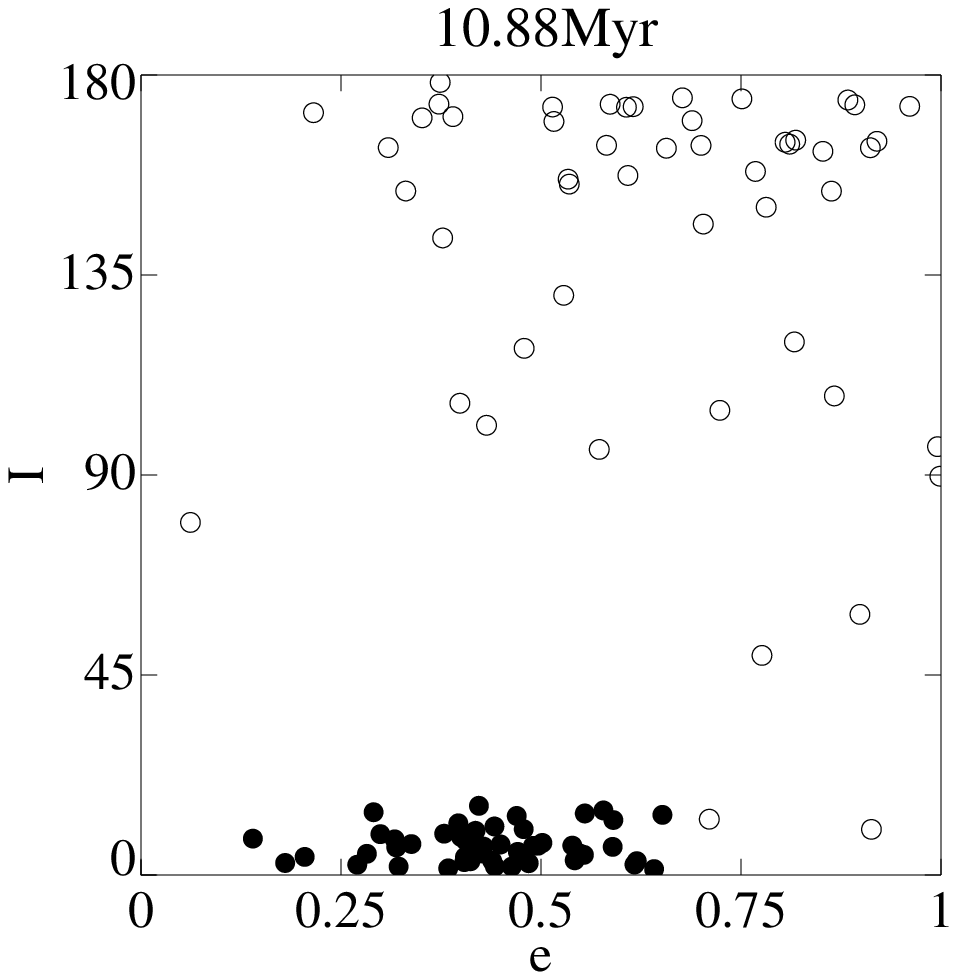,height=0.36\textwidth,   clip=} 
\end{tabular}
\end{narrow}
\caption[fig7]{\small The instability viewed in the $(e, i)$ plane. Prograde
  rings are represented by filled circles, while retrograde rings are open
  circles. As the time increases, from top left to bottom right, one witnesses
  the growth of eccentricity at near constant inclination, followed by the
  flip-over of the retrograde rings, then their attempted return to their
  initially coplanar configuration. The more massive prograde rings experience
  a significant net growth in their mean eccentricity, while remaining
  clustered close to their initial inclination.}
\label{fig:cr_ei}
\end{figure}

We follow the dynamics of a system of 100 rings, half of them prograde
with a total mass of $10^{-3}$ relative to the central point mass, and
the other half retrograde, with the same semi-major axis distribution
as the prograde rings, and $20$ per cent of their mass. In this illustrative
example, we picked the rings from uniform distributions in semi-major
axis (over the interval $[1,1.2]\pc$), eccentricity (over $[0,0.3]$),
inclination ($[0,10^{\circ}]$ for prograde and $[170^\circ,180^\circ]$
for retrograde), argument of node and periapsis (over $[0,2\pi]$). The
softening length was $b=0.1\pc$.

\begin{figure}
\begin{narrow}{-0.6in}{-0.6in}
\begin{tabular}{lll}
\epsfig{file=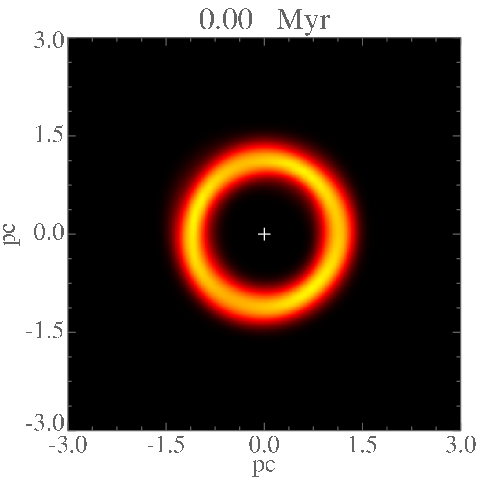,height=0.36\textwidth,   clip=} &
\epsfig{file=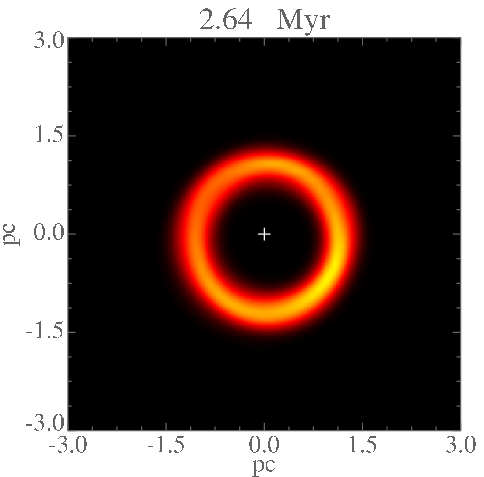,height=0.36\textwidth,   clip=} & 
\epsfig{file=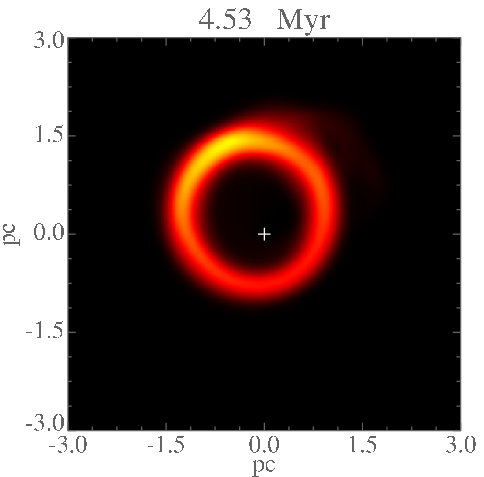,height=0.36\textwidth,   clip=} 
\end{tabular}
\end{narrow}
\begin{narrow}{-0.6in}{-0.6in}
\begin{tabular}{lll}
\epsfig{file=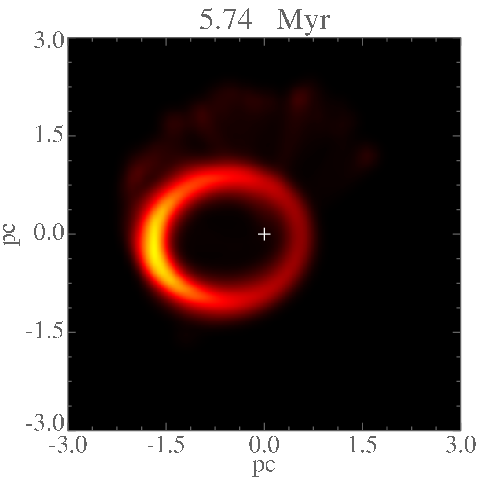,height=0.36\textwidth,   clip=} &
\epsfig{file=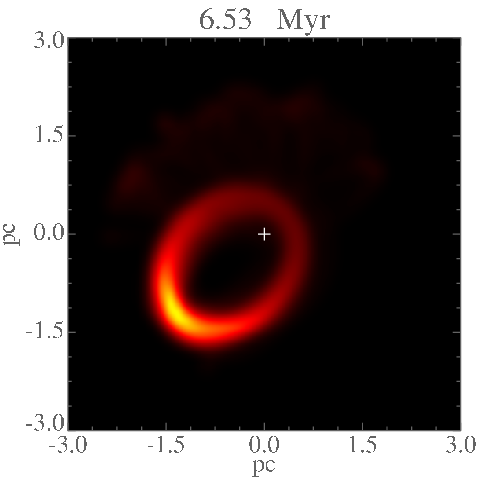,height=0.36\textwidth,   clip=} &
\epsfig{file=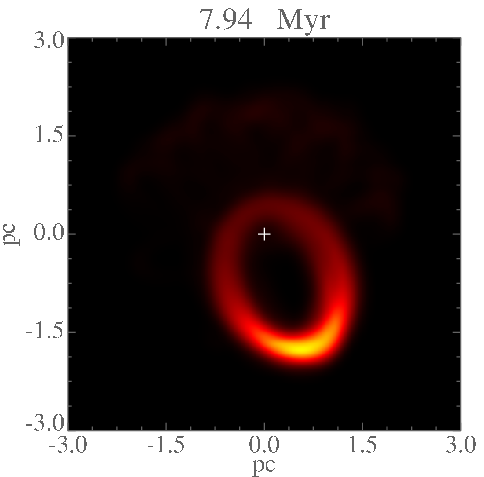,height=0.36\textwidth,   clip=}
\end{tabular}
\end{narrow}
\begin{narrow}{-0.6in}{-0.6in}
\begin{tabular}{lll}
\epsfig{file=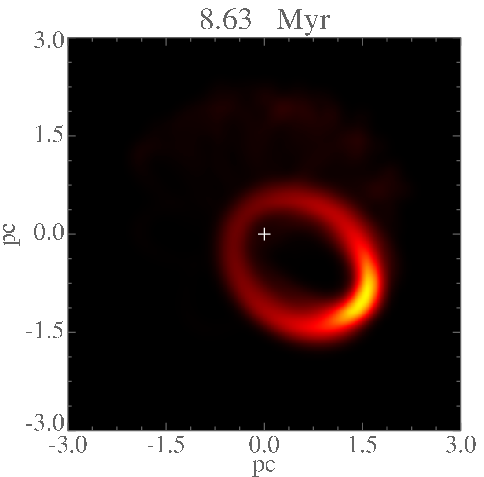,height=0.36\textwidth,   clip=} &
\epsfig{file=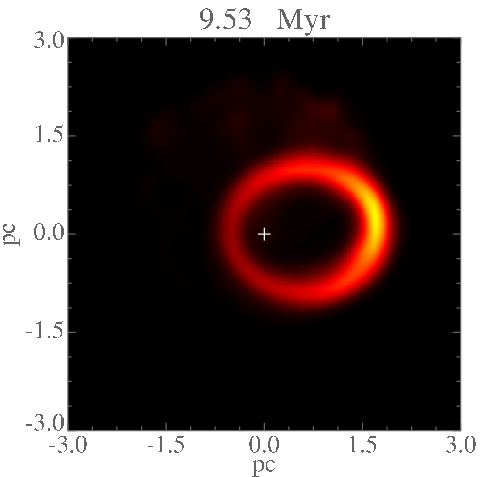,height=0.36\textwidth,   clip=} &
\epsfig{file=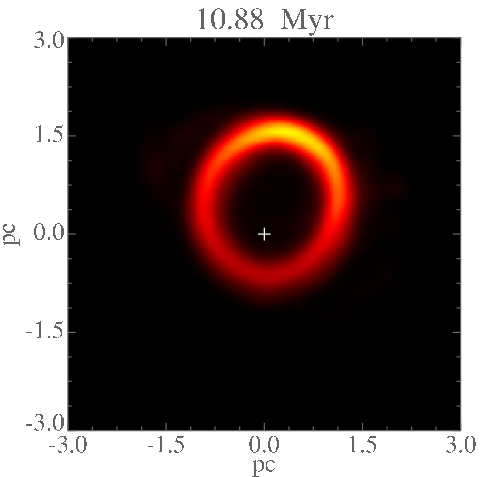,height=0.36\textwidth,   clip=} 
\end{tabular}
\end{narrow}
\caption[fig8]{\small Evolution of the projected and smoothed surface density
  of the full system.  The projection is on the reference $x$-$y$ plane,
  lumping together contributions from prograde and retrograde orbits. The
  central point mass is at the origin, marked by a cross. Color refers to
  density levels. An $m=1$ mode grows as a result of the counter-rotation
  instability, capturing rings in its way, with the prograde population
  maintaining the precessing single-lobed pattern, all the while the initially
  retrograde population prepares for disintegration.}
\label{fig:cr_dens}
\end{figure}

The integrator used a tolerance $\epsilon_{\rm int}=10^{-8}$, which
gave a mean timestep of $10^4$ yr; the timestep reached a maximum of
$6\times 10^4$ yr and dipped as low as 100 yr for brief periods.  The
numerical averaging was performed with a minimum of $K=128$ points per
ring and a maximum of $K=512$, enforcing an averaging error
$\epsilon_{\rm quad}=10^{-10}$. The system was followed for 30
Myr. During this period, the secular energy was conserved to within a
fractional error of $10^{-9}$, with ten times smaller error in the
total angular momentum.  The run took about two weeks on a single 3.2
Ghz processor. An order of magnitude improvement in performance can be
achieved if one is willing to live with a tolerance $\epsilon_{\rm
int}=10^{-6}$.

\begin{figure}
\begin{center}
\epsfxsize= 5 in
\epsfysize= 5 in
\begin{tabular}{cc}
\epsfig{file=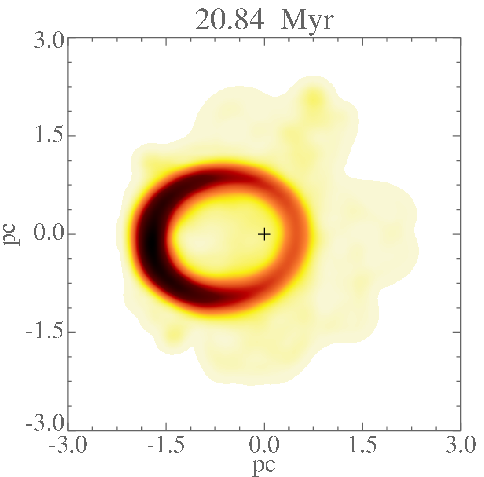,height=0.45\textwidth,   clip=} &
\epsfig{file=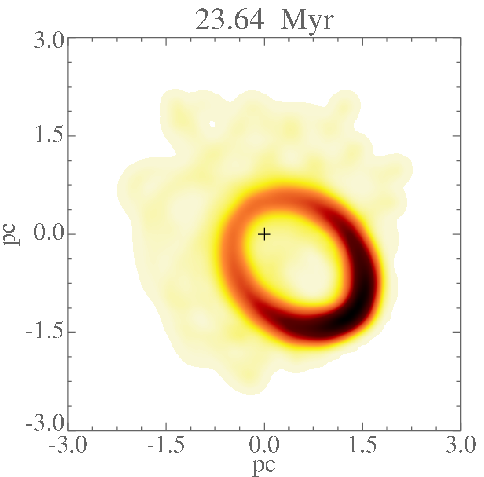,height=0.45\textwidth,   clip=} \\ 
\epsfig{file=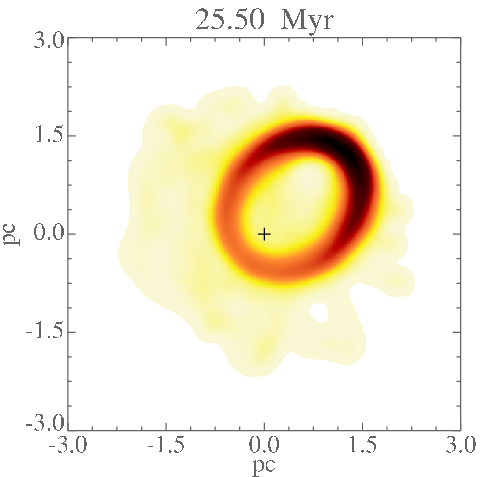,height=0.45\textwidth,   clip=} &
\epsfig{file=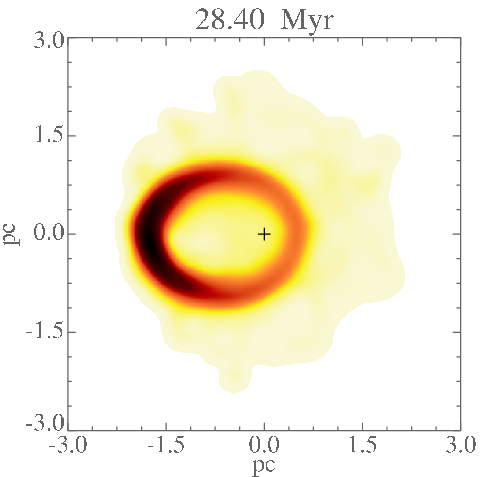,height=0.45\textwidth,   clip=}
\end{tabular}
\end{center}
\caption[fig9]{\small The projected surface density evolution of the
  rings over longer times, when viewed in negative color. The highly
  eccentric and dispersed population of initially retrograde
  rings projects onto a lightly hued halo, around the heavier lopsided
  configuration of prograde rings, captured in a precessing $m=1$ mode.}
\label{fig:disp_dens}
\end{figure}

\begin{figure}
\begin{center}
\epsfxsize= 8 in
\epsfysize= 8 in
\begin{tabular}{cc}
\epsfig{file=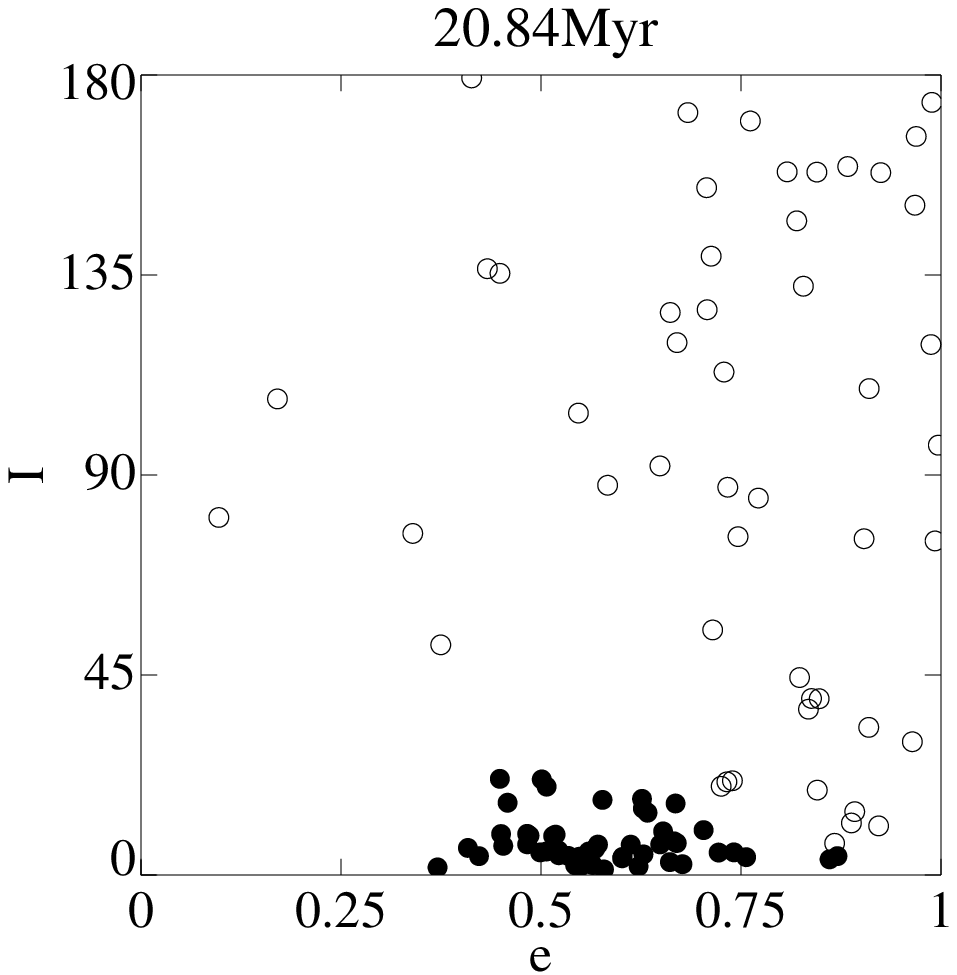,height=0.45\textwidth, clip=} &
\epsfig{file=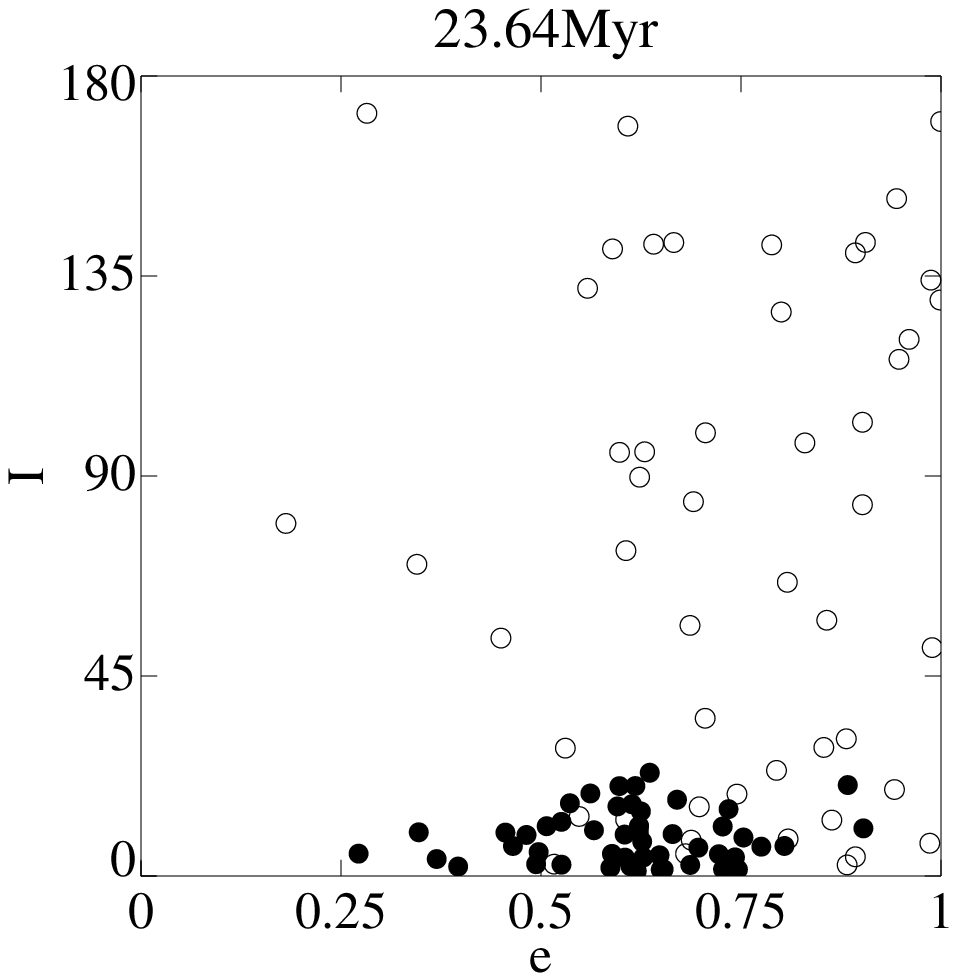,height=0.45\textwidth, clip=} \\ 
\epsfig{file=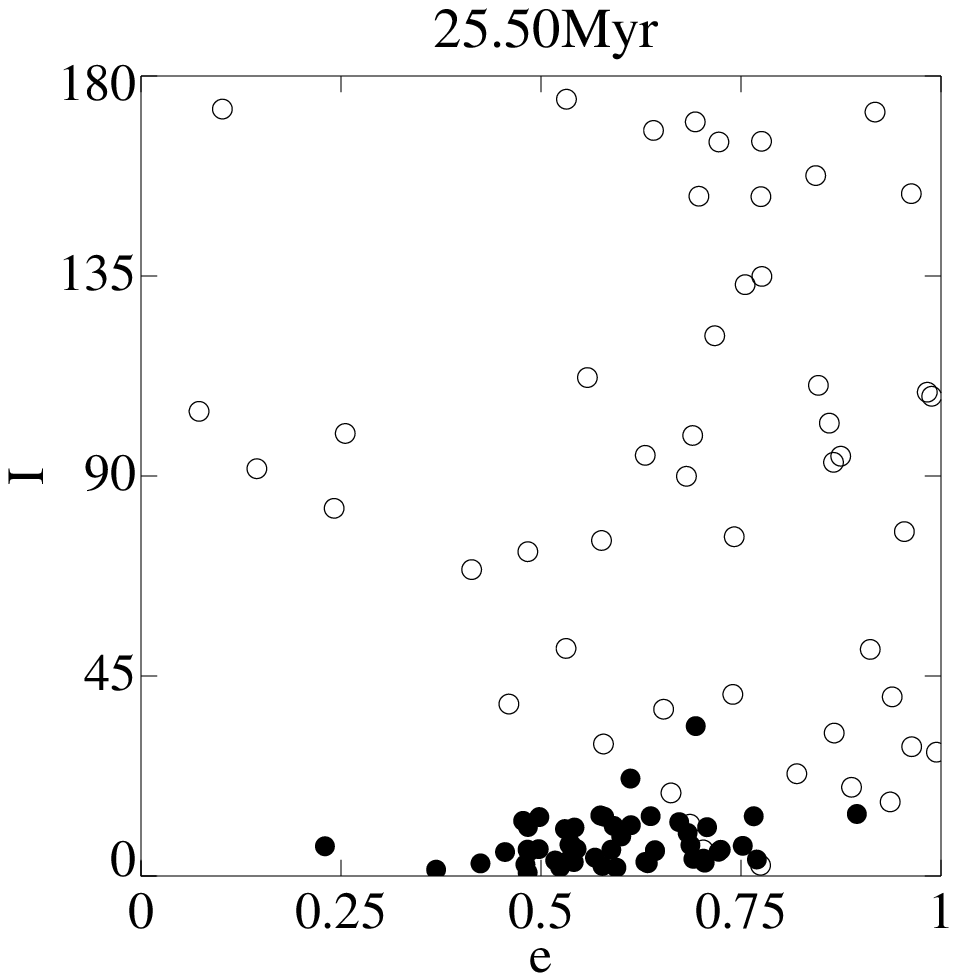,height=0.45\textwidth, clip=} &
\epsfig{file=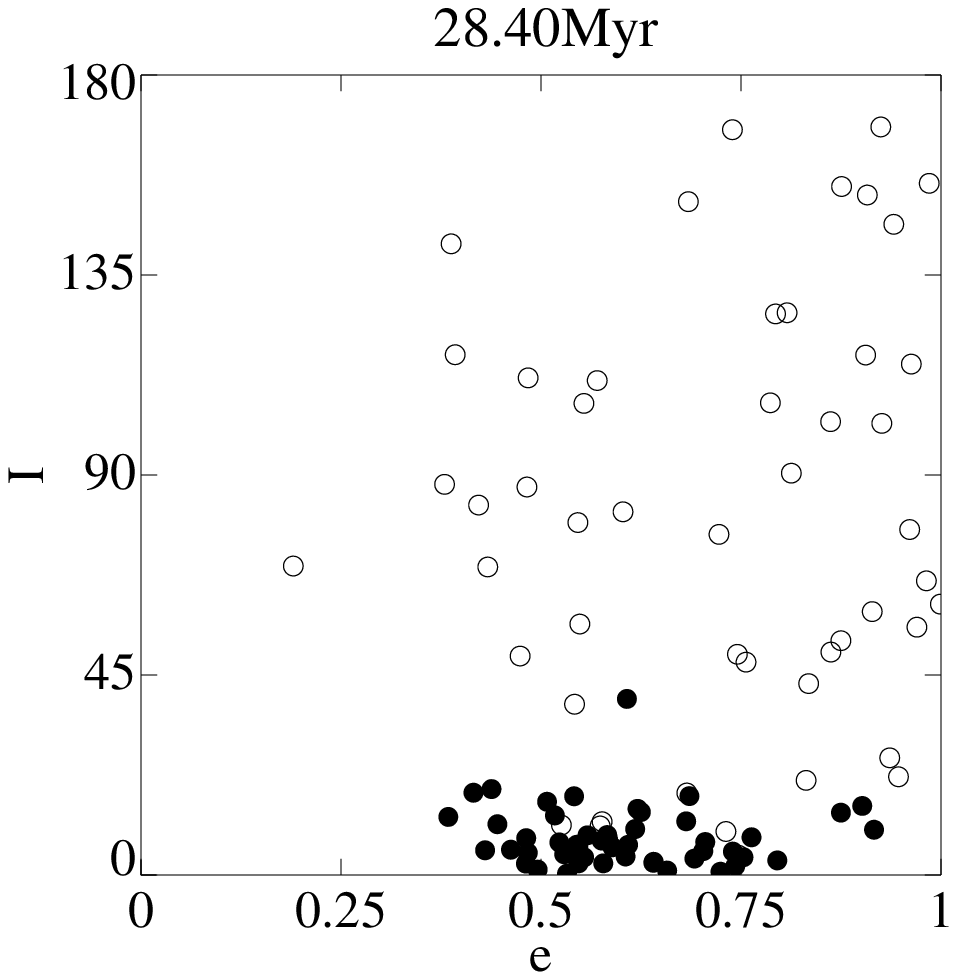,height=0.45\textwidth, clip=}
\end{tabular}
\end{center}
\caption[fig10]{\small The dispersal of the retrograde rings (open circles),
  and the increased cohesion of the prograde rings (filled circles), as seen
  in the $(e,i)$ plane.}
\label{fig:disp_ei}
\end{figure}

The evolution of the system in eccentricity and inclination is shown in Fig.\
\ref{fig:cr_ei}. In the first 5 Myr the system undergoes significant growth in
eccentricity, at near constant inclination; by 5 Myr the mean eccentricity of
the prograde population is around 0.6 and the mean eccentricity of the less
massive retrograde population is larger at 0.85. This result is not new: a
closely related planar system is known to be linearly unstable \citep{tou02},
with the lighter/retrograde component experiencing a more significant growth
of eccentricity than the heavier/prograde component. It is the later fully
non-linear behavior that we can confidently follow with the Gaussian ring
algorithm for the first time. Following the initial growth of eccentricity,
and in a manner reminiscent of the behavior witnessed in the previous section
for two rings, the rings in the retrograde/lighter component experience a
dramatic growth and spread in inclination, practically reversing orientation
by 7.8 Myr. The mean eccentricity of both populations increases slightly
during this flipover phase, with the retrograde rings reaching a mean of 0.9
around 6 Myr. Then, the initially retrograde rings try to regain their
home-base; they almost succeed around 10.3 Myr, but suffer significant losses
along the way. At this point their mean inclination is about
$140^{\circ}$, and their mean eccentricity has decreased to $0.6$; the
prograde rings are still in low-inclination orbits, with a mean eccentricity
of 0.45. The prograde rings are also bunched in the conjugate angles, the
longitude of periapsis in particular. This is apparent in the associated
spatial densities, shown in Fig.\ \ref{fig:cr_dens}. The non-linear saturation
of the instability leads to an oscillating, eccentric ring composed of
prograde particles, with a mean inclination of $7.5^{\circ}$, a mean
eccentricity of around $0.6$, and an increasingly steady prograde precession with a frequency
$0.8\hbox{\,rad Myr}^{-1}$ (Fig.\ \ref{fig:disp_dens}).  Meanwhile the
culprits behind the instability, namely the initially retrograde rings, are
dispersed into a population with a wide spread in inclination, node and
argument of periapsis, and a mean eccentricity of $0.7$ (Fig.\ \ref{fig:disp_ei}).

\begin{figure}
\begin{center}
\epsfxsize= 8 in
\epsfysize= 8 in
\begin{tabular}{cc}
\epsfig{file=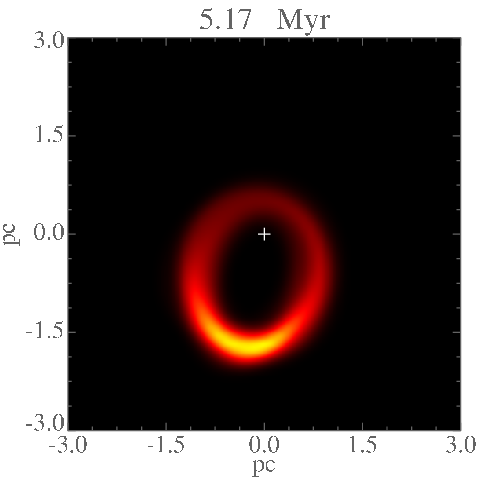,height=0.45\textwidth,   clip=} &
\epsfig{file=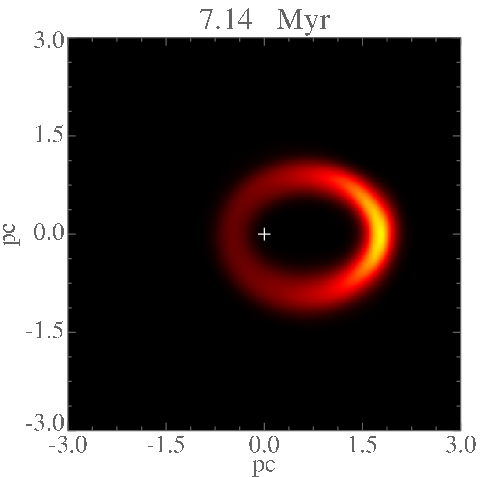,height=0.45\textwidth,   clip=} \\ 
\epsfig{file=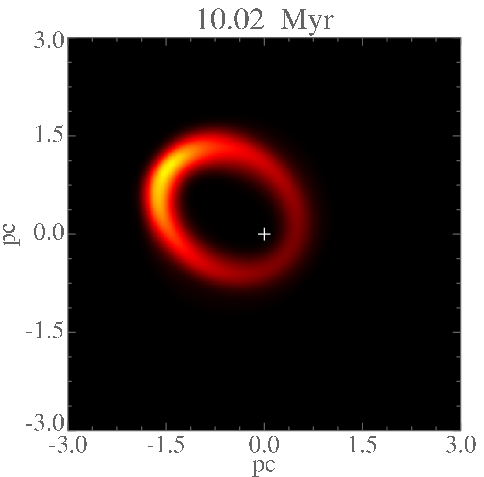,height=0.45\textwidth,   clip=} &
\epsfig{file=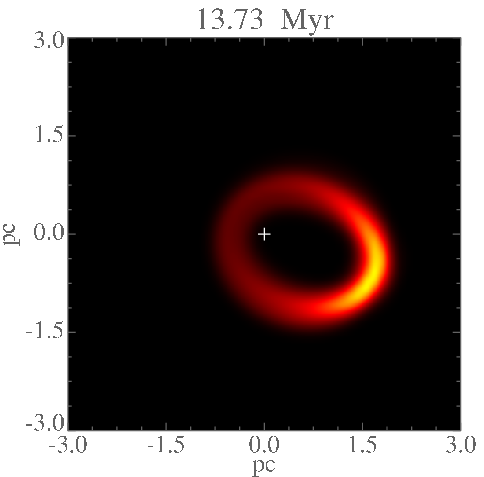,height=0.45\textwidth,   clip=}
\end{tabular}
\end{center}
\caption[fig11]{\small The evolution of the prograde rings after the
retrograde rings were removed, leaving a slightly
thickened, lopsided, and uniformly precessing annulus.}
\label{fig:prog_dens}
\end{figure}

In short, the mode, which is ignited by the instability,
grows through resonant capture of both prograde and retrograde
populations, then saturates as the retrograde population (which has a
lower angular momentum) escapes from resonance, and disperses through
(partially stochastic) phase mixing; this leaves a captured prograde
population which maintains a lopsided $m=1$ mode that is increasingly
steady in its (prograde) precession. Just for fun, and once the mode
was fully established, we removed all initially retrograde rings, and
left the prograde population to itself. Perhaps not surprisingly, the
captured prograde population, shown in Fig.\ \ref{fig:prog_dens},
maintained its cohesive state, now precessing uniformly with a
slightly higher frequency of $0.9\hbox{\,rad Myr}^{-1}$ for at least
four precession periods. At a mean timestep of $1.1\times10^{4}$ yr,
the integration kept fractional errors in energy and angular momentum
below $10^{-9}$.  One could think of the resulting mode as an analog
of the lopsided mode which was artificially excited by \cite{js01}.
The main differences are: (i) the Jacobs-Sellwood simulation was a
two-dimensional N-body simulation, while ours is a three-dimensional
N-ring simulation; (ii) the Jacobs-Sellwood simulation used
$N=$100,000 while ours used $N=100$ (each of our rings should be
regarded as representing many stars); (iii) the Jacobs-Sellwood
simulation formed an eccentric disc using eccentric initial
conditions, while we relied on counter-rotation to generate an
eccentric equilibrium through the non-linear evolution of an
instability.

A careful treatment of the growth and saturation of the
counter-rotating instability would take us too far afield. It is in
fact the subject of an upcoming study by Touma and Kazandjian (in
preparation), who perform sufficiently resolved N-body simulations to
permit a detailed phase-space dissection of the dynamics in question.
Their results, which are based on long term N-body simulations of up to
500,000 counter-rotating particles around a dominant central
body, are in remarkable qualitative agreement with the salient
features of the 100 ring simulations reported here. Still, the full
fledged confrontation of our algorithm with a conventional N-body code
will have to wait for ring simulations that are sufficiently resolved
($10^5$ rings, pursued over a hundred precession periods) to permit
detailed phase-space analysis, and comparison of the dynamics. These
simulations are surely within the reach of a parallel version of our
algorithm, running on a state-of-the-art cluster, and we hope to
report on such undertakings in the near future.

\section{Discussion}

\label{sec:disc}

\noindent
The algorithm we have described computes the secular or orbit-averaged
evolution of systems of particles in a Keplerian force field and
interacting via softened gravity.  It is accurate to first order in the masses
of the particles, and to all orders in eccentricity and inclination. The order
for a system of $N$ particles, which is $6N$ in a conventional $N$-body code
(assuming that momentum conservation is used to solve for the motion of the
central body), is reduced by $2N$ as a result of averaging and the consequent
conservation of semi-major axis. However, computing with angular momentum and
eccentricity vectors forces a redundancy as it resolves a degeneracy. In other
words, in order to facilitate the traversal of high eccentricity and
inclination regimes, we simulate a redundant $6N$-dimensional system of
equations. The algorithm uses analytic averaging over the orbit of the
perturbing particle (Gauss's method) and numerical averaging over the orbit of
the perturbed particle. The timesteps are now fractions of secular time-scales,
which are at least a factor of $M_\star/(Nm)$ longer than steps taken in
unaveraged simulations of equivalent accuracy. The price for this is that our
method cannot capture the effects of mean-motion resonances, which appear in
the secular dynamics at second order in the masses.

The Gaussian ring algorithm provides us with a powerful numerical tool for the
study of secular dynamics of nearly Keplerian N-body systems, especially those
with large eccentricities and inclinations (so that perturbation expansions
converge slowly or not at all) and/or crossing orbits. Our method resolves the
dominant, secular evolution of these particles with much greater accuracy and
speed than conventional N-body codes: in the examples we have examined the
typical timesteps are two to three orders of magnitude larger, depending on
the mass of the N-body cluster relative to the central mass (although the
calculations per timestep are more lengthy). Moreover a single ring should be
thought of as containing many stars, so the $N$ required for reliable results
should be much smaller.

We have discussed the implementation of the method
(\S\ref{sec:imp}). The accuracy goals of the algorithm are
encapsulated in two parameters, $\epsilon_{\rm quad}$ and
$\epsilon_{\rm int}$, the first expressing the desired fractional
accuracy of the numerical quadrature of the forces over the perturbed
ring, and the second expressing the desired accuracy of the
integration of the equations of motion. The accuracy of the quadrature
can be assessed by evaluating the rate of change of the semi-major
axis (\S\ref{sec:numav}), which should be zero in secular
dynamics. Provided one can live with moderate softening ($b\ga
0.01a$), between 16 and 128 force evaluations per ring should be
sufficient to yield $\epsilon_{\rm quad}\sim 10^{-11}$; 10 to 100 times that
number of force evaluations may be required for extreme
configurations or very small softening.

Much can be done to enhance the efficiency and accuracy of the method: (i)
adaptive quadrature routines can reduce the number of force evaluations
required for ring pairs experiencing a close approach; (ii) geometric integrators
for this most illustrious of Lie-Poisson systems should enhance the long-term
accuracy of the numerical integration; (iii) the cost of quadratures makes the
method eminently parallelizable in a distributed environment. 

JT acknowledges the generous hospitality of Stephane Colombi at the Institute
for Astrophysics in Paris, and a visiting position at the Institute for
Advanced Study. We also acknowledge support from NSF grants AST-0206038,
AST-0507401, and AST-0807432, and NASA grant NNX08AH24G. We thank Abdel
Hussein Mrou\'e, S.\ Sridhar, and Michael Balabane for discussions.

\appendix
\section{Halphen's cone: a delayed geometric interlude}

\label{sec:halphen}

\noindent
As \cite{H1901} put it: `Gauss first clearly indicated the role elliptic
functions play in this subject. Halphen has since presented the investigation
in an elegant manner.' Halphen's elegant approach is thoroughly geometrical,
based on the realization that the average force exerted by an eccentric
Keplerian ring at a given point depends only on the invariants of the
cone formed by ring and point. Halphen's approach has received numerous
renditions since; for example \cite{Mu1963} further simplifies Halphen's
computations by noting the role played by a natural dyadic in the
computation. Here we stick with the essentials of Hill's account, which is
closest in spirit (and date) to Halphen (and Gauss). We describe the
derivation for an unsoftened potential here, with the modifications required
by softening in Appendix \ref{app:halphensoft}.

The origin of the reference frame is best taken at the attracted
particle. Denote by $\bfr=(x, y, z)$ the attracting planet, and by
$\bfr_\star=(x_\star, y_\star, z_\star)$ the central body, with $r=|\bfr|$
(note the change in notation from the main text, where $\bfr$ denotes what is
here $\bfr-\bfr_\star$). The averaged direct force is
\beq 
[\bff']_{l} = \frac{Gm}{2 \pi}\oint \frac{\bfr(l) \, {\rm d}l}{{r}^3},
\eeq
where $l$ is the mean anomaly.  Equal area in equal time implies 
${\rm d}l/(2\pi) = {\rm d}\sigma/(\pi {a}^2\sqrt{1-{e}^2})$, $\sigma$ being the area of a
sector swept by the radius vector from the central body to the attracting
ellipse, and $\pi{a}^2\sqrt{1-{e}^2}$ the area of the ellipse. Let ${\bf
w}=(\bfr-\bfr_\star)\cross {\rm d}\bfr$, where ${\rm d}\bfr$ is the change in position of
the attracting body over some small time interval $dt>0$ as it travels around
its orbit.  Then ${\rm d}\sigma=\half|\bfw|$ and $h=\bfw\bcdot
\bfr_\star/|\bfw|$ is the perpendicular distance from the origin to the
plane of the orbit of the attracting body. Thus ${\rm d}\sigma = \half\bfw\bcdot
\bfr_\star/h$ and 
\beq 
[\bff']_{l} = \frac{Gm}{2\pi h{a}^2\sqrt{1-{e}^2}}\oint_C \frac{\bfr \, 
\bfr_\star\bcdot(\bfr-\bfr_\star)\cross {\rm d}\bfr}{{r}^3}
= \frac{Gm}{2\pi h{a}^2\sqrt{1-{e}^2}}\oint_C \frac{\bfr\,
  {\rm d}\bfr\bcdot\bfr_\star\cross \bfr}{{r}^3},
\label{eq:hal}
\eeq
where the curve $C$ is the orbit of the perturber. 

Consider for example the $x$-component of the force. Using Stokes's theorem
equation (\ref{eq:hal}) may be re-written
\beq 
[f_x]_{l} = \frac{Gm}{2\pi h{a}^2\sqrt{1-e^2}}\int {\rm d}{\bfs}\bcdot \del \cross 
  \left(\frac{x\,\bfr_\star\cross \bfr}{r^3}\right)=
  \frac{Gm}{2\pi ha^2\sqrt{1-e^2}}\int {\rm d}{\bfs}\bcdot\bfr\left({3x\,\bfr_\star\bcdot
  \bfr\over r^5}-{x_\star\over r^3}\right),
\eeq
where ${\rm d}{\bfs}$ is an element of surface area and the integral is over the
surface area bounded by the curve $C$, i.e., the area swept out by the vector
$\bfr-\bfr_\star$ in one orbit of the perturber. Now consider the cone having
vertex at the origin (the attracted particle) that passes through the curve
$C$, i.e., the set of all vectors parallel to $\bfr(l), 0\le l<2\pi$. The
curve $C$ lies on this cone, and if it is smoothly distorted to some other
curve on the cone the additional contribution to $F_x$ will be zero, because
the normal to the cone ${\rm d}{\bfs}$ at $\bfr$ is perpendicular to $\bfr$. We
conclude that the integral in equation (\ref{eq:hal}) is independent of the
curve $C$ so long as it lies on the cone. We may therefore choose a curve $C$
on the cone that simplifies the line integrals in (\ref{eq:hal}).

We write the equation of the cone in a reference frame with origin at the
attracted particle and axes parallel to the $\hat{\bfx}$, $\hat{\bfy}$,
$\hat{\bfz}$ axes defined in \S\ref{sec:regular} ($\hat{\bfx}$ points from the
central body to the periapsis of the orbit of the attracting particle;
$\hat{\bfz}$ is parallel to the angular momentum of this orbit, and
$\hat{\bfy}=\hat{\bfz}\cross\hat{\bfx}$). In these axes the position of the
central body is $\bfr_\star=x_\star\hat{\bfx}+y_\star\hat{\bfy}+ h
\hat{\bfz}$. The orbit of the attracting particle is described by the ellipse
$(1-e^2)(x-x_\star+ae)^2+(y-y_\star)^2=a^2(1-e^2)$ on the plane
$z=h$. Thus the equation of the cone is
\beq
(1-{e}^2)(xh-x_\star z+aez)^2+(yh-y_\star z)^2
-(1-e^2)a^2z^2=0,
\label{eq:hcone}
\eeq
or
\beq
{\bfr}^{\rm T}\bfssW\bfr=0,
\eeq
where $\bfr=x\hat{\bfx}=y\hat{\bfy}+z\hat{\bfz}$ and 
\beq
\bfssW\equiv \left(\begin{array}{ccc}
(1-{e}^2)h^2    & 0            & (1-{e}^2)h(ae-x_\star) \\
       0                  & h^2 & -y_\star h \\
(1-{e}^2)h(ae-x_\star) & -y_\star h & (1-e^2)(ae-x_\star)^2 +y_\star^2-(1-e^2){a}^2 \end{array}\right).
\eeq
This quadratic form can be diagonalized by an orthogonal
transformation $\bfssO$ to new Cartesian coordinates $\bfu=(u_1,u_2,u_3)$,
i.e., $\bfr=\bfssO\bfu$. Thus
\beq
{\bfssO}^{\rm T}\bfssW\bfssO=\mbox{\textsf{\textbf{diag}\,}}(\mu_i)
\eeq
and
\beq
{\bfr}^{\rm T}{\bfssW}\bfr={\bfu}^{\rm T}\hbox{\bf diag}\,
(\mu_1,\mu_2,\mu_3){\bfu}
\eeq
is the equation of the cone. The $\mu$'s are the eigenvalues of $\bfssW$ and
the roots of the function 
\bea
y(\mu)&=&\mu^3+\mu^2[(1-{e}^2){a}^2-{y_\star}^2-(2-{e}^2)h^2-(1-{e}^2)
({a}{e}-x_\star)^2] \nonumber \\
&&\qquad +\mu h^2(1-{e}^2)[{y_\star}^2+h^2
+({a}{e}-x_\star)^2-(2-{e}^2){a}^2] + (1-{e}^2)^2{a}^2h^4.
\label{eq:cubh}
\eea
We note that
\beq
y(0)=(1-e^2)^2a^2h^4>0, \quad y(h^2)=-e^2{y_\star}^2h^4<0,
\eeq
which implies that the roots are real, one negative and two
positive, and we may therefore assume that $\mu_1 > \mu_2 > 0 > \mu_3$.

The eigenvectors of $\bfssW$ may be written
\beq
{\bfh}^k=\theta_k\left({(1-{e}^2)h(ae-x_\star)\over 
 \mu_k-(1-{e}^2)h^2}, -{y_\star h\over\mu_k-h^2},1\right), 
\eeq
where $\theta_k$ is real, chosen so that ${\bfh}^k\bcdot{\bf
h}^k=1$.  We assume that $\theta_1$ and $\theta_2$ are positive and
then choose the sign of $\theta_3$ so that $\hat{\bfe}_1$, $\hat{\bfe}_2$,
$\hat{\bfe}_3$ form a right-handed triad, that is, $\hat{\bfe}_3=\hat{\bf
e}_1\cross\hat{\bfe}_2$. Then the orthogonal matrix $\bfssO$ is defined
by $O_{ij}=e_i^j$ and $\hbox{det}\,\bfssO=+1$.

The equation of the cone in the new coordinates is 
\beq
\mu_1u_1^2+\mu_2u_2^2=|\mu_3|u_3^2.
\eeq
We choose the curve $C$ to be the intersection of the plane $u_3=\hbox{const}$ 
with the cone. A parametric equation for $C$ is then
\beq
u_1=\alpha u_3\cos\psi,\quad u_2=\beta u_3\sin\psi,\quad 
\alpha=\sqrt{|\mu_3|/\mu_1},\ \beta=\sqrt{|\mu_3|/\mu_2}.
\label{eq:param}
\eeq
Then along $C$,
\beq
{r}^2=u^2=u_1^2+u_2^2+u_3^2=u_3^2(1+\alpha^2\cos^2\psi+\beta^2\sin^2\psi)
\eeq
and
\beq
{\rm d}\bfr\bcdot (\bfr_\star\cross\bfr)={\rm d}{\bfu}\bcdot ({\bfu}_\star\cross{\bfu})=
u_3^2 {\rm d}\psi[\alpha\beta u_{\star3}
-\alpha u_{\star2}\sin\psi-\beta u_{\star1}\cos\psi)].
\eeq

The curve $C$ as defined in equation (\ref{eq:hal}) is oriented
counter-clockwise as viewed by a distant observer on the positive
angular-momentum axis (the $z$ axis). This is the same orientation as the
curve defined by equation (\ref{eq:param}) if and only if $h$ and $u_3$ (the
distance of the integration plane from the origin) have the same sign. To
account for this we assume $u_3>0$ and replace the pre-factor $2\pi h$ in
equation (\ref{eq:hal}) by $2\pi|h|$. Then equation (\ref{eq:hal}) becomes
\beq 
[\bff']_{l} = \frac{Gm\alpha\beta}{2\pi |h|{a}^2\sqrt{1-{e}^2}}
\int_0^{2\pi}{\rm d}\psi{\hat{\bfu}_3u_{\star3}
-\hat{\bfu}_1u_{\star1}\cos^2\psi - \hat{\bfu}_2u_{\star2}\sin^2\psi\over 
(1+\alpha^2\cos^2\psi+\beta^2\sin^2\psi)^{3/2}}.
\eeq
The integral can be evaluated using equations (\ref{eq:ellint}), using the 
substitution $\psi=\half\pi-T$:
\beq 
[\bff']_{l} = \frac{2Gm}{\pi |h|{a}^2\sqrt{1-{e}^2}}{\mu_1^{3/2}
\sqrt{\mu_2-\mu_3}
\over(\mu_1-\mu_2)(\mu_1-\mu_3)}[(\kappa^2{\bfG}_a+{\bfG}_b)E(\kappa)
-(1-\kappa^2){\bfG}_bK(\kappa)]
\eeq
where
\beq
\kappa^2={|\mu_3|\over\mu_1}{\mu_1-\mu_2\over\mu_2-\mu_3}, \quad {\bfG}_a
=u_{\star3}\hat{\bfu}_3-u_{\star2}\hat{\bfu}_2, 
\quad {\bfG}_b=u_{\star2}\hat{\bfu}_2-u_{\star1}\hat{\bfu}_1.
\eeq
It is straightforward to demonstrate numerically that this expression for
$[\bff']_{l}$ is the same as that given by equation (\ref{eq:fav}), and
that the argument $k$ of the elliptic integrals in that equation
is equal to their argument here, $\kappa$.

\subsection{Halphen's cone, softened}

\label{app:halphensoft}

\noindent
At first look, it appears that Halphen's geometric treatment, which is
predicated on the homogeneity of the integrand, would fail when a length
scale, the softening length, is introduced in the problem.  However, the fact
that Gauss's algebra is flexible enough to handle softened interactions
suggests the opposite.  So one wonders whether, despite softening, Halphen's
derivation can be generalized to a softened potential. The answer is yes, and
following are the essentials of how.

With softening, equation (\ref{eq:hal}) is modified to 
\beq 
[\bff']_{l} = \frac{1}{2\pi  h{a}^2\sqrt{1-{e}^2}}\oint_C \frac{\bfr\,
  {\rm d}\bfr\bcdot(\bfr_\star\cross 
  \bfr)}{({r}^2+b^2)^{3/2}},
\label{eq:halsoft}
\eeq
where $b$ is the usual softening length. We now stretch the coordinates in the
directions $\hat{\bfx}$, $\hat{\bfy}$ parallel to the orbit plane of the
attracting particle, setting 
\beq
\bfr=x\hat{\bfx}+y\hat{\bfy}+z\hat{\bfz}=s\overline x\hat{\bfx}+
s\overline y\hat{\bfy}+z\hat{\bfz}, \qquad \overline{\bfr} = \overline
x\hat{\bfx}+ \overline y\hat{\bfy}+z\hat{\bfz}.
\eeq
Then along any small increment of the curve $C$ (the orbit of the attracting
particle) 
\beq
{\rm d}\bfr\bcdot(\bfr_\star\cross \bfr)=s^2
{\rm d}\overline{\bfr}\bcdot(\overline{\bfr}_\star\cross \overline{\bfr}).
\eeq
Furthermore on this curve 
\beq
{r}^2+b^2={x}^2+{y}^2+h^2+b^2=s^2(\overline x^2+\overline y^2)+h^2+b^2.
\eeq
If we now set
\beq
s^2=1+b^2/h^2\quad\hbox{then}\quad {r}^2+b^2=s^2\overline r^2.
\eeq
Thus equation (\ref{eq:halsoft}) becomes
\beq 
[\bff']_{l}=[\overline f_x\hat{\bfx}+\overline f_y\hat{\bfy}+ 
s^{-1}\overline f_z\hat{\bfz}]_{l}\quad\hbox{where}\quad 
[\overline{\bff}']_{l} = \frac{1}{2\pi
  h{a}^2\sqrt{1-{e}^2}}\oint_{\overline C} 
\frac{\overline{\bfr}\,{\rm d}\overline{\bfr}\bcdot(\overline{\bfr}_\star\cross 
  \overline{\bfr})}{\overline r^3};
\label{eq:halsofta}
\eeq
the contour $\overline C$ is obtained by shrinking the vectors from the attracted
particle to the star and from the star to the attracting particle by a factor
$s$ in the directions parallel to the orbit plane of the attracting particle. 

Since the integrals for $[\bff']_{l}$ in equation (\ref{eq:halsoft}) and
for $[\overline{\bff}']_{l}$ in equation (\ref{eq:halsofta}) are identical,
the derivation in the preceding section can be carried through with only minor
changes. The analog to equation (\ref{eq:hcone}) for the cone is 
\beq
(1-{e}^2)(xh-x_\star z/s+aez/s)^2+(yh-y_\star z/s)^2
-(1-{e}^2){a}^2{z}^2/s^2=0.
\eeq
The cubic equation (\ref{eq:cubh}) becomes 
\bea
y(\overline\mu)&=&\overline\mu^3+\overline\mu^2[(1-{e}^2){a}^2/s^2-
{y_\star}^2/s^2-(2-{e}^2)h^2 -(1-{e}^2)({a}{e}-x_\star)^2/s^2]
\nonumber \\
&&\qquad +\overline\mu h^2(1-{e}^2)[{y_\star}^2/s^2+h^2
+({a}{e}-x_\star)^2/s^2-(2-{e}^2){a}^2/s^2] + (1-{e}^2)^2{a}^2h^4. 
\eea
The eigenvectors of $\bfssW$ are
\beq
{\bfh}^k=\theta_k\left({(1-{e}^2)h(ae-x_\star)/s\over 
 \mu_k-(1-{e}^2)h^2}, -{y_\star h/s\over\mu_k-h^2},1\right), 
\eeq
The remaining formulae are the same, except that $[\bff']_{l}$ is to be
replaced by  $[\overline{\bff}']_{l}$, and $\mu_i$ is replaced by
$\overline\mu_i$. 

\section{General-relativistic precession}
\label{app:gr}

\noindent
We may correct Newton's equations with the dominant secular contribution from
general-relativistic effects. These effects are mainly felt by nearly
radial orbits on their close encounters with the central black hole. 
When averaged over the orbital period these corrections contribute a
Hamiltonian 
\beq
H_{\rm gr} = -\frac{3 (G{M}_{\star})^{2}}{a^2 c^2 \sqrt{1-e^2}}.
\eeq
The gradients are $\del_{\bfL} H_{\rm gr}=0$ and 
\beq
{\del}_{\bfA} H_{\rm gr} = -\frac{3 (G{M}_{\star})^{2}}{a^2 c^2}
\frac{\bfA}{{(1-e^2)}^{3/2}}.
\eeq

The secular equations of motion are \citep{tre08}
\beq
 \frac{{\rm d}\bfL}{{\rm d}t} = -\frac{1}{\sqrt{G M_{\star} a}} ({\bfA} \cross
\del_{\bfA} H+\bfL\cross\del_{\bfL} H), \quad 
 \frac{{\rm d} \bfA}{{\rm d}t} = -\frac{1}{\sqrt{G M_{\star} a}} ({\bfL} \cross
\del_{\bfA} H+\bfA\cross\del_{\bfL} H). 
\label{eq:eqmot} 
\eeq
To this order, the main effect is on precession of periapsis, leaving
eccentricity, orientation and semi-major axis of the orbit unaffected
\beq
\frac{{\rm d} {\bfA}}{{\rm d}t} = \frac{6\pi G{M}_{\star}}{c^2a
  P(1-e^2)^{3/2}}{\bfL}\cross{\bfA}.  
\eeq
where $P=2\pi a^{3/2}/(GM_\star)^{1/2}$ is the orbital period.

For a spinning (Kerr) black hole, the dominant secular effect is
Lense-Thirring precession. The relevant averaged Hamiltonian is \citep{llctf}
\beq
H_{\rm Kerr}={2(GM_\star)^{5/2}\over c^3a^{5/2}(1-e^2)^{3/2}}\bfS\bcdot\bfL,
\eeq
where $(GM_\star^2/c)\bfS$ is the spin angular momentum of the central black
hole and $|\bfS|\le 1$. Then
\beq
{\del}_{\bfL} H_{\rm Kerr}={2(GM_\star)^{5/2}\over
  c^3a^{5/2}(1-e^2)^{3/2}}\bfS,
\quad
{\del}_{\bfA} H_{\rm Kerr}={6(GM_\star)^{5/2}\over
  c^3a^{5/2}(1-e^2)^{5/2}}(\bfS\bcdot\bfL)\bfA.
\eeq
The equations of motion (\ref{eq:eqmot}) yield
\beq
\frac{{\rm d} {\bfL}}{{\rm d}t} =
\frac{4\pi(G{M}_{\star})^{3/2}}{c^3a^{3/2}P(1-e^2)^{3/2}}\bfS\cross\bfL, \quad 
\frac{{\rm d} {\bfA}}{{\rm d}t} =
\frac{4\pi(G{M}_{\star})^{3/2}}{c^3a^{3/2}P(1-e^2)^{3/2}}
    [\bfS-3\hat{\bfL}(\hat{\bfL}\bcdot\bfS)]\cross\bfA. 
\eeq

\end{document}